\pdfoutput=1
\documentclass[12pt]{article}

\usepackage{graphicx}
\usepackage{cite}

\newenvironment{sciabstract}{%
\begin{quote} \bf}
{\end{quote}}

\title{Molecular Assembly on Two-Dimensional Materials \footnote{This is slightly expanded version of the review published in \textbf{Nanotechnology 28(2017), 082001}}}

\author
{Avijit Kumar,$^{1\dagger}$ Kaustuv Banerjee,$^{1}$ Peter Liljeroth$^{1}$\\
\\
\normalsize{$^{1}$Department of Applied Physics}\\
\normalsize{Aalto University School of Science, P.O. Box 15100, 00076 Aalto, Finland}\\
\normalsize{$^\dagger$To whom correspondence should be addressed; e-mail:  avijit.kumar@aalto.fi.}
}

\date{}

\begin{document}

\maketitle

\begin{sciabstract}
Molecular self-assembly is a well-known technique to create highly functional nanostructures on surfaces. Self-assembly on two-dimensional (2D) materials is a developing field driven by the interest in functionalization of 2D materials in order to tune their electronic properties. This has resulted in the discovery of several rich and interesting phenomena. Here, we review this progress with an emphasis on the electronic properties of the adsorbates and the substrate in well-defined systems, as unveiled by scanning tunneling microscopy (STM). The review covers three aspects of the self-assembly. The first one focuses on non-covalent self-assembly dealing with site-selectivity due to inherent moir\'e pattern present on 2D materials grown on substrates. We also see that modification of intermolecular interactions and molecule-substrate interactions influences the assembly drastically and that 2D materials can also be used as a platform to carry out covalent and metal-coordinated assembly. The second part deals with the electronic properties of molecules adsorbed on 2D materials. By virtue of being inert and possessing low density of states near the Fermi level, 2D materials decouple molecules electronically from the underlying metal substrate and allow high-resolution spectroscopy and imaging of molecular orbitals. The moir\'e pattern on the 2D materials causes site-selective gating and charging of molecules in some cases. The last section covers the effects of self-assembled, acceptor and donor type, organic molecules on the electronic properties of graphene as revealed by spectroscopy and electrical transport measurements. Non-covalent functionalization of 2D materials has already been applied for their application as catalysts and sensors. With the current surge of activity on building van der Waals heterostructures from atomically thin crystals, molecular self-assembly has the potential to add an extra level of flexibility and functionality for applications ranging from flexible electronics and OLEDs to novel electronic devices and spintronics.
\end{sciabstract}

\newpage{}

\section{Introduction}
\label{sec:intro}

Molecular self-assembly on solid surfaces have been studied with an aim to create ultra-high density functional nanostructures which can be fabricated in a parallel fashion. Structure of such assembly is driven by the interplay of intricate intermolecular and molecule-substrate interactions. On weakly interacting homogeneous metallic surfaces, molecules typically interact with substrate only through weak van der Waals (vdW) interactions; the final structure of the molecular layer is then determined by intermolecular forces. For example, van der Waals forces result in close-packed assemblies where the symmetry is determined by the geometry of adsorbed molecules \cite{de2003two}, while slightly stronger and directional interactions (\textit{e.g.} hydrogen, metal-coordination or covalent bonds) lead to structure with more varied and open assembling motifs \cite{slater2011two,kudernac2009two}. This simple picture changes on interacting substrates, where the surface is not an innocent bystander \cite{rosei2003properties}. In extreme cases, native metal adatoms can be incorporated into the network \cite{stohr_prb_2014} or adsorbate induced reconstruction of the surface can lead to new self-assemblies \cite{Grobis2002,Bedwani2008}. Furthermore, the reactive $d_z$ orbitals of the metal often interact quite strongly with the molecules, adding an extra factor in the assembly process and changing the properties of the molecules compared to those in the gas phase. Lastly, in the case of bulk (metallic) substrates, molecular self-assembly cannot be used for a profound modification of the substrate electronic structure beyond doping or quenching of surface states.

Molecular self-assemblies can be fundamentally different in the case of two-dimensional (2D) materials, such as graphene, hexagonal boron nitride (hBN) and transition metal dichalcogenides (TMDCs), which have attracted enormous interest recently \cite{fiori2014electronics, mas20112d, das2015beyond}. The study of atomically thin crystals has been initially dominated by research on graphene. Graphene consists of $sp^2$ hybridized carbon atoms in a triangular lattice with a two-atom unit cell. The out-of-plane $p_z$ orbitals form delocalized bonding $\pi$ and antibonding $\pi^*$ bands. These bands touch at the $K$- and $K'$-points of the Brillouin zone (also known as the Dirac points) and gives the charge carriers a linear dispersion rendering graphene a zero band gap semiconductor \cite{neto2009electronic}. The rise of graphene has been closely followed by that of monolayer hBN. It is isostructural to graphene, with boron and nitrogen, instead of carbon, forming the basis of a similar triangular lattice (the lattice parameters of graphene and hBN differ by 1.8\%) \cite{pakdel2014nano}. Unlike graphene, hBN is an insulator with a band gap of $\sim$6 eV. Monolayers of TMDCs are the latest entrant to the family of 2D crystals and a lot of work is going on to unravel the unique properties of these atomically thin materials \cite{wang2012electronics}. Although the spectrum of 2D material is expanding rapidly \cite{gupta2015recent}, the study of molecular self-assembly has so far, to a large degree, been concentrated on graphene and hBN only.

Investigation of adsorbed single molecules and assemblies on 2D materials is important because of several reasons. Essentially all of the atoms on these materials are on the surface; hence, non-covalent and covalent functionalization can completely change their physical and electronic properties \cite{geim_sci_2009}. This has already been shown for graphene where covalent functionalization using hydrogen, oxygen, or fluorine has modified the electronic properties while the functionalized carbon atoms are rehybridized from $sp^2$ to $sp^3$ \cite{Hersam2013review,Geim2009graphane,Balog2010graphane}. Non-covalent functionalization using molecular adlayers has been suggested as means of tailoring graphene's band structure without compromising its excellent electronic properties, improving its potential application in electronics or sensors. It becomes increasingly important to study molecular assemblies on graphene on technologically relevant, insulating substrates \textit{e.g.}, silicon dioxide or bulk hBN for real device applications. Also at a very nascent stage of research, chemical functionalization of TMDCs have shown to alter their electronic properties \cite{hersam_acsn_2016}. In general, chemical functionalization is perceived as an important method to engineer metal-2D contacts in devices through dipole formation, charge transfer, energy band alignment, and orbital interactions. Further, molecular adlayers on 2D materials can bring novel applications that can be uniquely enabled by 2D materials. 	

Secondly, by the virtue of strong in-plane bonding, both graphene and hBN offer inert surfaces on which the fundamental principles of surface assembly can be verified. Close-packed molecular assemblies with weak intermolecular bonds or hydrogen-bonded networks have been studied extensively on graphene - reviews of such studies can be found in Refs.~\cite{macleod2014molecular,mali2015nanostructuring}. An important point here is that the bulk substrate supporting the atomically thin 2D material can have a significant effect on the molecular assembly. As we discuss in Sec.~\ref{sec:substrates}, the structural and electronic landscape of both epitaxial graphene and hBN is intimately dependent on the surface on which it is grown. Periodic, structural as well as electronic corrugation, on these surfaces can lead to highly site-selective molecular adsorption - examples will be given in Sec.~\ref{sec:site_adsorption}. Sec.~\ref{sec:cp_assembly} describes close-packed assembly on epitaxial graphene while Sec.~\ref{sec:assembly_engg} offers engineering aspects of assembly on 2D materials. The first layer of molecular network has often been used to form templates for multi-component, hierarchical growth of molecules - these phenomena will be discussed in Sec.~\ref{sec:template}. Studies on chemical reactions of molecules on graphene and hBN will be presented in Sec.~\ref{sec:reaction}.

Furthermore, due to their inertness, graphene and hBN decouple the adsorbed molecules from the supporting layer underneath. Similar effect has been observed earlier on ultrathin insulating layers of oxides and alkalihalides \cite{Ho2003fluor,repp2005molecules,Swart2011review} allowing the study of essentially isolated molecules. Low-temperature scanning tunneling microscopy (STM) studies make it possible to resolve intramolecular features; increased lifetime of injected electrons result in narrow spectral features and make it possible to visualize molecular orbitals in real space. In Sec.~\ref{sec:decoupling}, the rich field of the application of graphene and hBN as substrates in the study of single-molecules will be explored. Sec.~\ref{sec:orb_imaging} illustrates orbital imaging of molecules on 2D materials. On strongly site-selective assemblies the properties of the molecules are modulated according to the potential landscape of the underlying 2D material as will be shown in Sec.~\ref{sec:site_modulation}. Together with STM tip, molecules on 2D surfaces may constitute double barrier tunnel junction and molecules can be charged through tip-gating. This effect has been illustrated in Sec.~\ref{sec:tip_gating}. In certain cases, molecules physiosorbed on the surface of graphene and hBN can get charged, potentially resulting in quantum many-body effects \textit{e.g.} many-body excitations and the Kondo effect. The Kondo effect on epitaxial graphene will be the topic of discussion in Sec. ~\ref{sec:kondo}. 

After the discussion on molecular self-assembly on epitaxial layers of 2D materials, we turn to the technologically more relevant systems where the 2D-material has been deposited on insulating substrates. Direct, scanning-probe observation of molecular self-assembly on such substrates will be summarized in Sec.~\ref{sec:cp_assembly}. The effect on graphene on insulating substrate with the aim of final practical application has been studied using Photo-electron spectroscopy (PES), Raman spectroscopy, and current-voltage (I-V) measurements on graphene field-effect transistors (FETs). These results will be summarized in Sec.~\ref{sec:band_structure}.

This review will be limited to the studies of ordered adlayers of molecules on the surface of 2D material, primarily graphene and hBN. The emphasis is on the electronic effects of the assembled molecule and the surface underneath, principally investigated by STM in ultra-high vacuum conditions at low temperature. For a more general review of molecular assembly on graphene readers are directed to Refs.~\cite{macleod2014molecular,mali2015nanostructuring}. Non-covalent functionalization of graphene is a vast field and has been reviewed extensively in Refs.~\cite{liu2011chemical,zhang2011tailoring,mao2013manipulating,tang2013graphene,kong2014molecular}. Review of covalent approaches toward functionalizing graphene can be found in Refs.~\cite{Georgakilas2012review,Hersam2013review,Koehler2013review,Chua2013review,mali2015nanostructuring}. Finally, surface functionalization of TMDCs and two dimensional materials other than graphene has been reviewed in Refs.~\cite{Voiry2014covalent,Wang2015review,cai2015noncovalent,hersam_acsn_2016}.

\section{Two-dimensional materials}
\label{sec:substrates}

2D crystals are only one atomic layer thick and hence, for most practical purposes they have to be supported by a bulk substrate. The nature of the support and its interaction with the 2D material are important in determining the surface electronic structure and, through the molecule-substrate interactions, it also affects molecular self-assembly. Graphene, the first 2D material to be discovered, was first obtained by means of mechanical exfoliation of graphite onto silicon dioxide surface. Since then, other means of producing graphene (\textit{e.g.}, chemical exfoliation of graphite, chemical vapour deposition (CVD) on metals) and, transferring it onto insulating substrates (\textit{e.g.}, SiO$_2$, bulk hBN, polymers) have emerged \cite{bonaccorso2012production}. In recent years, graphene (produced by exfoliation or CVD) transferred onto bulk hBN has attracted a lot of attention as many properties of free-standing graphene are retained on hBN and it has emerged as the \emph{de facto} standard substrate in the highest performance graphene devices \cite{Dean2010hBN,yankowitz2014graphene}. The atomically smooth surface of graphene on hBN presents the ideal surface on which the assembling behaviour of various molecules can be tested. However, the transfer processes involved can result in contamination on the graphene surface, which naturally hinders precise studies of molecular self-assembly.

\begin{figure}[htb!]
    \centering
    \includegraphics[width=0.8\textwidth]{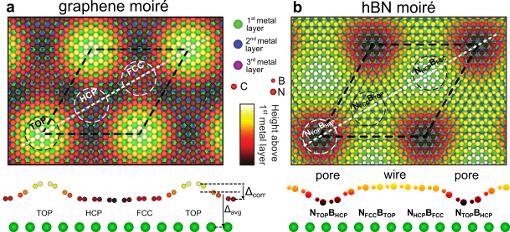}
    \caption{\textbf{2D epitaxial substrates.} (a) A schematic of moir\'e pattern in graphene due to lattice mismatch between graphene and underlying substrate. The lower panel shows corrugation profile of carbon atoms of the moir\'e along the white line. (b) A schematic of moir\'e pattern in hBN due to lattice mismatch between hBN and underlying substrate. The lower panel depicts corrugation profile along the while line across the moir\'e. Adapted with permission from \cite{Hamalainen2013LEED,Schulz2014BN}}
	\label{fig:overview}
\end{figure}

Graphene for studies on molecular self-assembly in well-defined and controlled environment (ultra-high vacuum, UHV) is mostly obtained \textit{in situ} via epitaxial growth on single crystals. Graphene can be grown by thermal sublimation of silicon from silicon carbide (6H-SiC(0001)) crystals \cite{norimatsu2014epitaxial} or by thermal decomposition of hydrocarbons on the hexagonal surfaces (FCC(111) or HCP(0001)) of transition metal single crystals \cite{batzill2012surface}. Recently, some methods for the direct growth of graphene on insulating substrates have also been presented \cite{Yang2013GhBN}. The substrate can have a significant effect on properties of the epitaxial graphene layer and the substrates can be classified into weakly and strongly interacting ones. Single layer graphene grown on SiC (G/SiC) is weakly coupled to the substrate with the graphene-substrate distance (3.3 \AA) \cite{varchon2007electronic} being close to the inter-layer separation in graphite. This indicates that the graphene interacts with the substrate only through van der Waals forces, which is reflected in the electronic properties being close to those of free-standing graphene \cite{ohta2007interlayer}. On the other hand, the interaction of graphene grown on metals depends on the substrate and has been discussed in detail in Refs.~\cite{wintterlin2009graphene,batzill2012surface}. Graphene-metal systems with a lattice mismatch gives rise to a long-range, periodic superstructure called the moir\'e pattern. The geometry and electronic properties of graphene vary over the moir\'e unit cell due to the periodically modulated carbon adsorption site w.r.t. the metal substrate. In the regions where the center of the carbon ring sits on top of the surface metal atom, the carbon-metal interaction is typically weak - at these points (so-called TOP-site) the graphene sheet is the farthest away from the metal surface. On the other hand, when alternate carbon atoms are positioned over the surface metal atoms, strong interaction between them leads to a lower adsorption height of the graphene sheet (valley site). In the low lying areas of the moir\'e further differentiation can be made on the basis of whether the carbon atoms not on top of metal atoms are on the FCC-hollow site (FCC-site), the HCP-hollow site (HCP-site), or in between these sites (bridge site) of the surface metal atoms. This is shown in Fig.~\ref{fig:overview}(a). Within the limit of this periodic variation, if the mean graphene-metal distance ($\Delta_{avg}$) is large, as in the case of graphene on iridium (G/Ir(111) $\sim$3.4 \AA) or platinum (G/Pt(111) $\sim$3.3 \AA), the resulting system is a weakly interacting \cite{Pletikosic2009mini,Sutter2009GPt,michely_prl_2011,Hamalainen2013LEED}. In these systems the topographic corrugation ($\Delta_{corr}$) over the moir\'e unit cell is small (\textless 0.5 \AA) and graphene retains its linear dispersion. Conversely, graphene grown on ruthenium (G/Ru(0001)) or rhodium (G/Rh(111)) interact strongly with the metal, and the graphene-metal distance varies from $\sim$2.1 \AA\ in the bridge sites (significant carbon-metal hybridization) to \textgreater 3.6 \AA\  at the TOP sites (free-standing graphene) \cite{Moritz2010GRu,wang2010coupling}. This is accompanied by a variation of work-function of 0.32 eV across the moir\'e unit cell and has profound implications in terms of adsorption energy of molecules at the moir\'e sites. The lattice matched graphene-nickel (G/Ni(111)) system which does not show any superstructure because of $1\times 1$ commensurate growth is another example of strongly interacting graphene \cite{Auwarter2003Ni,Dahal2014review,Drost2015Ni}. The electronic properties of graphene on strongly interacting metals are strongly modified, including the possible opening of a band-gap \cite{voloshina2012graphene}.

Epitaxial growth of hBN behaves in many ways similarly to that of graphene. The most widely used way to obtain hBN is by thermal decomposition of borazine (B$_3$H$_6$N$_3$) on transition metals; the surface properties of hBN are again dictated by the metal underneath \cite{diaz2013hexagonal}. Here too, lattice mismatch between the hBN and the metal substrate leads to moir\'e pattern as shown in Fig.~\ref{fig:overview}(b). In hBN-metal systems the region of strong interaction is limited to the sites where the N-atoms are on the top of the metal surface atoms. For other adsorption sites of the B and N atoms (\textit{i.e.}, TOP-FCC or HCP-FCC) the hBN-metal interaction is weak. Thus, the moir\'e of hBN on metals is characterized by periodically arranged, interacting ``pore'' areas separated by connected regions of almost free-standing ``wires''. The terminology of pores and wires stems from historical reasons \cite{Corso2004BN}; nevertheless, the hBN layer forms a full, atomically perfect layer without holes or other defects. The moir\'e topography is in contrast to graphene where connected valley sites separate isolated, periodic TOP sites \cite{brugger2009comparison} (compare Fig.~\ref{fig:overview}(a) to Fig.~\ref{fig:overview}(b)). The corrugation of the hBN ``nanomesh'' varies from metal to metal. If the mean hBN-metal distance is comparable to inter-layer distance in bulk boron-nitride ($\sim$3.33 \AA) as in the case of epitaxial hBN on Ir(111), Pt(111), or Cu(111), the topographic corrugation is small (\textit{e.g.}, $\sim$0.35 \AA\ for hBN/Ir(111)) \cite{Schulz2014BN,cavar2008single,Joshi2012BN}. For hBN grown on Ru(0001) or Rh(111) the minimum hBN-metal height is low ($\sim$2.5 \AA) and the resulting corrugation is quite high ($\sim$1 \AA) \cite{goriachko2007self,greber_sci_2008}. Irrespective of the supporting bulk metal, hBN shows a high electronic corrugation over the period of the moir\'e resulting in a significant difference in the local work-function between the wire and pore sites ($\sim$0.5 eV for hBN on Rh(111), Ru(0001) or, Ir(111) and, $\sim$0.3 eV for hBN/Cu(111)) \cite{diaz2013hexagonal}.

While the bulk support can have an effect on the properties of graphene and hBN, they are otherwise chemically relatively inert. Thus, molecules deposited on them interact with the 2D surface primarily through weak interactions like van der Waals forces, $\pi-\pi$  interactions or through surface dipoles in strongly interacting regime. This makes these surfaces quite attractive for studying the fundamentals of supramolecular self-organization. We will go through self-assembly on 2D materials in detail in the next section.

\section{Molecular assembly on 2D materials}

The strength of the molecule-substrate interaction (and its variation) is the key to tuning the molecular adsorption patterns. If the molecule-substrate interactions are weaker than the intermolecular attraction, a close-packed islands of molecules are readily observed also on surfaces exhibiting a moir\'e pattern. On the other hand, stronger site-specific molecule-substrate interaction leads to patterned adsorption. This competition between intermolecular and molecular-substrate interactions can be readily observed at low coverages.

\subsection{Site-selective adsorption}
\label{sec:site_adsorption}

\begin{figure}[h!]
    \centering
    \includegraphics[width=0.9\textwidth]{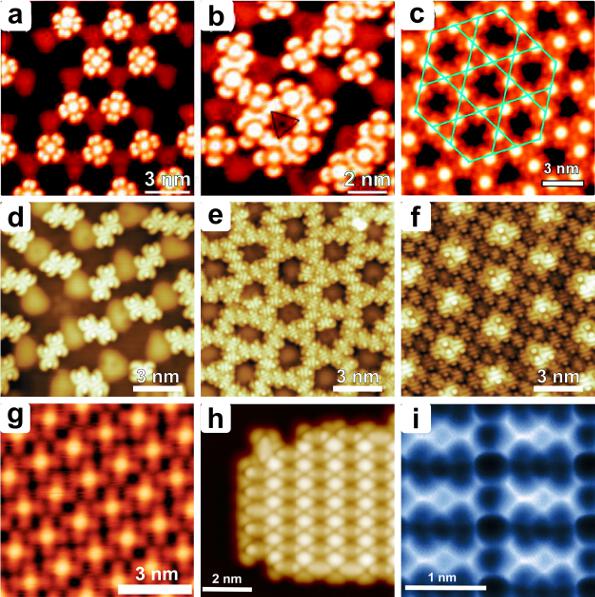}
    \caption{\textbf{Site-selective assembly on epitaxial graphene.} (a) Site-selective assembly of FePc molecules on G/Ru(0001) surface. The molecules occupy the FCC sites of the moir\'e. The TOP site of the moir\'e is brighter than the low lying area. (b) As the coverage is increased, the molecules go on to occupy the HCP sites while the TOP sites still remain occupied \cite{gao_prb_2011}. (c) At sufficiently high coverage, these molecules form Kagome lattice, highlighted in the image \cite{gao_jacs_2009}. (d) Similarly, TCNQ molecules adsorb on HCP and FCC areas at lower coverages and with increasing coverage they form (e) Kagome lattice and for even higher coverage they form (f) a continuous monolayer \cite{miranda_cm_2014}. (g) On the weakly interacting surface of G/Ir(111), CoPc form closed packed islands even at a sub-monolayer coverage \cite{liljeroth_jpcc_2014}. Due to reduced screening in graphene on hBN, F$_4$TCNQ assemble in a close-packed geometry at sub-monolayer coverage as evidenced by (h) STM and (i) non-contact atomic force microscope images \cite{Tsai2015}. Adapted with permission from \cite{gao_prb_2011, miranda_cm_2014, gao_jacs_2009, liljeroth_jpcc_2014, Tsai2015}.}
	\label{fig:image1}
\end{figure}

First we will focus on the site-selective adsorption on epitaxial graphene prepared on Ru(0001) which exhibits a strong work function modulation along with high geometric corrugation ($\sim$1.1 \AA) across the moir\'e pattern. Self-assembly of molecules such as free-base phthalocyanine(H$_2$Pc) \cite{gao_jacs_2009, gao_jpcc_2012_2}, FePc \cite{gao_jacs_2009,gao_prb_2011,gao_jpcc_2012_2,gao_jpcc_2012}, NiPc \cite{gao_jacs_2009,gao_jpcc_2012_2}, MnPc \cite{gao_jpcc_2012_2}, pentacene \cite{gao_prb_2011,zhou2013template}, PTCDA \cite{roos2011intermolecular,gao_apl_2011}, tetracyanoquinodimethane (TCNQ) \cite{Garnica2013,miranda_cm_2014} and fullerene (C$_{60}$) \cite{loh_acsn_2012,li2012self} have been studied on this surface. At lower coverage, the individual MPc molecules occupy FCC region of the moir\'e. As the molecular coverage is increased, all FCC regions are occupied, following which the molecules occupy edges of the TOP region, instead of the HCP region. For a certain coverage, this results in the formation of a Kagome lattice, where the TOP sites still remain unoccupied while the FCC and HCP sites are occupied. The coverage dependent assembly has been studied by Gao and coworkers \cite{gao_jacs_2009,gao_prb_2011} for FePc molecules (shown in Fig.~\ref{fig:image1}(a-c)). Other examples include H$_2$Pc, NiPc, FePc, MnPc molecules on G/Ru surface\cite{gao_jacs_2009,gao_jpcc_2012_2} which were shown to also form a Kagome lattice for right coverage.

Zhang \textit{et al.}\ \cite{gao_prb_2011} studied molecular self-assembly of FePc and pentacene molecules on G/Ru(0001) and put forward a mechanism related to the presence of surface dipoles. The authors used \textit{ab initio} calculations and found that the surface dipoles are formed because of the high corrugation present on G/Ru(0001) surface. Calculations suggest that the in-plane surface dipole (due to the work function modulation within the moir\'e unit cell) exists around the edge of the TOP region with the largest values along the TOP-FCC direction of the moir\'e. Therefore, molecules having significant polarizability tend to adsorb preferentially at FCC site. The enhanced stabilization of the molecules is the result of the interaction between the induced dipole moment of the molecule and the in-plane surface dipole \cite{nilius_jpcc_2008}. A large electric field in FCC-TOP direction due to the surface dipole is confirmed by the fact that pentacene molecules adsorbs exclusively along this direction at low coverage \cite{zhou2013template}. Density functional theory (DFT) calculations including vdW interactions demonstrated that the adsorption energy of the molecule was highest in this configuration. Increased deposition leads to sequential filling up of all FCC positions followed by the filling of the HCP sites -- even at high coverage of 0.7 monolayer (ML) the TOP regions are found to be unoccupied. The order of occupancy is again due to difference in adsorption energies, as confirmed by the DFT calculations. A site-dependence of the adsorption energy is also at the heart of hierarchical adsorption of C$_{60}$ molecules on G/Ru(0001) \cite{loh_acsn_2012}. Unlike the previously described molecules, the buckyballs start by populating the HCP valley sites to form heptamers before continuing onto the FCC sites and finally onto the TOP sites. A combination of decreasing adsorption energy and decreasing amount of charge transferred to C$_{60}$ from HCP to the TOP sites, as indicated by DFT calculations, was suggested as a reason for the hierarchical growth.

TCNQ molecules deposited on G/Ru(0001) also show site-selective adsorption \cite{Garnica2013,miranda_cm_2014}. Here, the adsorption mechanism is different than the dipole mediated adsorption as discussed earlier. TCNQ molecules are known to be strong electron acceptors with electron affinity of 2.8 eV \cite{Milian2004}. Therefore, such molecules tend to adsorb at the sites with higher electron density and lower work function \cite{Garnica2013}. As the work function is significantly lower for the HCP and FCC region compared to the TOP region, single TCNQ molecules adsorb on FCC and HCP regions at lower coverages.  It is also possible that higher reactivity of the FCC/HCP sites compared to the TOP regions also contributes to the patterned assembly. At slightly higher coverage all the FCC and HCP regions are covered while the TOP regions stay unoccupied creating a Kagome lattice. At full monolayer coverage the molecules occupy the TOP region as well. The coverage dependent site-selective adsorption is shown in Fig.~\ref{fig:image1}(d-f). 

\begin{figure}[h!]
    \centering
    \includegraphics[width=0.8\textwidth]{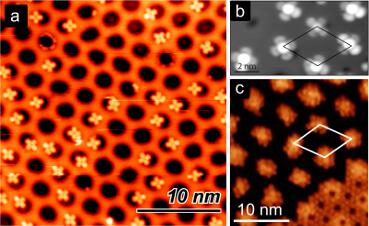}
    \caption{\textbf{Site-selective assembly on hBN.} (a) CuPc molecules adsorb preferentially on the pore site of hBN/Rh(111) nanomesh \cite{groning_pccp_2014}. (b) CoPc molecules preferentially adsorb on the wire sites of hBN/Ir(111) moir\'e \cite{liljeroth_acsn_2013}. (c) Extended self-assembled 2H-P islands form on hBN/Cu(111) confined to the wire site of moirons. The moir\'e unit cell is shown by the white rhombus. The islands can merge to form a continuous monolayer as seen in the bottom-right \cite{auwarter_acsn_2014}. Adapted with permission from \cite{groning_pccp_2014, auwarter_acsn_2014, liljeroth_acsn_2013}.}
    \label{fig:image2}
\end{figure}

Interestingly, close-packed assemblies at sub-monolayer coverage can be formed even on the surface of strongly corrugated graphene by post-deposition annealing. On the G/Ru(0001) surface, PTCDA has been observed to assemble into a herringbone arrangement with only some of the TOP sites remaining empty after heating the substrate \cite{roos2011intermolecular}. Similarly, pentacene \cite{zhou2013template} and C$_{60}$ \cite{li2012self} assemble into close-packed structures on G/Ru(0001) under high temperature deposition and annealing after deposition, respectively. Evidently, the higher temperature causes the molecules to be mobile, facilitating their assembly into a close-packed, ordered structure which is marked by a  higher adsorption energy per unit area than loosely-packed, disordered assembly.

Molecular self-assemblies on epitaxial hBN layer grown on metal surfaces also exhibit site-selectivity due to the presence of a moir\'e pattern. The presence of a significant geometric corrugation and periodic modulation of the work function across the hBN moir\'e also makes it an appealing substrate for templated molecular assembly. In addition, hBN is an insulator which implies that it can be used to decouple molecules from the underlying metal substrate \cite{liljeroth_acsn_2013,Ternes2013Mn12,liljeroth_natphy_2015}. A number of molecules such as C$_{60}$ \cite{Corso2004BN}, naphthalocyanine \cite{osterwalder_angc_2007}, CuPc \cite{greber_sci_2008,groning_pccp_2014}, H$_2$Pc \cite{groning_pccp_2014}, CoPc \cite{liljeroth_acsn_2013}, porphine \cite{auwarter_acsn_2014}, 5,5',5'',5''',5'''',5'''''-hexaiodocyclohexa-m-phenylene (I$_6$CHP) \cite{Groning2014radical}, TCNQ \cite{auwarter_acsn_2014} selectively adsorb on the pores of the hBN surface. Similar to the case of graphene, the adsorption mechanism is related to the presence of in-plane electric field due to the strong work function modulation. For  hBN grown on Rh(111) \cite{greber_sci_2008}, Ru(0001) \cite{goriachko2007self}, and Ir(111) \cite{Schulz2014BN} the work function difference amounts to $\sim$0.5 V while on Cu(111) it is 0.3 V. As discussed previously, the surface dipoles interact with the permanent or induced dipoles of the molecules thereby trapping them in the pores \cite{greber_sci_2008,groning_pccp_2014}. 

The maximum potential gradient exists at the edge of the pore where it can be as large as 1 V/nm. This can be a reason why molecules tend to adsorb close to the edge of the pore as shown in the Fig. \ref{fig:image2}(a-b). CoPc on  hBN/Ir(111) \cite{Schulz2014BN} and I$_6$CHP, H$_2$Pc, and CuPc on hBN/Rh(111) \cite{groning_pccp_2014} are examples of such off-center adsorption. For small molecules such as porphine and TCNQ on long-period hBN moir\'e on Cu(111) (periodicity 12 nm), it has been observed that small molecular aggregates nucleate at the pore site for low coverages \cite{auwarter_acsn_2014}. As the coverage is increased, the size of the molecular aggregates increases until they form a full monolayer. A similar observation has been made for Xe atoms on  hBN/Rh(111) surface \cite{greber_sci_2008,groning_nanosc_2010}. 

On systems with a shorter moir\'e period, preferential adsorption in the pore regions disturb the long-range order of the molecular layer. For example, deposition of nearly a full-monolayer of CoPc on hBN/Ir(111) results in short range order with a close-packed square lattice \cite{liljeroth_acsn_2013}. However, no long range order is observed due to the strong molecule-substrate interactions in the pores of the hBN moir\'e on Ir(111) and the mismatch in the preferred nearest-neighbour distance in the molecular layer and the moir\'e period.

While porphine exhibits site-selective adsorption on  hBN/Cu(111) surface \cite{auwarter_acsn_2014} as shown in the Fig.~\ref{fig:image2}(c), it is possible to tailor the intermolecular interactions by adding terminal groups to porphine molecules resulting in the loss of site-selectivity. In fact, carbonitrile-functionalized porphyrin (2H-TPCN) molecules deposited on hBN/Cu(111) \cite{Urgel2015coord} show close-packed islands as non-covalent intermolecular attraction (due to the presence of terminal cyano-bi-phenylene groups) dominates over the molecule-substrate interaction. Metallation of the macrocycle (adding Co atom at the center of macrocycle) does not change the molecular assembly despite the fact that the process includes heating the molecular assembly.

\subsection{Close-packed assembly} 
\label{sec:cp_assembly}

Site-selectivity is not expected on weakly interacting systems, where the work-function modulation and the geometric corrugation across the moir\'e unit cell are small \cite{Dedkov2014,Altenburg2014,liljeroth_prb_2011}. Precisely this has been observed for TCNQ \cite{miranda_cc_2010} and MPc molecules (e.g. CoPc\cite{sainio_jpcc_2012,liljeroth_jpcc_2014}, CuPc \cite{liljeroth_jpcc_2014}, and F$_{16}$CoPc \cite{liljeroth_jpcc_2014}) on G/Ir(111). An example of CoPc on G/Ir(111) is shown in Fig.~\ref{fig:image1}(g). While the work function modulation and geometrical corrugation are too small to cause site-selective assembly of molecules, the unit cell of the molecular lattice can differ slightly from ideal square packing \cite{sainio_jpcc_2012}. Close-packed molecular islands with the exact geometry being determined by the shape of the molecule is also observed on other weakly interacting metal-graphene systems \textit{e.g.}, square lattice of FePc on G/Pt(111) \cite{gao_jpcc_2012_2}, F$_{16}$CuPc on G/SiC \cite{wang2010selective}, CoPc on G/hBN \cite{liljeroth_nl_2013} and hexagonal packing of C$_{60}$ on G/Cu(111) \cite{jung2014atomically} and G/SiC \cite{guisinger_nl_2012,vsvec2012van}. Larger substrate corrugation can affect the lattice quality and domain size as demonstrated by CoPc self-assembly on G/SiO$_2$ and G/hBN. While CoPc forms a square lattice on both substrates, on G/SiO$_2$ the domain size is limited and there is disorder in the molecular ordering within the domains. On the geometrically smooth G/hBN substrate, the domain size is only limited by the size of the terraces of the underlying hBN \cite{liljeroth_nl_2013}.

On weakly interacting graphene stronger intermolecular interactions affect the self-assembly profoundly - \textit{e.g.}, PTCDA on G/Pt(111) \cite{rodriguez_jpcc_2014} and G/SiC \cite{huang2009structural,wang2009room,emery2011structural,alaboson2011seeding} and, PTCDI (perylene tetra-car\-box\-ylic di-imide) on G/SiC \cite{karmel2014self} assemble in close-packed herringbone structures to maximize intermolecular hydrogen bonds (C--H$\cdot\cdot\cdot$O and N--H$\cdot\cdot\cdot$O, respectively). Additionally, the close-packed islands of PTCDA \cite{wang2009room} and PCTDI \cite{karmel2014self} were shown to grow uninterrupted over step-edges and defects in the underlying graphene. These studies observed growth of multiple domains which were not aligned to the high-symmetry directions of graphene. These observations indicate that the molecular assembly on weakly coupled graphene is dictated by intermolecular interactions (vdW or H-bond) rather than molecule-substrate (vdW or $\pi$-$\pi$ or dipole) interactions.

In the case of graphene deposited on insulating substrates, the reduced screening can have an effect on the molecular self-assembly. This was demonstrated by Tsai \textit{et al.}, who showed that F$_4$TCNQ on G/hBN forms close-packed islands at sufficiently high coverages \cite{Tsai2015} (Fig.~\ref{fig:image1}(h-i)). The molecules pack in a head-to-tail fashion in compact islands, in contrast to F$_4$TCNQ on epitaxial G/Ru(0001) or G/Ir(111), where intermolecular repulsion dominates and neighboring molecules adopt a staggered geometry \cite{miranda_cc_2010,martin_ns_2014}.

Interestingly, close-packed assemblies at sub-monolayer coverage can be formed even on the surface of strongly corrugated graphene by post-deposition annealing. On the G/Ru(0001) surface, PTCDA has been observed to assemble into a herringbone arrangement with only some of the TOP sites remaining empty after heating the substrate \cite{roos2011intermolecular}. Similarly, pentacene \cite{zhou2013template} and C$_{60}$ \cite{li2012self} assemble into close-packed structures on G/Ru(0001) under high temperature deposition and annealing after deposition, respectively. The higher temperature causes the molecules to be mobile, facilitating their assembly into a close-packed, ordered structures corresponding to a higher adsorption energy per unit area.

\subsection{Assembly engineering}
\label{sec:assembly_engg}

\begin{figure}[h!]
    \centering
    \includegraphics[width=0.8\textwidth]{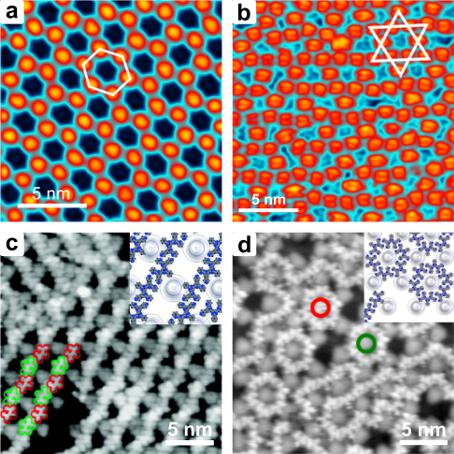}
    \caption{\textbf{Assembly engineering.} (a) CoPc molecules form honeycomb lattice on G/Ir(111) intercalated with Co. Here the moir\'e periodicity is same as that of pristine G/Ir. (b) On G/Co/Ir(111) surface with larger moir\'e periodicity (2.85 nm compared to $\sim$2.5 nm), the structure changes to Kagome lattice \cite{wiesendanger_acsn_2013}. (c) 2,4’-BTP molecules form long parallel chains along the symmetry direction of G/Ru \cite{roos2011hierarchical}. (d) On the other hand, 3,3’-BTP molecules form rings and short chains on the surface \cite{roos2011intermolecular}. The difference comes from the different position of the N atoms on the isostructural molecules which modifies the directional H-bonding. Adapted with permission from \cite{roos2011hierarchical, roos2011intermolecular, wiesendanger_acsn_2013}.}
    \label{fig:assembly_engg}
\end{figure}

It is possible to tailor molecular self-assembly by controlling the molecule-substrate interactions, which is also evident for site-selective assemblies on 2D materials. For example, the molecule-substrate interaction of MPc can influence the self-assembly even on the strongly corrugated G/Ru(0001) surface. Yang \textit{et al.}\ \cite{gao_jpcc_2012_2} demonstrated that for comparable coverage, a stronger (FePc), intermediate (MnPc), and weaker (NiPc and H$_2$Pc) molecule-substrate interaction leads to selective adsorption at FCC site only, molecular chains in the valley regions, and Kagome lattice, respectively. The molecule-substrate interaction can further be changed through metal intercalation in otherwise weakly interacting G/Ir(111). The intercalation of Co and Fe leads to an enhancement of the moir\'e corrugation (to $\sim$120-150 pm) \cite{Decker2013,Decker2014}. Strong moir\'e corrugation and modified chemical binding between graphene and the substrate result in increased molecule-substrate interactions. Bazarnik \textit{et al.}\ \cite{wiesendanger_acsn_2013} demonstrated that for low temperature deposition and at low coverages both CoPc and CuPc form one dimensional molecular chains in the lower lying region of the moir\'e formed on Fe (or Co) intercalated G/Ir(111). At slightly higher coverage, the molecules begin to form hexagonal rings around the TOP region as shown in Fig.~\ref{fig:assembly_engg}(a). For room temperature deposition at a higher coverage, the molecules assemble into slightly distorted honeycomb lattices with empty TOP sites. Interestingly, when the periodicity of the moir\'e increases due to intercalation of a different rotational domain of graphene on Ir(111), the assembled structure changes too. For intercalated graphene with a periodicity of 2.85 nm, the honeycomb lattice transforms into a Kagome lattice. An STM image of this assembly is shown in Fig.~\ref{fig:assembly_engg}(b).

Further, self-assembled structure can also by modified by tuning intermolecular interactions on a given surface. For assemblies stabilized by directional H-bond between the molecules, tuning the location of the electronegative species in the molecule can lead to completely different assembly. An example of this behaviour is that of the assembly of 2,4'-bis(terpyridine) (2,4'-BTP) \cite{roos2011hierarchical} (Fig.~\ref{fig:assembly_engg}(c)) and 3,3'-bis(terpyridine) (3,3'-BTP) \cite{roos2011intermolecular} (Fig.~\ref{fig:assembly_engg}(d)) on G/Ru(0001). The molecule 2,4'-BTP assembles exclusively into linear, parallel chains lying along the symmetry axis of graphene; on the other hand, 3,3'-BTP forms rings and short chains centered around the TOP site . In both cases, adjacent molecules are seen to lie in a anti-parallel fashion, indicating the formation of C--H$\cdot\cdot\cdot$N hydrogen bonds. The large disparity in the assembly is simply due to the difference in the position of the nitrogen atoms in these two molecules.

\subsection{Molecular templates}
\label{sec:template}

\begin{figure}[h!]
    \centering
    \includegraphics[width=0.9\textwidth]{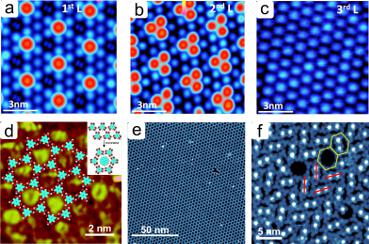}
    \caption{\textbf{Multi-layer templated assembly.} (a) C$_{60}$ molecular assembly on G/Ru(0001) showing contrast in accordance with the moir\'e pattern; the molecule lying on TOP region of the moir\'e is the brightest. (b) The contrast is still visible in the second layer; however, three C$_{60}$ molecules find places at hollow sites next to the brightest C$_{60}$ of first monolayer and appear bright. (c) On the third layer the influence of moir\'e is insignificant \cite{loh_acsn_2012}. (d) An otherwise close-packed assembly of trimesic acid (TMA) molecules on graphene on SiO$_2$ transforms into hexagonal structure on introduction of coronene molecules which settle in the pore of the assembly \cite{zhou2014switchable}. (e) Assembly of 1,3,5-benzenetribenzoic acid (BTB) molecules on G/Ir(111) showing long-range ordered network of hexagonal pores stabilized by H-bonding. (f) The flexible network hosts two CoPc molecules in each pore and CoPc assembles in a herringbone pattern; the red lines highlight the arms of the herringbone \cite{banerjee2016flexible}. Adapted with permission from \cite{loh_acsn_2012, zhou2014switchable, banerjee2016flexible}.}
    \label{fig:image3}
\end{figure}

The templating effect of graphene and hBN is not confined to a monolayer of molecules. 2nd layer of C$_{60}$ molecules deposited on G/Ru(0001) surface shows highly-ordered assembly following the corrugation of the moir\'e pattern shown in Fig. \ref{fig:image3}(a-b). However, the 3rd layer is no longer corrugated and reaches the bulk structure of C$_{60}$ as shown in Fig. \ref{fig:image3}(c) \cite{loh_acsn_2012}. The molecules in the second layer are electronically more decoupled from the substrate than the first layer. Naturally, molecules in the third layer will be even further decoupled. As another example, the molecules of the second layer of CoPc on hBN/Ir(111) align themselves with the pores of the underlying moir\'e pattern \cite{liljeroth_acsn_2013}. 

Multi-layer growth on weakly coupled graphene is dependent on the assembling motif of the first adlayer. As mentioned in Sec.~\ref{sec:site_adsorption} PTCDA on G/SiC assembles in a herringbone pattern, lying flat on the surface. The molecules in 2nd and 3rd layer too, lie flat and mimic the herringbone arrangement of the first layer \cite{huang2009structural}. The morphology of the thin film follows that of the underlying epitaxial graphene. For pentacene, on the other hand, although the first layer of molecules lie flat with their long-axis aligned along the graphene zigzag direction, the second layer prefers to stand upright in a herringbone fashion \cite{ahn_apl_2014} similar to on HOPG \cite{Chen2008}. 

On weakly interacting surfaces (graphite and close-packed noble metals) strong, directional inter-molecular interactions such as hydrogen-bonds and metal coordination bonds have been frequently utilized to obtain networks with periodic pores \cite{kudernac2009two}. Furthermore, such porous Molecular assembly with porous networks have often been used as a template to steer the assembly of other molecular species \cite{kudernac2009two,slater2011two}. Similar assemblies have been replicated on weakly coupled graphene to obtain nano-porous networks and use them in the so-called ``host-guest" architectures. For example, sequential deposition of PTCDI and the 3-fold symmetric molecule melamine has been used to form hexagonal network stabilized by intermolecular hydrogen bonds on G/SiC \cite{karmel2013self}. Hexagonal porous network stabilized by hydrogen bonds have also been realized by 3-fold symmetric tricarboxylic acids \textit{e.g.}, trimesic acid (TMA) and benzene-tribenzoic acid (BTB), which have been studied in detail on graphite and noble metal surfaces \cite{lackinger2009carboxylic}. In a comparative study of TMA assembly at the liquid-solid interface on G/SiC and graphite, the molecules were observed to assemble into a hexagonal network with a periodicity of $\sim$1.6 nm and identical epitaxy on both surfaces \cite{macleod2015substrate}. In another study carried out at the solid-liquid interface on exfoliated graphene on silicon dioxide \cite{zhou2014switchable}, deposition of TMA resulted in a close-packed structure. On subsequent deposition of coronene, the assembly changes to a honeycomb arrangement of TMA (periodicity $\approx$ 1.7 nm) with the coronene occupying the pores, as shown in Fig.~\ref{fig:image3}(d). This is an interesting example of the incorporation of the guest molecule causing a phase-change of the host network. 

After deposition in UHV, 1,3,5-benzenetribenzoic acid (BTB) assembles into an extended hexagonal mesh with a periodicity of $\sim$3.2 nm on G/Ir(111) \cite{banerjee2016flexible}. The network is quite stable and robust (Fig.~\ref{fig:image3}(e)) with an estimated bonding energy of 1.95 eV per molecule, extending for hundreds of nanometers over step-edges of the underlying substrate. The periodic pores of the nanomesh can be used to pattern molecular assembly as demonstrated with CoPc. Remarkably two CoPcs could be hosted in a single pore without destroying the network - the strong yet flexible hydrogen bonds holding the network together undergo stretching to accommodate the guest molecules. When all pores of the BTB network are uniformly occupied with two CoPcs, the guest molecules are seen to arrange in a herringbone pattern as shown in Fig.~\ref{fig:image3}(f). The formation of the herringbone pattern is driven by the reduction of the energy cost associated with stretching and twisting the hydrogen bonds between the BTB molecules.

As mentioned in the previous section, the high site-selectivity of molecules on strongly corrugated graphene leads to formation of open networks. Periodic, porous networks such as the Kagome lattice formed by assembly of MPc's on the surface of G/Ru(0001) \cite{gao_jacs_2009} can, and have been used as a template for deposition of guest molecules. Zhang et al. \cite{gao_jpcc_2012} used the FePc Kagome lattice to host tert-butyl zinc phthalocyanine ((t-Bu)4−ZnPc). On strongly interacting graphene, a complete phase change of the host network is observed if the guest molecule interacts more strongly with the substrate then the host ones. A Kagome lattice consisting of H$_2$Pc which interacts weakly with the underlying G/Ru(0001) cannot accommodate a strongly interacting FePc at the hollow sites; rather FePc molecules adsorb on the FCC site by displacing and distorting the existing Kagome lattice\cite{gao_jpcc_2012_2}. This again highlights the nature of these soft, deformable molecular templates, where the template symmetry can be altered in accommodating the guest molecules.

\subsection{Covalently bonded structures}
\label{sec:reaction}

So far, we have only discussed non-bonded or at most weakly-bonded (hydrogen bonding) assemblies. We will now focus on strongly bonded molecular structures on 2D-materials. These are interesting systems on several accounts. First of all, these again present opportunities to modify the electronic and other properties of the 2D-materials in a controlled and tuneable fashion. Secondly, there is a strong fundamental interest in understanding chemical reactions on weakly interacting, non-reactive surfaces in contrast to the well-studied metal substrates (e.g. coinage metals). Finally, the 2D-material can form a gateable substrate that yields an entirely different angle into tuning the doping in molecular materials \cite{riss2014imaging}. 
 
In addition to covalent and coordination chemistry on 2D-materials, there has been enormous attention also on covalent chemistry with the 2D materials. One of the initial motivations was in opening a band gap in graphene by hydrogenation, i.e. turning graphene into graphane \cite{Geim2009graphane,Balog2010graphane}. Even before that, a big part of the graphene effort was generated by graphene synthesis from graphene oxide by chemical reduction \cite{Georgakilas2012review}. We will not cover these topics in detail in this review and the reader is directed to existing literature on covalent functionalization of graphene and other 2D materials \cite{Georgakilas2012review,Hersam2013review,Koehler2013review,Chua2013review,Voiry2014covalent,Wang2015review}.
 
There has been intense interest on on-surface synthesis of covalently linked molecular assemblies and large macromolecules. Several different coupling mechanisms have been demonstrated on metal single crystal substrates, where the catalytic activity of the metal substrate is important. One of the better studied mechanisms is Ullmann-type coupling, where first a halogen - carbon bond in a precursor is cleaved on the substrate, and the resulting radicals react through the formation of carbon-carbon bonds \cite{Xi1992Ullmann,Hla2000Ullmann,Grill2007covalent,Cai2010GNR}. This has  been used to carry out on-surface polymerization \cite{Grill2007covalent,Laffarenz2009polymer}, and in particular, synthesize atomically well-defined graphene nanoribbons of different widths and edge terminations \cite{Cai2010GNR,Chen2013GNR,Cai2014hetero,Kimouche2015GNR,Chen2015mixed,Ruffiex2016zigzag,Talirz2016review}. In the case of graphene nanoribbons, the formation process proceeds in two thermally activated steps on a metal substrate (typically Au(111) or Ag(111)). In the first step, a covalently bonded polymer is formed, which is converted into a fully aromatic nanoribbon through a cyclodehalogenation step at a higher temperature\cite{Cai2010GNR}. The activation energies (at which temperatures the reaction proceeds) of both of these steps are influenced by the substrate, i.e. its catalytic properties are important.

\begin{figure}
	\centering
	\includegraphics[width=0.9\textwidth]{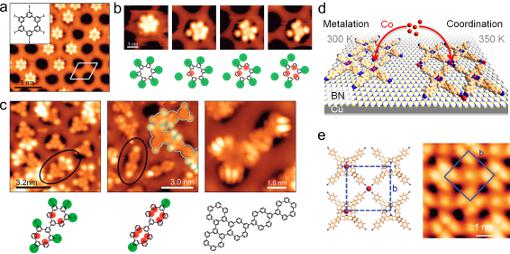}
	\caption{\textbf{Covalent reaction on surface.} (a) I$_6$CHP adsorption on hBN on Rh(111). (b) Step-by-step dehalogenation of a single I$_6$CHP induced by voltage pulses from the STM tip. (c) Dehalogenation and subsequent cross-linking following thermal annealing to 550 K (left), 800 K (middle) and 850 K (right) \cite{Groning2014radical}.  (d) Schematic showing two ways of metal complexation of 2H-TPCN. At room temperature the Co atoms go to the tetrapyrrole pocket whereas on annealing to 350K they coordinate to the --CN peripheral group giving a metal-organic coordination network. (e) Structure (left) and small-scale STM image (right) of the metallosupramolecular network \cite{Urgel2015coord}. Adapted with permission from \cite{Groning2014radical,Urgel2015coord}. }
\label{fig:I6-CHP}
\end{figure}

The difficulties in using the radical coupling on surfaces other than bulk metals are highlighted by Dienel \textit{et al.}\ \cite{Groning2014radical}. They describe the reactivity of 5,5',5'',5''',5'''',5'''''-hexaiodocyclo- hexa-m-phenylene (I$_6$CHP) on epitaxial hexagonal boron nitride on Rh(111) (Fig.~\ref{fig:I6-CHP}(a-c)). I$_6$CHP is typically very reactive on metal surfaces and dissociates all its iodine atoms on Cu(111), Ag(111), and even Au(111) already at room temperature \cite{Bieri2010CHP}. The situation is different on hBN-covered metal surface as shown in Fig.~\ref{fig:I6-CHP}(a), where is can be seen that following deposition at room-temperature, I$_6$CHP remains undissociated. The reactivity of the I$_6$CHP was tested on the single molecule level using voltage pulses from the tip of the STM as shown in Fig.~\ref{fig:I6-CHP}(b) , which yielded two important observations: the iodine atoms dissociate at alternating positions around the molecule and not more than three iodine atoms can be removed. The formed radicals are immediately stabilized by a bond formation of the radical site with the underlying boron atom \cite{Bacle2014I6CHP}. This locks the molecule in position and consequently favours the remaining alternating sites for dehalogenation.

Dehalogenation can also be induced by thermal annealing \cite{Groning2014radical}. After relatively mild annealing (up to 550 K for 20 min), the molecules still remain mostly oriented along a specific crystallographic direction of the underlying hBN surface and again, at most three iodines are cleaved Fig.~\ref{fig:I6-CHP}(c) (left panel). After annealing to even higher temperatures (800 K and 850 K for 30 minutes, middle and right panels of Fig.~\ref{fig:I6-CHP}(c)), more iodine can be removed. Only at annealing temperatures of 850 K was it possible to produce iodine-free oligomers. However, the resulting structures do not yield similar long-range order that can be observed on metal substrates after similar annealing \cite{Bieri2010CHP}. This highlights the advantages on metal surfaces, where the radicals are stabilized by bonding with the metallic substrate, while the precursor molecules can still retain sufficient mobility to yield high quality structures. Nevertheless, aryl-aryl coupling similar to the metal-mediated Ullmann coupling can be achieved also on non-metallic substrates \cite{Groning2014radical}.

Dehalogenation with brominated precursor molecule 1,3,5-tris(4-bromophenyl)-benzene (or, TBB) on hBN and graphene on Ni(111) has also been attempted \cite{Kern2015hBN}. Here, the inherent reactivity of TBB precursor is smaller than that of the iodinated I$_6$CHP. Nevertheless, annealing the sample to 520-570 K results in debromination of the precursor molecules, which subsequently form covalently bound assemblies. However, these lack the long range order typically observed on assemblies on coinage metals. There, the same precursor molecule forms extended polymeric networks presumably due to dehalogenation and C--C coupling occurring at lower temperatures. Similarly to I$_6$CHP, limited mobility of the dehalogenated molecules due to a strong interaction with the hBN and graphene surfaces is likely to be responsible for the lack of long-range order. The strong interaction was further substantiated by the observation of single molecules pinned to the surface and DFT calculations that also show a strong interaction between the phenyl unit and graphene/Ni(111) explaining the reduced diffusion of debrominated TBB molecules \cite{Kern2015hBN}. Finally, the authors argue that the order and spatial extension of the cross-linked molecules can be improved by a suitable choice of supporting metal surface for graphene and hBN, which would be more weakly interacting than Ni(111).

In contrast to the covalently bonded assemblies, metal coordination bonds have been demonstrated to yield molecular assemblies with long-range order on non-reactive hBN \cite{Urgel2015coord}. Urgel and co-workers investigated carbonitrile-functionalized porphyrin derivatives (2H-TPCN) cross-linked with Co atoms on epitaxial hBN on Cu(111) (Fig.~\ref{fig:I6-CHP}(d-e)).  Here, the molecule self-assembles into a densely-packed layer with a square unit cell. After evaporation of cobalt atoms onto the substrate at room temperature, metalation of the porphyrins is observed \textit{i.e.}, Co-TPCN is formed. Coordination networks can be formed once submonolayer coverages of TPCN is exposed to Co atoms at 350 K \cite{Urgel2015coord}. Two different packing schemes are formed at this temperature: a densely-packed array representing the pure molecular phase and a network structure exhibiting a larger square unit cell and a central protrusion (blue square in Fig.~\ref{fig:I6-CHP}(e)); the latter structure was assigned to a Co-directed assembly of a metal−organic coordination network. Surprisingly, they have 4-fold coordination in contrast to what is observed on metal substrates. This study demonstrates that it is possible to form metal-coordination networks with long-range order on non-metallic substrates, in this case hBN on Cu(111). 

Single-molecule reactions have also been investigated, for example, dehydrogenation of free-base phthalocyanine (H$_2$Pc) on epitaxial graphene \cite{Neel2016H2Pc}. In this particular case, graphene was used as an inert substrate that does not participate in the removal of the inner core hydrogens. Importantly, the formed radical did not bond with the underlying graphene, which allowed the study of the molecule in the same adsorption geometry before and after dehydrogenation. Similar reaction has been demonstrated on a metal surface, where the radical formation was associated with a molecular reconfiguration \cite{Auwarter20124level}. Another single molecule reaction, this time with the underlying graphene was the novel cycloaddition reaction observed to occur between various different phthalocyanines and epitaxial graphene on Ir(111) \cite{Berndt2015FePc}. This reaction could be driven reversibly by voltage pulses from the tip of a low-temperature STM. The reaction only occurs on a certain part of the G/Ir(111) moir\'e, which suggests that the bonding of the graphene with the underlying Ir(111) results in dangling bonds on the graphene surface.

\section{Tunneling spectroscopy on 2D materials}

\label{sec:decoupling}

Molecular adsorption on metal surfaces is a complex and interesting phenomenon. The surface chemistry of molecules on metals depend on the nature of the bond between the molecules and the surface. For molecules physisorbed on the surface through vdW interactions, the electronic properties of the molecule are similar to that of the molecule in the gas phase . However, for molecules adsorbing on metal surfaces through strong chemical bonds (\emph{e.g.} covalent, metallic, and ionic bonds), the electronic properties are drastically modified \cite{bligaard_pnas_2011}. An extensive review on molecule-metal bonds has been provided by Nilsson \textit{et al.}\ \cite{moody_ssr_2004}. Molecular orbitals of molecules adsorbed on metal surfaces are strongly hybdridized (resulting in broadening, shifting, and mixing of the orbitals) because of the interaction with the continuum of electronic states in the metal \cite{altibelli_cpl_1993,bligaard_pnas_2011}. Consequently, a collection of molecular orbitals contributes to the tunneling current in STM/STS studies despite the fact that the pristine molecular orbitals are well-separated from each other as well as from the substrate Fermi level \cite{joachim_ss_2009}. This complicates the interpretation of STM images of molecules on metal substrates and typically, the spectroscopic signatures of the molecular orbitals are too broad to be reliably detected \cite{altibelli_cpl_1993}. Therefore, it is imperative to decouple the molecules electronically from the metal substrate in order to investigate their pristine electronic properties\cite{joachim_ss_2009}. While STM/STS cannot be carried out on insulating substrates, the use of ultrathin insulating layers between the molecules and the metal substrate allows combining insulating surfaces and STM. The insulating layer decouples the molecules electronically from the metal surface while the layer ($\sim$ few \AA) is still sufficiently thin to allow STM operation without difficulties. Ultrathin layers of conventional insulating materials, for example, alkali-halides \cite{Repp2005,persson_sci_2006,meyer_sci_2007}, metal-oxides \cite{Ho2003fluor,ho_jcp_2004,nilius_jpcc_2008,kawai_prl_2009}, metal-nitrides \cite{heinrich_sci_2007,pasa_jpcc_2013}, passivated semiconductors \cite{joachim_nl_2009,szymonski_acsn_2013} have been used to study the properties of single atoms, single molecules, and various aspects of molecular electronics in general. Usage of two-dimensional films of noble gases \cite{repp_prl_2013,repp2010coherent}, molecules \cite{millo_nat_1999,kahn_cpl_2002,zhu_prl_2005,zhu_jcp_2006,berndt_jpcc_2011} also serve similar purpose of decoupling molecules from the metal surface.

The decoupling makes it possible to study molecules whose properties closely match those found in the gas phase (\emph{e.g.} magnetic moment \cite{Ternes2013Mn12}, fluorescence \cite{Ho2003fluor}). In addition, it helps more precise tunneling spectroscopy experiments as the increasing electron residence time leads to sharper molecular resonance peaks \cite{ho_pnas_2005,persson_prl_2005,zhu_jcp_2006}, and the observation of vibronics \cite{ho_pnas_2005,repp2010coherent} in the STM/STS measurements. In a seminal experiment, Repp \textit{et al.}\ \cite{Repp2005} demonstrated that few layers of NaCl is sufficient to electronically decouple pentacene molecules from the underlying Cu(111) surface. STS on the molecule leads sharp resonances in the differential conductance (d$I$/d$V$) spectra corresponding to temporary electron addition (positive sample bias) and electron removal (negative sample bias) to/from the molecule. These peaks are also termed positive (PIR) and negative ion resonances (NIR) referring to the transient occupation of the molecular orbitals \cite{Repp2005,Swart2011review}. The first PIR at negative and the first NIR peak at positive bias correspond to the electron tunneling through the Highest Occupied Molecular Orbital (HOMO) and the Lowest Unoccupied Molecular Orbital (LUMO) of the molecule, respectively. While these peaks are often referred to HOMO and LUMO, energy difference between the PIR and NIR is not equal to the HOMO-LUMO gap. This so-called transport gap is larger than the HOMO-LUMO gap due to the Coulomb charging energies associated with adding/removing electrons to/from the molecule \cite{pascal_cpl_2000,van2010charge}. It is more correct to think of the PIR and NIR as the ionization potential and the electron affinity of the molecule, respectively. In the simple case when the molecules remain neutral on the surface, PIR and NIR values extracted from the STS and the gap between them depends on the energy level alignment of the molecular orbitals with respect to the substrate Fermi level and screening effects due to the polarization of the substrate and the neighbouring molecules (see  \cite{seki_advmat_1999,kahn_cpl_2002,louie_prl_2006,fahlman_advmat_2009} for details). 

The very exciting aspect of STM is that it allows - in addition to measuring the energy positions of the molecular resonances - mapping out the molecular orbitals in real space. STM topographic images at bias values of PIR and NIR on a molecule deposited on a decoupling layer resemble the HOMO and the LUMO orbitals of gas phase molecule as shown by Repp \textit{et al.}\ \cite{Repp2005}. These orbitals can be imaged with very high spatial resolution, especially using molecular modified tips \cite{Repp2005,persson_prl_2011,meyer_sci_2007}. Depending on the nature of the insulating film, the peak widths of the molecular resonances can be significantly larger than the expected broadening due to finite life-time or temperature. For example, on ionic films, the tunneling electrons couple strongly with the optical phonons of the insulator, which leads to significant broadening of the resonances \cite{Repp2005,persson_prl_2005,ho_pnas_2005}.

\begin{figure}[h!]
    \centering
    \includegraphics[width=0.9\textwidth]{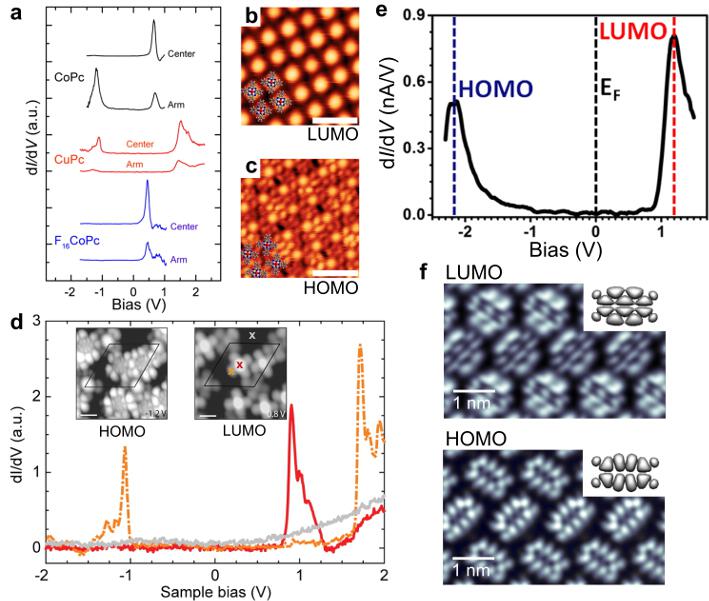}
    \caption{\textbf{Molecular orbital imaging.} (a) dI/dV spectra of CoPc, CuPc, and F$_{16}$CoPc molecules on G/Ir(111) recorded at the center and arm of the molecules. Peaks at the positive and negative biases correspond to tunneling into LUMO and HOMO, respectively. STM images recorded at the respective bias show LUMO and HOMO orbitals shown in (b) and (c), respectively. Scale bars are 3 nm. \cite{liljeroth_jpcc_2014} (d) STS recorded on CoPc molecule adsorbed on the wire site of hBN/Ir(111) moir\'e. Insets show STM images at -1.2 V and 0.8 V resembling HOMO and LUMO orbitals, respectively. Peak at 1.9 V may correspond to tunneling into LUMO+1 orbital. \cite{liljeroth_acsn_2013}. (e) dI/dV spectrum of PTCDA molecules on G/Pt(111) showing molecular resonances of LUMO and HOMO. (f) STM images recorded at 1.34 V and -2.3 V show respectively HOMO and LUMO orbitals resembling the calculated orbitals \cite{rodriguez_jpcc_2014}. Adapted with permission from \cite{liljeroth_jpcc_2014, liljeroth_acsn_2013,rodriguez_jpcc_2014}.}
    \label{fig:image4}
\end{figure}

\subsection{Orbital decoupling and orbital imaging on 2D materials}
\label{sec:orb_imaging}

Similar to the above mentioned insulating layers, epitaxial graphene and hBN are also able to efficiently decouple the molecules from the underlying metal substrates by virtue of their chemical inertness and low density of states around the Fermi level. They can be used as a supporting surface to study the properties of single atoms and molecules for various aspects of molecular electronics. A number of molecules such as C$_{60}$ \cite{guisinger_nl_2012,jung2014atomically}, TCNQ \cite{Garnica2013}, MPcs \cite{liljeroth_jpcc_2014, kroger_jcp_2014}, and PTCDA \cite{huang2009structural,gao_apl_2011,rodriguez_jpcc_2014} adsorbed on epitaxial graphene indicate that it can indeed be used to decouple adsorbed molecules from the underlying substrate. J\"{a}rvinen \textit{et al.}\ \cite{liljeroth_jpcc_2014} have carried out STS and orbital imaging of CuPc, CoPc and F$_{16}$CoPc molecules deposited on G/Ir(111). For each of the molecule, they observed a HOMO and a LUMO peak in the dI/dV spectra as shown in Fig.~\ref{fig:image4}(a). The LUMO orbital is located at 750 mV, 1400 mV, and 500 mV while HOMO orbitals at -1000 mV, -1300 mV, and -1400 mV for CoPc, CuPc, and F$_{16}$CoPc molecules, respectively. The value of this transport gap (HOMO-LUMO gap measured on G/Ir(111)) should be smaller than in thick molecular films due to the additional screening from the graphene and the underlying metal substrate. As expected, the STM gap for CoPc (1.75 eV), CuPc (2.7 eV) on G/Ir(111) are smaller than the transport gap of $\sim$3.2 eV for thick films of each of CoPc and CuPc measured using PES and IPES \cite{peter_jpcm_2009}. As discussed above, imaging at the bias voltages corresponding to the molecular resonances can be used to study the molecular orbital structure of each of the molecules. Fig.~\ref{fig:image4}(b-c) shows LUMO and HOMO images of CoPc which closely resemble the DFT calculated orbitals for the gas-phase molecule. For CoPc molecule, the HOMO orbital is delocalized over the $\pi$ system of the carbon backbone while the LUMO orbital is localized on the central Co atom. While the molecules interact only weakly with the graphene surface, (local) variations of the graphene Fermi level and chemical reactivity can lead to charge transfer and shift the molecular resonances \cite{michely_prl_2011,riss2014imaging}. For example, the moir\'e structure of G/Ir(111) leads to a modulation of molecular resonance peaks by 200 mV \cite{liljeroth_jpcc_2014}. In an extreme case, the charge transfer can result in permanent charging of the molecule if its gas phase LUMO is below the graphene Fermi level. This does not prevent imaging the molecular orbitals; for example, TCNQ on G/Ru(0001) molecular orbitals can still be easily resolved even though a charge transfer takes place between the molecule and G/Ru(0001) substrate \cite{miranda_nl_2014}.

Similarly, Mart\'{i}nez-Galera \textit{et al.}\  \cite{rodriguez_jpcc_2014} demonstrated that graphene decouples PTCDA from Pt(111) surface as seen from the sharp HOMO and LUMO resonance peaks (recorded at -2.2 eV and 1.2 eV, respectively as shown in Fig.~\ref{fig:image4}(e)). Here too, each of HOMO and LUMO orbital images closely resemble the respective calculated orbital images as shown in Fig.~\ref{fig:image4}(f). Zheng \textit{et al.}\ \cite{zheng2016heterointerface} observed that the HOMO-LUMO gap of PTCDA molecule increases in size from Au(111) (3.1 eV) to HOPG (3.5 eV) to semi-conducting WSe$_2$ (3.7 eV) surface. The experimental observations were substantiated using DFT based first-principle calculations that take image charges at the substrate into account, thereby verifying the importance of substrate-induced screening in STM studies of single molecule. Cho \textit{et al.}\ \cite{guisinger_nl_2012} studied STS on C$_{60}$ molecules deposited on G/SiC surface. dI/dV spectra on monolayer G/SiC indicates the presence of charge neutrality point at -450 mV. However, on C$_{60}$ molecules the feature is not present and instead three peaks at -2.7 V, 0.8 V, and 1.6 V are observed. These peaks are likely to correspond to HOMO, LUMO, and LUMO+1; here too, the HOMO-LUMO gap of 3.5 eV is higher than that for C$_{60}$ molecules adsorbed on Au(111) surface (2.7 eV \cite{crommie_prb_2004}) in line with enhanced screening on metal surfaces. 

The role of 2D materials in decoupling molecules of the adlayer from the underlying metal has also been studied using UHV-based, non-STM methods. Using electron energy loss spectroscopy (EELS), Endlich \textit{et al.}\ \cite{kroger_jcp_2014} also found that H$_2$Pc molecules are decoupled from underlying Ir(111) surface because of the graphene layer. EELS spectra showing many vibrational features H$_2$Pc molecules on Ir(111) surface - compared to that on G/Ir(111) - have been suggested due to deformation of molecular structure upon adsorption on Ir(111) surface. In contrast the molecule retains planar geometry on G/Ir(111) and therefore the out-of-plane vibrational features remain absent. The authors also performed DFT calculations to show that molecules are adsorbed at a distance 3.4 $\AA$ away from the graphene surface compared to 2.3 $\AA$ from the Ir(111) surface. Since graphene grows on Ir(111) at a distance of 3.4 \AA \cite{michely_prl_2011}, the H$_2$Pc molecule is lying 6.8 $\AA$ away from the Ir(111) surface. Such a large separation from the underlying metal surface is sufficient to prevent the molecule from hybridizing strongly with the metal surface. Similarly, FePc molecules on G/Ir(111) adsorb in a flat configuration, as revealed by Near Edge X-Ray Adsorption Spectroscopy (NEXAFS) \cite{scardamaglia2011metal}. Furthermore, from X-Ray Photoemission Spectroscopy (XPS) studies on the same system \cite{scardamaglia2013graphene}, it was observed that the electronic structure of the FePc is completely preserved. 

In two separate reports, Schulz \textit{et al.}\ \cite{liljeroth_acsn_2013} and Joshi \textit{et al.}\ \cite{auwarter_acsn_2014} have shown that CoPc molecules deposited on hBN/Ir(111) surface and 2H-P molecules deposited on hBN/Cu(111) surface are also decoupled from the respective metal surfaces due to the epitaxial hBN layer. Fig.~\ref{fig:image4}(d) shows HOMO (-1.1 V) and LUMO (0.8 V) peaks in the STS spectra recorded on CoPc molecules adsorbed on the wire region of the hBN moir\'e and the STM images recorded at the corresponding bias voltages. These sharp resonances have life-time limited lorentzian shape and enable high energy resolution spectroscopy of isolated molecules \cite{liljeroth_acsn_2013,liljeroth_natphy_2015}. Similarly, sharp HOMO and LUMO peaks were observed for the 2H-P molecules on the hBN/Cu(111) surface \cite{auwarter_acsn_2014}. In both cases, the corresponding orbital images match with DFT calculated orbitals. Finally, the presence of work function modulation over the moir\'e of hBN on metal surface leads to site-selective charging of the molecules which will be discussed in the next section.

hBN has also been used to decouple single-molecule magnets (SMMs) from the metallic substrate. SMMs are a type of metal-organic compounds with a magnetic core surrounded by organic ligands. They combine the properties of bulk magnetic materials with the quantum effects seen at a molecular level and have garnered a lot of interest in recent years \cite{bogani2008molecular}. Studies of SMMs assemblies on 2D materials are challenging as SMMs may not have the sufficient thermal stability required for sublimation. This can be overcome by using a gentle electrospray deposition \cite{graham_sci_2003, oshea_nanotech_2007}, which has been demonstrated for Fe$_4$(L)$_2$(dpm)$_6$ (where H$_3$L is of the general form R−C(CH$_2$OH)$_3$ and Hdpm = dipivaloylmethane) with a Fe$_4$ core \cite{erler2015highly} and Mn$_{12}$O$_{12}$(CH$_3$COO)$_{16}$ with Mn$_{12}$ core molecules \cite{Ternes2013Mn12} on epitaxial hBN on Ru(0001). For Mn$_{12}$ based molecule, Kahle \textit{et al.} \cite{Ternes2013Mn12} performed a comparative study on metal surfaces such as Au(111) and Cu(001) along with hBN/Rh(111) using STM. They found out that the spin state of the SMMs is preserved only on hBN \cite{Ternes2013Mn12} surface. This was demonstrated using inelastic spin-flip spectroscopy on isolated molecules trapped in pore sites of hBN moir\'e which revealed symmetric steps (peaks) in the  $\textrm{d}I/\textrm{d}V$ ($\textrm{d}^2I/\textrm{d}V^2$) signal around zero bias. Additionally, the position of the peaks shift with applied external magnetic field. These features, which are due to spin-flip excitation of the molecule caused by the tunneling electrons, are absent on Au(111) and Cu(001) surfaces due to quenching of the spin.

\subsection{Site-selective gating and charging}
\label{sec:site_modulation}

\begin{figure}[h!]
    \centering
    \includegraphics[width=0.8\textwidth]{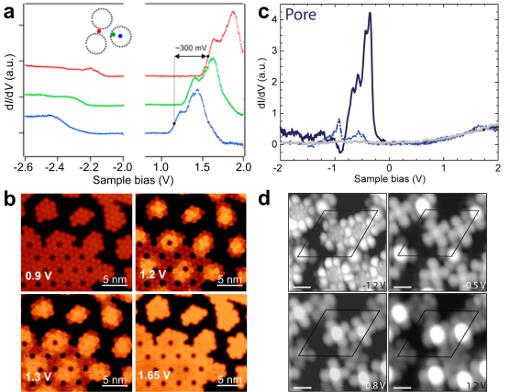}
    \caption{\textbf{Site-selective modulation of molecular orbitals.} (a) dI/dV spectra recorded on porphines molecules located at pore (blue), rim(green), and bridge (red) sites of the moire. Across the moire, the energy of LUMO varies by about 300 mV and the HOMO-LUMO gap increases from pore to the bridge site. (b) Bias dependent imaging shows that the molecules at the pore region are brighter at 1.2 V and the brightness extends away as the bias is increased. At 1.65 V all the molecules have similar brightness \cite{auwarter_acsn_2014}. (c) dI/dV spectra recorded at the center of CoPc molecules on the pore site of the hBN/Ir(111) moire shows peak at -0.4 V. For dI/dV spectra on molecules on wire site see Fig.~\ref{fig:image4}(d). The molecule on the pores are singly negatively charged and LUMO (of the neutral CoPc on wire site) has shifted below Fermi level. (d) Bias dependent STM imaging show different orbitals participating at different biases. Black rhombus shows the unit cell of hBN/Ir(111) moir\'e. While the molecules at the pore resemble LUMO, at the wire site they resemble HOMO at -1.2 V. As the bias is increased to 0.8 V and further to 1.2 V, the molecules at the wire site resemble LUMO, the molecules at the pore site resemble higher order orbitals \cite{liljeroth_acsn_2013}. Adapted with permission from \cite{auwarter_acsn_2014, liljeroth_acsn_2013}.}
    \label{fig:image5}
\end{figure}

So far, we saw site-selective adsorption of molecules on 2D materials which is related to periodic modulation of electronic and chemical properties across the moir\'e pattern. Here, we will discuss how these properties modulate molecular orbitals of the adsorbed molecules and in some cases to such an extent that molecules get charged selectively. Already on a weakly interacting system, G/Ir(111), J\"{a}rvinen \textit{et al.}\ \cite{liljeroth_jpcc_2014} observed that the LUMO resonance peak of CoPc and F$_{16}$CoPc is modulated by as much as 200 mV across the moir\'e. The molecules adsorbed on the TOP region of the moir\'e showed the LUMO peak at higher energy in line with the fact that the TOP region has a larger work function. However, the HOMO did not shift as much, i.e. the HOMO-LUMO gap is increased on the TOP region. Thus the rigid band shift due to work function difference (estimated 100 meV across the moir\'e) alone is not sufficient to explain these orbital shifts. One possible explanation is the change in the screening due to modulation of graphene-iridium distance across the moir\'e, which together with work function modulation, would result in the observed variation of the transport gap. The authors also mention that modulation of bonding character of graphene across the moir\'e \cite{michely_prl_2011} and spatially different symmetry of molecular resonances may play additional role.  

Compared to G/Ir(111), epitaxial hBN on different metals exhibit stronger work function modulation and geometric corrugation across moir\'e pattern, which would be expected to lead to more drastic effects on the orbital energies. A number of molecules deposited on different hBN surfaces \cite{liljeroth_acsn_2013, auwarter_acsn_2014, Urgel2015coord, groning_acsn_2015} show such site-specific modulation of molecular orbitals. Therefore, many interesting physical effects become visible. Joshi \textit{et al.}\ \cite{auwarter_acsn_2014} report modulation of the position of the molecular orbitals and the HOMO-LUMO gap across the moir\'e of hBN/Cu(111). Fig.~\ref{fig:image5}(a) shows dI/dV spectra recorded on 2H-P molecules adsorbed on different sites of the moir\'e. The LUMO peak is shifted by as much as 300 mV for the molecules lying on the bridge compared to the pore site. This shifts is smaller for the molecules lying on the rim of the moir\'e. The shift of the LUMO matches the work function difference between the pore and the bridge position ($\sim$300 meV), indicating that the change in the orbital energy could simply reflect the local vacuum level alignment. However, HOMO-LUMO gap was also found to be modulated with respect to the moir\'e as shown in Fig.~\ref{fig:image5}(a). The smallest gap is observed for the molecules on pore region. As discussed above, work function difference alone is not sufficient to explain these orbital shifts: charge screening due to hBN and Cu(111) surface needs to be taken into account. The energy shift of the LUMO orbital is also observed in the bias dependent STM imaging as shown in Fig.~\ref{fig:image5}(b). The molecules on the pore region have lowest lying LUMO and hence, already at a bias of 1.2 V, start to appear bright in STM images. LUMO orbitals of other molecules get inside the bias window as the sample bias is increased to 1.3 V and subsequently at 1.65 V, the whole monolayer has similar brightness. These results are not particular to 2H-P; 2H-TPCN and Co-TPCN molecules \cite{Urgel2015coord} adsorbed on on hBN/Cu(111) show similar modulation of LUMO peak and HOMO-LUMO gap. 

The effect of work function modulation across moir\'e is even stronger for CoPc molecules adsorbed on hBN/Ir(111) surface. Schulz \textit{et al.}\ \cite{liljeroth_acsn_2013} performed STS on CoPc molecules adsorbed on the pore and on wire site of the moir\'e. While the molecules on the wire site shows peaks at -1.1 V (HOMO), 0.8 V (LUMO) and 1.7 V (LUMO+1), the molecules on pore site shows only one peak at -0.4 V in the voltage range of -2 V to 2 V. As the orbital mapping at -0.4 V resembles the LUMO of the wire site (neutral CoPc), the peak is attributed to the LUMO shifted below the Fermi level due to electron transfer from the substrate to the molecule. Work function of pore site is smaller by 0.5 eV from the wire regions which is sufficient to induce charging. The  spectra recorded on the lobes of the charged molecule display additional resonance that cannot be explained within the simple single-particle picture \cite{liljeroth_natphy_2015}.

The d$I$/d$V$ spectra measured on the lobes of the charged molecules reveal peaks which can be understood using many-body excitations of the negatively charged molecule \cite{liljeroth_natphy_2015}. According to the single-particle picture, the spectrum of the negatively charged molecule should have the same orbitals as the neutral molecule, simply shifted down in energy due to the charging. The measured spectra contain three elastic peaks at -0.1 V, -0.3 V, and -0.8 V; however, orbital mapping shows that these states do not follow the single particle ordering of the orbitals. To explain these peaks, the authors invoke many-body effects that account for the reordering of molecular orbitals due to the added electron and correlation effects. Using time-dependent DFT (TDDFT), they are able to fully explain the observed resonances. These results rely on the possibility of carrying out high-resolution STS and fact that different charge states of the same molecule can be probed in chemically equivalent environment, which are made possible by the hBN insulating film. 

The decoupling by hBN results in sufficiently narrow widths of the molecular resonances which allows resolving vibronic features in the spectra. The vibronic peaks are the inelastic features resulting from elastic tunneling through one of the molecular resonances while simultaneously exciting a molecular vibration \cite{ho_pnas_2005,repp2010coherent}. This is different from inelastic electron tunneling spectroscopy which correspond to inelastic scattering in non-resonant tunneling. If the tunneling electrons have energy in excess of a molecular resonance, they can excite a molecular vibration and the corresponding vibronic modes appear as satellite peaks around that molecular resonance peak. For detailed information on vibronics, the readers are directed to Refs.~\cite{wingreen1989inelastic,gadzuk1991inelastic,galperin2007molecular,galperin2008nuclear}. The vibronic progressions of LUMO orbitals of CoPC molecules on the wire and pore sites of the moir\'e formed by hBN/Ir(111) are evident in Fig.~\ref{fig:image4}(d) and Fig.~\ref{fig:image5}(c) \cite{liljeroth_acsn_2013}. Line widths of 37 meV and 60 meV of the elastic peak have been estimated for the molecules on the wire and the pore site, respectively. This indicates that the electronic coupling between molecules and the metal substrates is significantly reduced due to the interposed hBN layer and this coupling is nearly two times smaller for the molecules on the wire site than on the pore site on hBN. 

\subsection{Tip-gated charging}
\label{sec:tip_gating}

\begin{figure}[h!]
    \centering
    \includegraphics[width=0.8\textwidth]{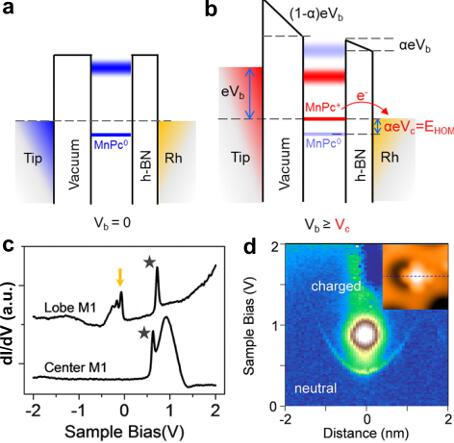}
    \caption{\textbf{Tip-gated charging.} (a) Energy level diagram for neutral MnPc in STM junction depicting the formation of double-barrier tunnel junction. The HOMO (shown in blue) of the molecule is very close the Fermi level. (b) Energy level diagram corresponding to tip-gated charging of the MnPc molecule at positive sample bias, V$_b$. Here, the HOMO crosses the Fermi level (shown in red) and the molecule loses an electron to form MnPc$^+$. $\alpha$ is the lever-arm factor. (c) dI/dV spectrum recorded on the lobe shows HOMO close to Fermi level (indicated by yellow arrow) and a sharp peak at positive bias (indicated by asterisk) corresponding to the tip-gated charging. Spectrum at the center of the molecule shows broad LUMO peak along with the charging peak. (d) Stacked dI/dV spectra taken across the molecule shows locus of charging peak and LUMO orbital. Adapted with permission from \cite{groning_acsn_2015}.}
    \label{fig:tip_gating}
\end{figure}

Molecules adsorbed on thin insulating layers probed by STM tip constitute a double-barrier junction, where a part of the voltage applied between the STM tip and the substrate drops between the molecule and the substrate. The ratio of the voltage drop across the tip-molecule and tip-substrate junctions depends on their respective capacitances and is called the lever-arm factor (denoted by $\alpha$) -- it can be tuned by varying the tip-surface distance. A double-barrier junction can giver rise to charging, when a molecular orbital close to the Fermi level of the substrate crosses the Fermi level due to finite voltage drop across the insulating layer \cite{ho_prl_2005}. In other words, the orbital position is tweaked across the Fermi level due to the voltage applied across the tip and substrate; therefore, the tip acts as a gate. The schematic is shown in Fig.~\ref{fig:tip_gating}(a-b). The charging results in a modification of the potential barrier profile of the junction \cite{wiesendanger_prb_2008} and subsequently results in a sharp change in the d$I$/d$V$ spectrum. Ho and his group used ultrathin Al$_2$O$_3$ layer on NiAl substrate to create the double-barrier junction and investigated tip-gated charging of the molecules \cite{wu2004control}. Liu \textit{et al.}\ shows that monolayer hBN/Rh(111) can serve the same purpose of forming a double-barrier junction during STM investigations \cite{groning_acsn_2015}. They demonstrate that MnPc molecules adsorbed site-selectively on the hBN/Rh(111) surface can be charged through tip-gating. At submonolayer coverage, the molecules adsorb preferentially on the pore-site of the moir\'e with three adsorption types. Out of the three types, the molecules with a bright centre are in a neutral state but have their HOMO level close ($\sim$100 meV) to the Fermi energy (shown by yellow arrow in Fig.~\ref{fig:tip_gating}(c)). This orbital is brought above the Fermi energy at positive bias and a sharp peak appears in d$I$/d$V$ spectrum along with LUMO orbital of the molecule. The bias value at which the charging peak appears depends on: (i) the exact location of the HOMO, (ii) the location of the STM tip, and (iii) the lever-arm factor. For molecules having HOMO level 200 mV below Fermi level, the charging peak was not seen within the 2 V. Fig.~\ref{fig:tip_gating}(d) shows a stacked dI/dV spectra recorded across the molecule with a topographic image shown in the inset. The locus of the charging peak (large d$I$/d$V$ feature) seems to follow parabolic arc along the bright spot at the centre, which corresponds to the LUMO of the molecule. Depending on the bias and the tip position, the molecule gets charged when a threshold electric field is established across the molecule and substrate. Therefore, above the parabola, the molecule is in the charged state and below it is neutral. The locus of the charging peak strongly depends on the tip geometry \cite{wiesendanger_prb_2008}. Further, by changing the set-point current, value of lever-arm factor can be varied which changes the position of the charging peak. As the set-point current was varied from 25 pA to 400 pA, the charging peak position changed from 0.90 V to 0.67 V.

\subsection{Kondo effect on G/Ru(0001)}
\label{sec:kondo}

Coupling of a magnetic moment to the conduction electrons of metal gives rise to many-body spin-singlet state known as Kondo effect \cite{kondo_physrev_1968}. This has been studied extensively in the last decades \cite{Jamneala2000,Madhavan2001,Wahl2004} using STM for transition metal atoms and also for several molecules\cite{Zhao2005,Iancu2006,Iancu2006a,Gao2007,Mugarza2011,DiLullo2012,Wu2015, Fernandez-Torrente2008} on noble metal surfaces. In case of the graphene, the Kondo effect has been proposed to be different than the metals because of (i) linear dispersion and (ii) the vanishing density of states at the Dirac point \cite{vojta_rpp_2013}. This may lead to interesting Kondo behaviour, for example, orbital dependent\cite{Wehling2010}, electron-hole asymmetry dependent \cite{Vojta2010}, and single- or multi-channel Kondo resonances \cite{baskaran_prb_2008,Zhu2010} as predicted theoretically. Despite the extensive theoretical work, a clean and unambiguous experimental signature of the Kondo effect has been elusive on graphene \cite{Chen2011,heiko_nphy_2012,fuhrer_natphys_2012}. 

Out of the pool of MPcs \cite{Zhao2005, Gao2007, Mugarza2011} showing Kondo resonances on metal surfaces, none have been reported to exhibit the same effect on epitaxial or transferred graphene surfaces. On the other hand, TCNQ \cite{Garnica2013} and F$_4$TCNQ molecules \cite{martin_ns_2014,miranda_ss_2014,miranda_nl_2014} exhibit Kondo effect when adsorbed on strongly coupled epitaxial G/Ru(0001) surface. TCNQ molecules have high electron affinity (2.8 eV\cite{Milian2004}), and therefore they get negatively charged  \cite{Garnica2013,miranda_ss_2014}. Orbital mappings confirm that the additional electron occupies the LUMO, splitting it into singly occupied molecular orbital (SOMO) and singly unoccupied molecular orbital (SUMO) with a Coulomb charging energy of 1.8 eV. The charge state of TCNQ molecules on G/Ru(0001) stays the same irrespective of their adsorption site \cite{Garnica2013} on the moir\'e. However, the Kondo resonance peak show dependence on the moir\'e pattern: the resonance is observed only when the molecules are adsorbed on the FCC and the HCP-FCC bridge region of the moir\'e \cite{miranda_nl_2014} with higher Kondo temperature for the molecules on the HCP-FCC bridge region. A similar effect has been observed for F$_4$TCNQ molecules and for Co atoms adsorbed on specific regions of the moir\'e on the G/Ru(0001) surface \cite{gao_nl_2014}. The Kondo resonance peak was observed for Co adatoms only on the edge of the TOP regions with different Kondo temperatures for the TOP-HCP edge (12.1 K) and TOP-FCC edge (5.4 K).

 \section{Band structure engineering and doping of graphene}
\label{sec:band_structure}

The electronic properties of monolayer graphene are described by its conical band structure at the \textit{K} and \textit{K'}-points of the Brillouin zone. The linear dispersion of the charge carriers near the Dirac points gives graphene many of its fascinating electronic properties \textit{e.g.}, high electron mobility and ambipolar transport \cite{sarma2011electronic}. However, application of graphene in practical electronic devices has still not been realized as it requires two things: precise control over the type and concentration of charge-carriers and opening a sizeable band gap in the graphene band-structure \cite{schwierz2010graphene}. Furthermore, these goals should be achieved through methods that are inexpensive and scalable and do not compromise the wonderful electronic properties of graphene. Many schemes have been used to introduce a band-gap in graphene. Quantum confinement of charge carriers in lithographically defined graphene nanostructures and unzipped carbon nanotubes can be used to create a band gap \cite{li2008chemically,han2007energy,eroms2009weak,giesbers2012charge,kosynkin2009longitudinal,tao2011spatially}, but these methods cannot be used to achieve atomically well-defined nanoribbon edges. Bottom-up on-surface synthesis offers unprecedented control over the GNR width and edge termination \cite{Cai2010GNR,Ruffiex2016zigzag,Kimouche2015GNR,Talirz2016review}. However, contacting these very narrow ribbons is still a major technological challenge \cite{llinas2016short}. Recently, experiments on graphene on hBN devices have shown band gaps of up to 300 meV depending on the misalignment angle of the two lattices \cite{hunt2013massive}. The exact mechanism of the gap opening is still under debate, with both breaking of the graphene sublattice symmetry due to the underlying hBN \cite{hunt2013massive} and biaxial strain \cite{woods2014commensurate} have been proposed.

Covalent functionalization can also be used for both gap opening and doping and band gaps greater than an eV have been reached through partial rehybridization of the sp$^2$  carbon atoms to sp$^3$ \textit{e.g.}, by hydrogen \cite{Geim2009graphane} or fluorine \cite{nair2010fluorographene}. Substitutional doping with boron and nitrogen atoms to obtain hole (p-type) or electron (n-type) rich graphene has also been demonstrated \cite{panchakarla2009synthesis,wei2009synthesis,telychko2014achieving,sforzini2016structural}. However, these methods potentially disrupt the extended $\pi$-conjugation of graphene and lead to a degradation of graphene's exceptional electronic properties. On the other hand, physisorbed molecules interacting with graphene through weak vdW interactions are expected to offer a route to modify graphene's electronic properties without degrading its desirable properties, even under high charge carrier density.

\begin{figure}[h!]
    \centering
    \includegraphics[width=0.9\textwidth]{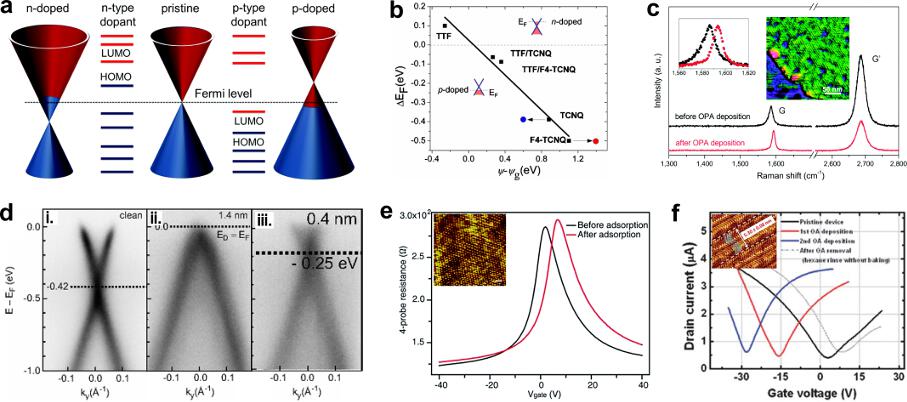}
    \caption{\textbf{Charge donor and acceptors on graphene.} (a) Schematic of doping of graphene by molecules: molecules with LUMO lying at energy lower than graphene Fermi level $E_\mathrm{F}$ spontaneously accept electrons from graphene making the latter p-doped (right side). Conversely, molecules with the HOMO above graphene $E_\mathrm{F}$ spontaneously donate electrons to graphene making the latter n-doped (left side) \cite{mali2015nanostructuring}. (b) Theoretical data points (black squares) on acceptors (TCNQ, F$_4$TCNQ), donors (TTF) and their salts on graphene. The abscissa gives change in $E_\mathrm{F}$ and, ordinate gives change in work-function of graphene after deposition of molecules \cite{sun2010linear}. (c) Raman measurements showing p-doping effect of OPA on graphene. Inset shows AFM image of OPA assembly on G/SiO$_2$ \cite{prado2010two}. (d) ARPES data of (i) G/SiC ($E_\mathrm{F}$ = -0.42 eV) which is intrinsically n-doped, (ii) after deposition of 0.8 nm of F$_4$TCNQ on it ($E_\mathrm{F}$ saturates at 0 eV) and, (iii) after deposition of 0.4 nm of TCNQ on it ($E_\mathrm{F}$ saturates at -0.25 eV) \cite{ Coletti2010}. (e) Electrical transport measurements on a G/SiO$_2$ FET before and after treatment with TMA indicating p-doping. Inset shows STM image of TMA assembly at solid-liquid interface on G/SiC. \cite{zhou2014switchable}. (f) $I_\mathrm{SD}-V_\mathrm{G}$ curves of G/SiO$_2$ FET before and after repeated cycles of treatment with OA indicating increased n-doping and improved mobility with each cycle. Inset shows STM image of OA assembly on HOPG \cite{li2013toward}. Adapted with permission from \cite{mali2015nanostructuring,sun2010linear, prado2010two,Coletti2010,zhou2014switchable, li2013toward}}
	 \label{fig:bandengg1}
\end{figure}

Doping of graphene by physisorbed molecules can be achieved easily by means of interfacial charge transfer, which can be tuned through the energy level alignment of the molecular frontier orbitals with respect to graphene's Fermi level ($E_\mathrm{F}$, which is $\approx$4.5 eV for free-standing graphene \cite{sque2007transfer}) as shown in Fig.~\ref{fig:bandengg1}(a). If the EA  of the molecule is larger than graphene work function, the molecule will accept electrons from graphene and make it hole-doped (\textit{i.e.}, p-doped). Conversely, if the IP of the molecule is smaller than graphene work function, the molecule will donate electrons to graphene, making it electron-doped (\textit{i.e.}, n-doped). The EA and IP of the molecule will be modified from their gas-phase values by screening due to the graphene surface and DFT calculations including vdW corrections are the best available theoretical means to estimate the amount of charge transfer between a given molecule and free graphene surface \cite{wehling2008molecular,rochefort2008interaction,hu2013theoretical}. In addition to doping, periodic potential modulation has been proposed as means of breaking the sub-lattice symmetry in graphene \cite{park2008new}. An adlayer with the correct symmetry and period can potentially lead to a band-gap in graphene.

To date, many small gas molecules and hydrocarbons have been used to dope graphene with both donor- (\textit{e.g.}, NH$_3$ \cite{romero2009adsorption}, polyethyleneimine \cite{farmer2008chemical}, naphthalenediamine, dimethylanthracene \cite{dong2009doping}) and acceptor-type dopants (\textit{e.g.}, H$_2$O \cite{moser2008environment}, O$_2$ \cite{sato2011electrically}, diazonium salts \cite{farmer2008chemical}, pyrenetetrasulfonic acid, dibromoanthracene \cite{dong2009doping}). These have resulted in changes in the charge carrier density without degradation of electron mobility. Moreover, this type of doping is reversible as the weakly absorbed molecules can desorbed by gentle annealing. While the there is an enormous body of literature on functionalization of graphene with molecules (see \cite{liu2011chemical,zhang2011tailoring,mao2013manipulating,tang2013graphene,kong2014molecular} for reviews), here we only concentrate on well-defined systems with molecules forming large-scale, periodic lattices on graphene.

Measuring doping level of graphene with tunneling spectroscopy is non-trivial in particular for graphene buried under a molecular adlayer with conflicting results in the literature for \emph{e.g.}, PTCDA adsorption \cite{lauffer2008molecular,wang2009room}. Assessing the effect of molecular layers on the band structure of the underlying 2D material is perhaps better carried out using photoelectron spectroscopy (PES), in particular angle-resolved photoelectron spectroscopy (ARPES). An easier way of doing it without the need for UHV conditions is Raman spectroscopy. Studies show that doping of graphene causes upshift and broadening of graphene's G-peak, and a decrease in the ratio of intensity of 2D peak and G peak (I$_{2D}$/I$_{G}$) \cite{das2008monitoring}. An effective way of testing the effect of molecular adlayers on graphene in a technologically relevant setting is through $I-V$ experiments on graphene-on-insulator field effect transistors. A shift of the charge neutrality point ($V_\mathrm{CNP}$, the back-gate voltage corresponding to maximum graphene resistivity) towards positive values on deposition of molecules indicate p-doping, and vice-versa

Synchrotron based high-resolution PES shows that the work function of G/SiC shifts up by $\sim$0.25 eV upon deposition of 1 ML of PCTDA \cite{huang2009structural} indicating weak hole transfer from the molecule. Differential reflectance spectroscopy (DRS) and angle resolved ultraviolet PES (ARUPS) study on a similar system, confirmed negligible charge transfer at room temperature \cite{meissner2012highly}. The fact that the graphene band structure remains unaltered after deposition of PTCDA has been utilized in using them as seeding layer for atomic layer deposition (ALD) \cite{alaboson2011seeding} of dielectric layers. High-$\kappa$ dielectric layers (\textit{e.g.}, aluminum oxide) are generally used as ultra-thin insulating layer on graphene for the deposition of a metallic top-gate on it. However, owing to the hydrophobicity and chemical inertness of graphene, these layers are usually of a bad quality. Using a PTCDA buffer layer enables the growth of uniform and stable dielectric layer. Similarly, a combined PES and ARPES study showed that pentacene does not affect graphene's band structure; in fact, a monolayer of penatcene protects the surface of graphene from exposure to the ambient and the band structure is completely recovered after removing the adlayer by annealing it \cite{jee2009pentacene}. 

C$_{60}$ is a well known weak electron acceptor (EA $\approx$ 3.7 eV) and assembles in hexagonal close-packed geomtery on graphene (refer Sec.~\ref{sec:site_adsorption}). High resolution PES data \cite{Chen2007} on G/SiC show a $\sim$0.15 eV upshift of the work function of graphene upon deposition of a monolayer of C$_{60}$; this indicates very weak p-doping of graphene. Raman spectroscopy and terahertz conductivity measurement on graphene on SiO$_2$ confirm a weak p-doping effect due to C$_{60}$ deposition (estimated as $\sim$0.04e transferred to each molecule) \cite{jnawali2015observation}. Additionally, the carrier mobility increases by $\sim$1500 cm$^2$/Vs - the authors hypothesize that it is due to screening of long-range scatterers (charged impurities at G/SiO$_2$ interface) by the molecules. The molecule trimesic acid, in its solution phase, assembles in a close-packed geometry on graphene and p-dopes it \cite{zhou2014switchable}. $I-V$ measurements on graphene FETs decorated with TMA show that the V$_{CNP}$ upshifts by $\sim$ 4.5 V (Fig.~\ref{fig:bandengg1}(e)) and each molecule is calculated to have gained 0.003e from graphene. Another organic molecule octadecylphosphonic acid (OPA), assembles into rippled domains on graphene on deposition (in solution phase), and as shown by Raman measurements, causes hole type doping of graphene (Fig.~\ref{fig:bandengg1}(c)) \cite{prado2010two}. 

We now turn our attention to organic molecules which are strong electron acceptors \textit{e.g.}, TCNQ, F$_4$TCNQ, and tetracyanoethylene (TCNE). Doping of graphene due to interfacial charge transfer by these molecules has been topic of many theoretical studies \cite{pinto2009p,manna2009tuning}. Sun \textit{et al.}\ \cite{sun2010linear} performed DFT calculations showing that F$_4$TCNQ, which has a higher electron affinity than TCNQ \cite{kanai2009determination}, is expected to accept more charge (0.4e) than TCNQ (0.33e) (Fig.~\ref{fig:bandengg1}(b)). This prediction is confirmed by the ARPES experiments by Coletti \textit{et al.}\ on G/SiC \cite{Coletti2010}. G/SiC is intrinsically n-doped with the Dirac point 0.42 eV below the $E_\mathrm{F}$(Fig.~\ref{fig:bandengg1}(d) (i)). Deposition of F$_4$TCNQ drives the Dirac point towards $E_\mathrm{F}$; it saturates at $E_\mathrm{F}$ after deposition of the molecular layer with a nominal thickness of 0.8 nm (Fig.~\ref{fig:bandengg1}(d) (ii)). This indicates a neutralization of the graphene surface due to p-type doping from the molecule. For TCNQ however, the Dirac point saturates at 0.25 eV below $E_\mathrm{F}$ (Fig.~\ref{fig:bandengg1}(d) (iii)). Thus, the p-type doping from TCNQ is weaker than its fluorinated counterpart. XPS data of F$_4$TCNQ on G/SiC further reveals that the charge transfer takes place exclusively through one pair of the terminal cyano groups with the molecules adsorbing in an upright configuration at high coverage. Another synchrotron-based PES study indicates that the work function of G/SiC indeed increases with increasing surface coverage of F$_4$TCNQ \cite{Chen2007}. This is associated with a 0.5 eV shift of the graphene related C1s peak to lower binding energy. The authors concluded that this is due to charge accumulation at the G/SiC interface due to strong hole doping by the molecule. 

Theoretical studies of the electron donating tetrathiafulvalene (TTF) molecule on graphene have predicted that it donates 0.1e to graphene, making the latter n-doped \cite{manna2009tuning,sun2010linear}. UHV based XPS data of chemically exfoliated graphene dispersed with TTF indicates the presence of both uncharged and positively charged sulphur \cite{choudhury2010xps}. This indicates electron donation to graphene by TTF. Raman spectroscopy on a similar system shows downshift of graphene's G-peak with increasing concentration of TTF \cite{voggu2008effects}. As-obtained graphene is p-doped due to exposure to ambient \cite{xu2012investigating}; n-doping by TTF neutralizes the excess holes thereby causing a downshift of G peak towards values corresponding to undoped graphene.

Vanadyl phthalocyanine (VOPc), a member of the phthalocyanine family which assembles into close-packed geometry on graphene, is seen to dope it with electrons \cite{wang2011quantitative}. Kelvin probe force microscopy (KPFM) on exfoliated graphene on SiO$_2$ reveals that deposition of VOPc results in an increase of the contact potential difference (CPD), indicating transfer of electrons from the molecules to graphene. Mono- and bilayer graphene are more strongly electron-doped by the molecules compared to thicker graphene layers. $I-V$ measurements on graphene FETs show a downshift of the $V_\mathrm{CNP}$ by $\sim$9 V, without any degradation of carrier mobility, confirming the n-doping. Based on the experimental data, the authors concluded that each molecule donates about 0.04e to graphene. n-doping of graphene has also been found with adsorption of oleylamine which assemble in lamellar nanostrips on G/SiC \cite{li2013toward}. The $V_\mathrm{CNP}$ of G/SiO$_2$ FET shifts down by 19 V after deposition of the molecules from solution; this is associated with a two-fold increase in mobility (Fig.~\ref{fig:bandengg1}(f)). 

\begin{figure}[h!]
    \centering
    \includegraphics[width=0.9\textwidth]{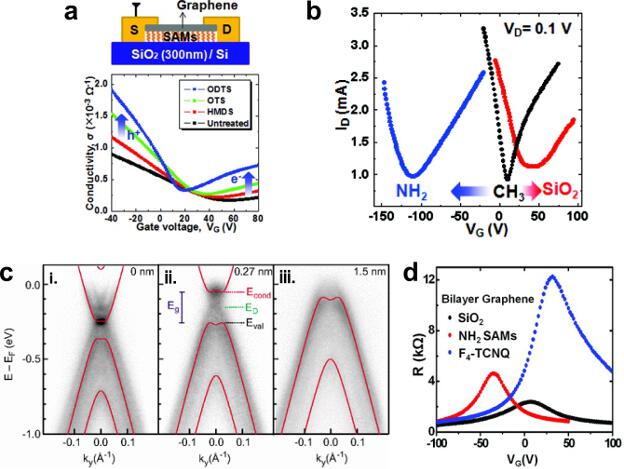}
    \caption{\textbf{Self-assembled monolayers and band gap opening in bilayer graphene.} (a) (top) Schematic of a graphene FET on SiO$_2$ treated with a SAM. (bottom) Transfer characteristics of graphene on untreated SiO$_2$ and graphene on SiO$_2$ treated with a SAM of increasing alkyl chain length (HMDS \textless OTS \textless ODTS) showing a concomitant decrease of $V_\mathrm{CNP}$ towards 0 V and increasing mobility \cite{lee2011control}. (b) $I_\mathrm{SD}-V_\mathrm{G}$ curves of graphene on untreated SiO$_2$ (heavily p-doped), and on CH$_3$-SAM (almost neutral) and NH$_2$-SAM treated SiO$_2$ (heavily n-doped) \cite{park2011work}. (c) ARPES data of (i) BLG/SiC ($E_\mathrm{F}$ = -0.3 eV, $\Delta$E$_g$ = 116 meV), (ii) after deposition of 0.27 nm of F$_4$TCNQ and, (iii) after deposition of 1.5 nm of molecule. $\Delta$E$_g$ saturates at 275 meV and $E_\mathrm{F}$ moves in-gap \cite{Coletti2010}. (d) Transfer characteristics of a BLG FET on SiO$_2$ treated with NH$_2$-SAM shows n-doping and two-fold increase in $R_\mathrm{CNP}$ (compared to BLG on native dioxide). Deposition of 10 \AA\  of F$_4$TCNQ on it results in p-doping and about six-fold increase in $R_\mathrm{CNP}$ (compared to initial state) \cite{park2012single}. Adapted with permission from \cite{lee2011control,park2011work,Coletti2010,park2012single}.}
	 \label{fig:bandengg2}
\end{figure}

The discussion so far has concerned molecular adlayers on graphene; now we turn to the effects of self-assembled monolayers (SAMs) under graphene. Silane-based molecules with a long alkyl chain capped with functional groups have been used to treat SiO$_2$ surface to tune its properties. The silane group bonds covalently with the silicon dioxide surface neutralizing the Si- and O- dangling bonds and passivating the surface \cite{kim2014infrared}. These dangling bonds facilitate p-type doping of graphene deposited on SiO$_2$ and act as long-range Coulomb scatterers reducing the electron mobility \cite{lafkioti2010graphene}. Graphene on a SAM passivated SiO$_2$ surface should therefore be less p-doped and have better mobility, which was confirmed by Lee \textit{et al.}\ \cite{lee2011control}. Three silane based molecules with -CH$_3$ functional group and increasing alkyl chain length were chosen for the study. XPS data show that the thickness and the packing density of the SAMs on SiO$_2$ increase with increasing alkyl chain length. Raman spectroscopy of graphene transferred onto untreated and SAM passivated surfaces suggests a decrease of surface induced p-doping with increasing alkyl chain length. The results are confirmed from ultraviolet photoemission (UPS) data which show a concomitant decrease of the work-function of graphene to almost free-standing values. Finally, $I-V$ measurements on graphene FETs show devices with SAMs have $V_\mathrm{CNP}$ close to zero back-gate voltage and an increased electron mobility; again a better packing density of the SAM leads to better performance of the device (Fig.~\ref{fig:bandengg2}(a)). 

SAMs offer the possibility to tune the reactivity of the surface by different functional groups. For example, SAMs with electron donating NH$_2$- groups n-dope graphene \cite{park2011work}. This was demonstrated by Raman and UPS measurements, where CVD-graphene transferred onto NH$_2$-SAM modified surface was n-doped compared to untreated or CH$_3$-SAM modified surface. $I-V$ data in Fig.~\ref{fig:bandengg2}(b), shows that NH$_2$-SAM causes reduction of $V_\mathrm{CNP}$ to large negative voltages, confirming the n-doping effect; in comparison, CH$_3$-SAM reduces the $V_\mathrm{CNP}$ to near-zero back-gate voltage by surface passivation. Based on a study of SAMs with several different functional groups, Yokota \textit{et al.}\ \cite{yokota2011carrier} concluded that the doping effect can be related to the induced dipole moment in the SAM. From angle resolved XPS it was found that for certain functional groups (NH$_2$C$_6$H$_4$- and C$_8$-SAM), the dipole points towards the graphene while for others (NH$_2$C$_3$H$_6$- and the fluorinated F$_{13}$-SAM), the direction of the dipole is opposite. From Raman measurments, it was found that the SAMs of first kind n-dope graphene, bringing the carrier-density to near pristine levels while the latter p-dope graphene. In another study of $I-V$ characteristics of CVD-graphene on SiO$_2$ passivated with different SAMs, the authors \cite{yan2011controlled} concluded that for some SAMs (strongly p-doping CF$_3$-SAM and slightly n-doping CH$_3$-SAM), the doping is due to the dipole moment. But for SAMs with strong electron donating (strongly n-doping NH$_2$-SAM) or accepting groups (weakly p-doping H$_3$N$^+$-SAM), the doping is due to interfacial charge transfer.

While opening a band-gap in monolayer graphene by self-assembled molecular layers has not yet been demonstrated, this has been achieved in the case of bilayer graphene (BLG). Although, pristine A-B stacked BLG (exactly half of the carbon atoms of the top graphene layer sits on the centre of the hexagonal ring of the bottom graphene layer) does not show any band-gap, it is possible to break the equivalence of the individual layers by application of an external electric field perpendicular to the layers \cite{neto2009electronic}, which has been shown by external electrostatic gates \cite{zhang2009direct}. However, this approach requires application of complex lithographic processes and high electric fields. An alternative way to achieve this would be to generate an interlayer charge asymmetry by individually tuning the polarity and concentration of charge carriers in the top and bottom graphene layers through chemical doping. Doping by F$_4$TCNQ has been seen to open a band-gap in epitaxial BLG on SiC \cite{Coletti2010} (Fig.~\ref{fig:bandengg2}(c)). BLG/SiC is intrinsically n-doped with a band-gap caused by electric dipoles at G/SiC interface which break the interlayer symmetry. This manifests in ARPES data as a gap of about 0.116 eV; the Dirac point lies mid-gap, about -0.3 eV below the $E_\mathrm{F}$. Increasing F$_4$TCNQ coverage drives the Dirac point close to $E_\mathrm{F}$ (\textit{i.e.}, a reduction in n-doping) with an increase in the band-gap. Finally at about 2 ML coverage, the effects saturates giving a final band-gap of 0.275 eV. The authors proposed that an additional dipole at the molecule-graphene interface increases the electrostatic asymmetry, resulting in an increase of the band-gap. Deposition of F$_4$TCNQ is also seen to open a band-gap in exfoliated BLG \cite{park2012single} (Fig.~\ref{fig:bandengg2}(d)) on NH$_2$-SAM functionalized SiO$_2$ substrate. As seen from $I-V$ measurements, SAM functionalization leads to n-doping of the graphene and compared to an unpassivated surface, the resistance at $V_\mathrm{CNP}$ is higher. Deposition of F$_4$TCNQ is seen to drive $V_\mathrm{CNP}$ to larger positive values with an accompanying increase of resistance at $V_\mathrm{CNP}$. Based on a model calculation, the authors estimated the largest achieved gap as 0.124 eV. Accompanying infrared absorption data gave an estimate of the maximum optical band-gap of 0.210 eV. In both cases, the gap increased with increasing F$_4$TCNQ surface coverage. The authors concluded that that NH$_2$-SAM gives stable n-doping to the bottom graphene layer and F$_4$TCNQ deposition leads to p-type doping of the top layer: the increased asymmetry results in a concomitant increase in the gap. Single molecule doping strategy where one of the layers of BLG is n-doped by a donor molecule leaving the other layer to be p-doped by ambient conditions (top layer) or trapped impurities (bottom layer) has also been used to open band-gap \cite{joshi2010intrinsic}. Functionalizing BLG with electron donating triazine molecules, which only n-dopes the top layer, causes a band-gap of upto 0.111 eV to open up \cite{zhang2011opening}. Conversely, depositing BLG on a dioxide surface treated with donor type benzyl-viologen molecule n-dopes the bottom layer and a band-gap of upto 0.132 eV opens up on exposure to ambient \cite{lee2015chemically}.

\section{Summary and outlook}
\label{sec:summary}

Molecular self-assembly has been typically studied on metal surfaces or on graphite, where the focus has been on understanding the fundamentals and to investigate in detail the structure of the assemblies. In the case of bulk substrates, molecular self-assembly cannot be used for a profound modification of the substrate electronic structure beyond doping or quenching surface states. Alternatively, molecular layers can be used as templates for patterning the deposition of subsequent layers.

Two-dimensional materials consist of only surface atoms, and hence, adsorption of a molecular layer can completely change their electronic response. This is a strong motivation to study molecular self-assembly on 2D materials. Understanding the formation of molecular assemblies, which can differ strongly from the corresponding systems on a metal substrate, is progressing especially in the case of epitaxial graphene and hexagonal boron nitride substrates. Close-packed molecular assemblies without intermolecular bonds or hydrogen-bonded networks are relatively well-understood. More strongly linked assemblies, in particular covalently bonded systems are still a virtual \textit{terra incognita} and we lack even basic understanding of chemical reactions where the substrate is involved.

Compared to the close-packed metal surfaces, 2D materials present an inert surface on which the electronic properties arising from the self-assembled molecules are preserved. Beyond its obvious importance in fundamental research on the electronic properties of molecular materials, this presents an unique opportunity to test monolayer thick, all-organic electronic devices. These could also be incorporated into van der Waals heterostructures (\textit{i.e.}, stacked layers of 2D materials), for additional stability and functionality. The immense flexibility in choosing the constituents of such hybrid materials makes a strong case for further studies in this field.

Molecular adlayers can help enable novel applications that are only made possible through the unique properties of 2D materials. For example, molecular layers can enable strong and precise doping profiles, which are important in formation of sharp p-n-junctions. Further-reaching potential applications include profound modifications of the band structure of 2D materials by, \emph{e.g.} imposing a periodic potential modulation through molecular self-assembly to generating totally novel materials where the properties of the 2D substrate and the molecular layer strongly hybridize. Beyond the application of functionalized 2D materials in the fields of catalysis, sensors, and flexible and transparent electronic materials, the combination of molecular adlayers and 2D materials in hybrid heterojunctions can open a hitherto unknown vista of materials with exciting electronic properties.

\section{Acknowledgements}
This research was supported by the European Research Council (ERC-2011-StG No. 278698 ”PRECISE-NANO”), the Academy of Finland (Centre of Excellence in Low Temperature Quantum Phenomena and Devices No. 284594).

\newpage{}



\begin{thebibliography}{100}
\expandafter\ifx\csname url\endcsname\relax
  \def\url#1{\texttt{#1}}\fi
\expandafter\ifx\csname urlprefix\endcsname\relax\def\urlprefix{URL }\fi
\providecommand{\bibinfo}[2]{#2}
\providecommand{\eprint}[2][]{\url{#2}}

\bibitem{de2003two}
\bibinfo{author}{De~Feyter, S.} \emph{et~al.}
\newblock \bibinfo{title}{Two-dimensional supramolecular self-assembly probed
  by scanning tunneling microscopy}.
\newblock \emph{\bibinfo{journal}{Chemical Society Reviews}}
  \textbf{\bibinfo{volume}{32}}, \bibinfo{pages}{139--150}
  (\bibinfo{year}{2003}).

\bibitem{slater2011two}
\bibinfo{author}{Slater, A.~G.} \emph{et~al.}
\newblock \bibinfo{title}{Two-dimensional supramolecular chemistry on
  surfaces}.
\newblock \emph{\bibinfo{journal}{Chemical Science}}
  \textbf{\bibinfo{volume}{2}}, \bibinfo{pages}{1440--1448}
  (\bibinfo{year}{2011}).

\bibitem{kudernac2009two}
\bibinfo{author}{Kudernac, T.} \emph{et~al.}
\newblock \bibinfo{title}{Two-dimensional supramolecular self-assembly:
  nanoporous networks on surfaces}.
\newblock \emph{\bibinfo{journal}{Chemical Society Reviews}}
  \textbf{\bibinfo{volume}{38}}, \bibinfo{pages}{402--421}
  (\bibinfo{year}{2009}).

\bibitem{rosei2003properties}
\bibinfo{author}{Rosei, F.} \emph{et~al.}
\newblock \bibinfo{title}{Properties of large organic molecules on metal
  surfaces}.
\newblock \emph{\bibinfo{journal}{Progress in Surface Science}}
  \textbf{\bibinfo{volume}{71}}, \bibinfo{pages}{95--146}
  (\bibinfo{year}{2003}).

\bibitem{stohr_prb_2014}
\bibinfo{author}{Matena, M.} \emph{et~al.}
\newblock \bibinfo{title}{On-surface synthesis of a two-dimensional porous
  coordination network: unraveling adsorbate interactions}.
\newblock \emph{\bibinfo{journal}{Physical Review B}}
  \textbf{\bibinfo{volume}{90}}, \bibinfo{pages}{125408--}
  (\bibinfo{year}{2014}).

\bibitem{Grobis2002}
\bibinfo{author}{Grobis, M.} \emph{et~al.}
\newblock \bibinfo{title}{Local electronic properties of a molecular monolayer:
  {C}$_{60}$ on {A}g(001)}.
\newblock \emph{\bibinfo{journal}{Physical Review B}}
  \textbf{\bibinfo{volume}{66}}, \bibinfo{pages}{161408--}
  (\bibinfo{year}{2002}).

\bibitem{Bedwani2008}
\bibinfo{author}{Bedwani, S.} \emph{et~al.}
\newblock \bibinfo{title}{Strongly reshaped organic-metal interfaces:
  tetracyanoethylene on {C}u(100)}.
\newblock \emph{\bibinfo{journal}{Physical Review Letters}}
  \textbf{\bibinfo{volume}{101}}, \bibinfo{pages}{216105--}
  (\bibinfo{year}{2008}).

\bibitem{fiori2014electronics}
\bibinfo{author}{Fiori, G.} \emph{et~al.}
\newblock \bibinfo{title}{Electronics based on two-dimensional materials}.
\newblock \emph{\bibinfo{journal}{Nature Nanotechnology}}
  \textbf{\bibinfo{volume}{9}}, \bibinfo{pages}{768--779}
  (\bibinfo{year}{2014}).

\bibitem{mas20112d}
\bibinfo{author}{Mas-Balleste, R.} \emph{et~al.}
\newblock \bibinfo{title}{2{D} materials: to graphene and beyond}.
\newblock \emph{\bibinfo{journal}{Nanoscale}} \textbf{\bibinfo{volume}{3}},
  \bibinfo{pages}{20--30} (\bibinfo{year}{2011}).

\bibitem{das2015beyond}
\bibinfo{author}{Das, S.} \emph{et~al.}
\newblock \bibinfo{title}{Beyond graphene: progress in novel two-dimensional
  materials and van der {W}aals solids}.
\newblock \emph{\bibinfo{journal}{Annual Review of Materials Research}}
  \textbf{\bibinfo{volume}{45}}, \bibinfo{pages}{1--27} (\bibinfo{year}{2015}).

\bibitem{neto2009electronic}
\bibinfo{author}{Neto, A.~C.} \emph{et~al.}
\newblock \bibinfo{title}{The electronic properties of graphene}.
\newblock \emph{\bibinfo{journal}{Reviews of modern physics}}
  \textbf{\bibinfo{volume}{81}}, \bibinfo{pages}{109} (\bibinfo{year}{2009}).

\bibitem{pakdel2014nano}
\bibinfo{author}{Pakdel, A.} \emph{et~al.}
\newblock \bibinfo{title}{Nano boron nitride flatland}.
\newblock \emph{\bibinfo{journal}{Chemical Society Reviews}}
  \textbf{\bibinfo{volume}{43}}, \bibinfo{pages}{934--959}
  (\bibinfo{year}{2014}).

\bibitem{wang2012electronics}
\bibinfo{author}{Wang, Q.~H.} \emph{et~al.}
\newblock \bibinfo{title}{Electronics and optoelectronics of two-dimensional
  transition metal dichalcogenides}.
\newblock \emph{\bibinfo{journal}{Nature Nanotechnology}}
  \textbf{\bibinfo{volume}{7}}, \bibinfo{pages}{699--712}
  (\bibinfo{year}{2012}).

\bibitem{gupta2015recent}
\bibinfo{author}{Gupta, A.} \emph{et~al.}
\newblock \bibinfo{title}{Recent development in 2{D} materials beyond
  graphene}.
\newblock \emph{\bibinfo{journal}{Progress in Materials Science}}
  \textbf{\bibinfo{volume}{73}}, \bibinfo{pages}{44--126}
  (\bibinfo{year}{2015}).

\bibitem{geim_sci_2009}
\bibinfo{author}{Geim, A.~K.}
\newblock \bibinfo{title}{Graphene: status and prospects}.
\newblock \emph{\bibinfo{journal}{Science}} \textbf{\bibinfo{volume}{324}},
  \bibinfo{pages}{1530--1534} (\bibinfo{year}{2009}).

\bibitem{Hersam2013review}
\bibinfo{author}{Johns, J.~E.} \emph{et~al.}
\newblock \bibinfo{title}{Atomic covalent functionalization of graphene}.
\newblock \emph{\bibinfo{journal}{Accounts of Chemical Research}}
  \textbf{\bibinfo{volume}{46}}, \bibinfo{pages}{77--86}
  (\bibinfo{year}{2013}).

\bibitem{Geim2009graphane}
\bibinfo{author}{Elias, D.~C.} \emph{et~al.}
\newblock \bibinfo{title}{Control of graphene's properties by reversible
  hydrogenation: evidence for graphane}.
\newblock \emph{\bibinfo{journal}{Science}} \textbf{\bibinfo{volume}{323}},
  \bibinfo{pages}{610--613} (\bibinfo{year}{2009}).

\bibitem{Balog2010graphane}
\bibinfo{author}{Balog, R.} \emph{et~al.}
\newblock \bibinfo{title}{Bandgap opening in graphene induced by patterned
  hydrogen adsorption}.
\newblock \emph{\bibinfo{journal}{Nature Materials}}
  \textbf{\bibinfo{volume}{9}}, \bibinfo{pages}{315--319}
  (\bibinfo{year}{2010}).

\bibitem{hersam_acsn_2016}
\bibinfo{author}{Ryder, C.~R.} \emph{et~al.}
\newblock \bibinfo{title}{Chemically tailoring semiconducting two-dimensional
  transition metal dichalcogenides and black phosphorus}.
\newblock \emph{\bibinfo{journal}{ACS Nano}} \textbf{\bibinfo{volume}{10}},
  \bibinfo{pages}{3900--3917} (\bibinfo{year}{2016}).

\bibitem{macleod2014molecular}
\bibinfo{author}{MacLeod, J.} \emph{et~al.}
\newblock \bibinfo{title}{Molecular self-assembly on graphene}.
\newblock \emph{\bibinfo{journal}{Small}} \textbf{\bibinfo{volume}{10}},
  \bibinfo{pages}{1038--1049} (\bibinfo{year}{2014}).

\bibitem{mali2015nanostructuring}
\bibinfo{author}{Mali, K.~S.} \emph{et~al.}
\newblock \bibinfo{title}{Nanostructuring graphene for controlled and
  reproducible functionalization}.
\newblock \emph{\bibinfo{journal}{Nanoscale}} \textbf{\bibinfo{volume}{7}},
  \bibinfo{pages}{1566--1585} (\bibinfo{year}{2015}).

\bibitem{Ho2003fluor}
\bibinfo{author}{Qiu, X.~H.} \emph{et~al.}
\newblock \bibinfo{title}{Vibrationally resolved fluorescence excited with
  submolecular precision}.
\newblock \emph{\bibinfo{journal}{Science}} \textbf{\bibinfo{volume}{299}},
  \bibinfo{pages}{542--546} (\bibinfo{year}{2003}).

\bibitem{repp2005molecules}
\bibinfo{author}{Repp, J.} \emph{et~al.}
\newblock \bibinfo{title}{Molecules on insulating films: Scanning-tunneling
  microscopy imaging of individual molecular orbitals}.
\newblock \emph{\bibinfo{journal}{Physical Review Letters}}
  \textbf{\bibinfo{volume}{94}}, \bibinfo{pages}{026803}
  (\bibinfo{year}{2005}).

\bibitem{Swart2011review}
\bibinfo{author}{Swart, I.} \emph{et~al.}
\newblock \bibinfo{title}{Single-molecule chemistry and physics explored by
  low-temperature scanning probe microscopy}.
\newblock \emph{\bibinfo{journal}{Chemical Communications}}
  \textbf{\bibinfo{volume}{47}}, \bibinfo{pages}{9011--9023}
  (\bibinfo{year}{2011}).

\bibitem{liu2011chemical}
\bibinfo{author}{Liu, H.} \emph{et~al.}
\newblock \bibinfo{title}{Chemical doping of graphene}.
\newblock \emph{\bibinfo{journal}{Journal of Materials Chemistry}}
  \textbf{\bibinfo{volume}{21}}, \bibinfo{pages}{3335--3345}
  (\bibinfo{year}{2011}).

\bibitem{zhang2011tailoring}
\bibinfo{author}{Zhang, Z.} \emph{et~al.}
\newblock \bibinfo{title}{Tailoring electronic properties of graphene by
  $\pi$--$\pi$ stacking with aromatic molecules}.
\newblock \emph{\bibinfo{journal}{The Journal of Physical Chemistry Letters}}
  \textbf{\bibinfo{volume}{2}}, \bibinfo{pages}{2897--2905}
  (\bibinfo{year}{2011}).

\bibitem{mao2013manipulating}
\bibinfo{author}{Mao, H.~Y.} \emph{et~al.}
\newblock \bibinfo{title}{Manipulating the electronic and chemical properties
  of graphene via molecular functionalization}.
\newblock \emph{\bibinfo{journal}{Progress in Surface Science}}
  \textbf{\bibinfo{volume}{88}}, \bibinfo{pages}{132--159}
  (\bibinfo{year}{2013}).

\bibitem{tang2013graphene}
\bibinfo{author}{Tang, Q.} \emph{et~al.}
\newblock \bibinfo{title}{Graphene-related nanomaterials: tuning properties by
  functionalization}.
\newblock \emph{\bibinfo{journal}{Nanoscale}} \textbf{\bibinfo{volume}{5}},
  \bibinfo{pages}{4541--4583} (\bibinfo{year}{2013}).

\bibitem{kong2014molecular}
\bibinfo{author}{Kong, L.} \emph{et~al.}
\newblock \bibinfo{title}{Molecular adsorption on graphene}.
\newblock \emph{\bibinfo{journal}{Journal of Physics: Condensed Matter}}
  \textbf{\bibinfo{volume}{26}}, \bibinfo{pages}{443001}
  (\bibinfo{year}{2014}).

\bibitem{Georgakilas2012review}
\bibinfo{author}{Georgakilas, V.} \emph{et~al.}
\newblock \bibinfo{title}{Functionalization of graphene: covalent and
  non-covalent approaches, derivatives and applications}.
\newblock \emph{\bibinfo{journal}{Chemical Reviews}}
  \textbf{\bibinfo{volume}{112}}, \bibinfo{pages}{6156--6214}
  (\bibinfo{year}{2012}).

\bibitem{Koehler2013review}
\bibinfo{author}{Koehler, F.~M.} \emph{et~al.}
\newblock \bibinfo{title}{Organic synthesis on graphene}.
\newblock \emph{\bibinfo{journal}{Accounts of Chemical Research}}
  \textbf{\bibinfo{volume}{46}}, \bibinfo{pages}{2297--2306}
  (\bibinfo{year}{2013}).

\bibitem{Chua2013review}
\bibinfo{author}{Chua, C.~K.} \emph{et~al.}
\newblock \bibinfo{title}{Covalent chemistry on graphene}.
\newblock \emph{\bibinfo{journal}{Chemical Society Reviews}}
  \textbf{\bibinfo{volume}{42}}, \bibinfo{pages}{3222--3233}
  (\bibinfo{year}{2013}).

\bibitem{Voiry2014covalent}
\bibinfo{author}{Voiry, D.} \emph{et~al.}
\newblock \bibinfo{title}{Covalent functionalization of monolayered transition
  metal dichalcogenides by phase engineering}.
\newblock \emph{\bibinfo{journal}{Nat. Chem.}} \textbf{\bibinfo{volume}{7}},
  \bibinfo{pages}{45--49} (\bibinfo{year}{2015}).

\bibitem{Wang2015review}
\bibinfo{author}{Wang, H.} \emph{et~al.}
\newblock \bibinfo{title}{Physical and chemical tuning of two-dimensional
  transition metal dichalcogenides}.
\newblock \emph{\bibinfo{journal}{Chemical Society Reviews}}
  \textbf{\bibinfo{volume}{44}}, \bibinfo{pages}{2664--2680}
  (\bibinfo{year}{2015}).

\bibitem{cai2015noncovalent}
\bibinfo{author}{Cai, B.} \emph{et~al.}
\newblock \bibinfo{title}{Noncovalent molecular doping of two-dimensional
  materials}.
\newblock \emph{\bibinfo{journal}{ChemNanoMat}} \textbf{\bibinfo{volume}{1}},
  \bibinfo{pages}{542--557} (\bibinfo{year}{2015}).

\bibitem{bonaccorso2012production}
\bibinfo{author}{Bonaccorso, F.} \emph{et~al.}
\newblock \bibinfo{title}{Production and processing of graphene and 2{D}
  crystals}.
\newblock \emph{\bibinfo{journal}{Materials Today}}
  \textbf{\bibinfo{volume}{15}}, \bibinfo{pages}{564--589}
  (\bibinfo{year}{2012}).

\bibitem{Dean2010hBN}
\bibinfo{author}{Dean, C.~R.} \emph{et~al.}
\newblock \bibinfo{title}{Boron nitride substrates for high-quality graphene
  electronics}.
\newblock \emph{\bibinfo{journal}{Nature Nanotechnology}}
  \textbf{\bibinfo{volume}{5}}, \bibinfo{pages}{722--726}
  (\bibinfo{year}{2010}).

\bibitem{yankowitz2014graphene}
\bibinfo{author}{Yankowitz, M.} \emph{et~al.}
\newblock \bibinfo{title}{Graphene on hexagonal boron nitride}.
\newblock \emph{\bibinfo{journal}{Journal of Physics: Condensed Matter}}
  \textbf{\bibinfo{volume}{26}}, \bibinfo{pages}{303201}
  (\bibinfo{year}{2014}).

\bibitem{norimatsu2014epitaxial}
\bibinfo{author}{Norimatsu, W.} \emph{et~al.}
\newblock \bibinfo{title}{Epitaxial graphene on {SiC}$\{$0001$\}$: advances and
  perspectives}.
\newblock \emph{\bibinfo{journal}{Physical Chemistry Chemical Physics}}
  \textbf{\bibinfo{volume}{16}}, \bibinfo{pages}{3501--3511}
  (\bibinfo{year}{2014}).

\bibitem{batzill2012surface}
\bibinfo{author}{Batzill, M.}
\newblock \bibinfo{title}{The surface science of graphene: metal interfaces,
  {CVD} synthesis, nanoribbons, chemical modifications, and defects}.
\newblock \emph{\bibinfo{journal}{Surface Science Reports}}
  \textbf{\bibinfo{volume}{67}}, \bibinfo{pages}{83--115}
  (\bibinfo{year}{2012}).

\bibitem{Yang2013GhBN}
\bibinfo{author}{Yang, W.} \emph{et~al.}
\newblock \bibinfo{title}{Epitaxial growth of single-domain graphene on
  hexagonal boron nitride}.
\newblock \emph{\bibinfo{journal}{Nature Materials}}
  \textbf{\bibinfo{volume}{12}}, \bibinfo{pages}{792--797}
  (\bibinfo{year}{2013}).

\bibitem{varchon2007electronic}
\bibinfo{author}{Varchon, F.} \emph{et~al.}
\newblock \bibinfo{title}{Electronic structure of epitaxial graphene layers on
  {SiC}: effect of the substrate}.
\newblock \emph{\bibinfo{journal}{Physical Review Letters}}
  \textbf{\bibinfo{volume}{99}}, \bibinfo{pages}{126805}
  (\bibinfo{year}{2007}).

\bibitem{ohta2007interlayer}
\bibinfo{author}{Ohta, T.} \emph{et~al.}
\newblock \bibinfo{title}{Interlayer interaction and electronic screening in
  multilayer graphene investigated with angle-resolved photoemission
  spectroscopy}.
\newblock \emph{\bibinfo{journal}{Physical Review Letters}}
  \textbf{\bibinfo{volume}{98}}, \bibinfo{pages}{206802}
  (\bibinfo{year}{2007}).

\bibitem{wintterlin2009graphene}
\bibinfo{author}{Wintterlin, J.} \emph{et~al.}
\newblock \bibinfo{title}{Graphene on metal surfaces}.
\newblock \emph{\bibinfo{journal}{Surface Science}}
  \textbf{\bibinfo{volume}{603}}, \bibinfo{pages}{1841--1852}
  (\bibinfo{year}{2009}).

\bibitem{Pletikosic2009mini}
\bibinfo{author}{Pletikosic, I.} \emph{et~al.}
\newblock \bibinfo{title}{Dirac cones and minigaps for graphene on {I}r(111)}.
\newblock \emph{\bibinfo{journal}{Physical Review Letters}}
  \textbf{\bibinfo{volume}{102}}, \bibinfo{pages}{056808}
  (\bibinfo{year}{2009}).

\bibitem{Sutter2009GPt}
\bibinfo{author}{Sutter, P.} \emph{et~al.}
\newblock \bibinfo{title}{Graphene on {P}t(111): growth and substrate
  interaction}.
\newblock \emph{\bibinfo{journal}{Physical Review B}}
  \textbf{\bibinfo{volume}{80}}, \bibinfo{pages}{245411}
  (\bibinfo{year}{2009}).

\bibitem{michely_prl_2011}
\bibinfo{author}{Busse, C.} \emph{et~al.}
\newblock \bibinfo{title}{Graphene on {I}r(111): physisorption with chemical
  modulation}.
\newblock \emph{\bibinfo{journal}{Physical Review Letters}}
  \textbf{\bibinfo{volume}{107}}, \bibinfo{pages}{036101}
  (\bibinfo{year}{2011}).

\bibitem{Hamalainen2013LEED}
\bibinfo{author}{H\"am\"al\"ainen, S.~K.} \emph{et~al.}
\newblock \bibinfo{title}{Structure and local variations of the graphene
  moir\'e on {I}r(111)}.
\newblock \emph{\bibinfo{journal}{Physical Review B}}
  \textbf{\bibinfo{volume}{88}}, \bibinfo{pages}{201406}
  (\bibinfo{year}{2013}).

\bibitem{Moritz2010GRu}
\bibinfo{author}{Moritz, W.} \emph{et~al.}
\newblock \bibinfo{title}{Structure determination of the coincidence phase of
  graphene on {R}u(0001)}.
\newblock \emph{\bibinfo{journal}{Physical Review Letters}}
  \textbf{\bibinfo{volume}{104}}, \bibinfo{pages}{136102}
  (\bibinfo{year}{2010}).

\bibitem{wang2010coupling}
\bibinfo{author}{Wang, B.} \emph{et~al.}
\newblock \bibinfo{title}{Coupling epitaxy, chemical bonding, and work function
  at the local scale in transition metal-supported graphene}.
\newblock \emph{\bibinfo{journal}{ACS Nano}} \textbf{\bibinfo{volume}{4}},
  \bibinfo{pages}{5773--5782} (\bibinfo{year}{2010}).

\bibitem{Auwarter2003Ni}
\bibinfo{author}{Auw\"arter, W.} \emph{et~al.}
\newblock \bibinfo{title}{Defect lines and two-domain structure of hexagonal
  boron nitride films on {N}i(111)}.
\newblock \emph{\bibinfo{journal}{Surface Science}}
  \textbf{\bibinfo{volume}{545}}, \bibinfo{pages}{L735--L740}
  (\bibinfo{year}{2003}).

\bibitem{Dahal2014review}
\bibinfo{author}{Dahal, A.} \emph{et~al.}
\newblock \bibinfo{title}{Graphene-nickel interfaces: a review}.
\newblock \emph{\bibinfo{journal}{Nanoscale}} \textbf{\bibinfo{volume}{6}},
  \bibinfo{pages}{2548--2562} (\bibinfo{year}{2014}).

\bibitem{Drost2015Ni}
\bibinfo{author}{Drost, R.} \emph{et~al.}
\newblock \bibinfo{title}{Synthesis of extended atomically perfect zigzag
  graphene - boron nitride interfaces}.
\newblock \emph{\bibinfo{journal}{Scientific Reports}}
  \textbf{\bibinfo{volume}{5}}, \bibinfo{pages}{16741} (\bibinfo{year}{2015}).

\bibitem{voloshina2012graphene}
\bibinfo{author}{Voloshina, E.} \emph{et~al.}
\newblock \bibinfo{title}{Graphene on metallic surfaces: problems and
  perspectives}.
\newblock \emph{\bibinfo{journal}{Physical Chemistry Chemical Physics}}
  \textbf{\bibinfo{volume}{14}}, \bibinfo{pages}{13502--13514}
  (\bibinfo{year}{2012}).

\bibitem{Schulz2014BN}
\bibinfo{author}{Schulz, F.} \emph{et~al.}
\newblock \bibinfo{title}{Epitaxial hexagonal boron nitride on {I}r(111): a
  work function template}.
\newblock \emph{\bibinfo{journal}{Physical Review B}}
  \textbf{\bibinfo{volume}{89}}, \bibinfo{pages}{235429}
  (\bibinfo{year}{2014}).

\bibitem{diaz2013hexagonal}
\bibinfo{author}{D{\'\i}az, J.~G.} \emph{et~al.}
\newblock \bibinfo{title}{Hexagonal boron nitride on transition metal
  surfaces}.
\newblock \emph{\bibinfo{journal}{Theoretical Chemistry Accounts}}
  \textbf{\bibinfo{volume}{132}}, \bibinfo{pages}{1--17}
  (\bibinfo{year}{2013}).

\bibitem{Corso2004BN}
\bibinfo{author}{Corso, M.} \emph{et~al.}
\newblock \bibinfo{title}{Boron nitride nanomesh}.
\newblock \emph{\bibinfo{journal}{Science}} \textbf{\bibinfo{volume}{303}},
  \bibinfo{pages}{217--220} (\bibinfo{year}{2004}).

\bibitem{brugger2009comparison}
\bibinfo{author}{Brugger, T.} \emph{et~al.}
\newblock \bibinfo{title}{Comparison of electronic structure and template
  function of single-layer graphene and a hexagonal boron nitride nanomesh on
  {R}u(0001)}.
\newblock \emph{\bibinfo{journal}{Physical Review B}}
  \textbf{\bibinfo{volume}{79}}, \bibinfo{pages}{045407}
  (\bibinfo{year}{2009}).

\bibitem{cavar2008single}
\bibinfo{author}{{\'C}avar, E.} \emph{et~al.}
\newblock \bibinfo{title}{A single {h-BN} layer on {P}t(111)}.
\newblock \emph{\bibinfo{journal}{Surface Science}}
  \textbf{\bibinfo{volume}{602}}, \bibinfo{pages}{1722--1726}
  (\bibinfo{year}{2008}).

\bibitem{Joshi2012BN}
\bibinfo{author}{Joshi, S.} \emph{et~al.}
\newblock \bibinfo{title}{Boron nitride on {C}u(111): an electronically
  corrugated monolayer}.
\newblock \emph{\bibinfo{journal}{Nano Letters}} \textbf{\bibinfo{volume}{12}},
  \bibinfo{pages}{5821--5828} (\bibinfo{year}{2012}).

\bibitem{goriachko2007self}
\bibinfo{author}{Goriachko, A.} \emph{et~al.}
\newblock \bibinfo{title}{Self-assembly of a hexagonal boron nitride nanomesh
  on {R}u(0001)}.
\newblock \emph{\bibinfo{journal}{Langmuir}} \textbf{\bibinfo{volume}{23}},
  \bibinfo{pages}{2928--2931} (\bibinfo{year}{2007}).

\bibitem{greber_sci_2008}
\bibinfo{author}{Dil, H.} \emph{et~al.}
\newblock \bibinfo{title}{Surface trapping of atoms and molecules with dipole
  rings}.
\newblock \emph{\bibinfo{journal}{Science}} \textbf{\bibinfo{volume}{319}},
  \bibinfo{pages}{1824} (\bibinfo{year}{2008}).

\bibitem{gao_jacs_2009}
\bibinfo{author}{Mao, J.} \emph{et~al.}
\newblock \bibinfo{title}{Tunability of supramolecular kagome lattices of
  magnetic phthalocyanines using graphene-based moire\'e patterns as
  templates}.
\newblock \emph{\bibinfo{journal}{Journal of the American Chemical Society}}
  \textbf{\bibinfo{volume}{131}}, \bibinfo{pages}{14136â€“14137}
  (\bibinfo{year}{2009}).

\bibitem{gao_jpcc_2012_2}
\bibinfo{author}{Yang, K.} \emph{et~al.}
\newblock \bibinfo{title}{Molecule âˆ’substrate coupling between metal
  phthalocyanines and epitaxial graphene grown on {R}u(0001) and {P}t(11)}.
\newblock \emph{\bibinfo{journal}{Journal of Physical Chemistry C}}
  \textbf{\bibinfo{volume}{116}}, \bibinfo{pages}{14052âˆ’14056}
  (\bibinfo{year}{2012}).

\bibitem{gao_prb_2011}
\bibinfo{author}{Zhang, H.~G.} \emph{et~al.}
\newblock \bibinfo{title}{Assembly of iron phthalocyanine and pentacene
  molecules on a graphene monolayer grown on {R}u(0001)}.
\newblock \emph{\bibinfo{journal}{Physical Review B}}
  \textbf{\bibinfo{volume}{84}}, \bibinfo{pages}{245436}
  (\bibinfo{year}{2011}).

\bibitem{gao_jpcc_2012}
\bibinfo{author}{Zhang, H.} \emph{et~al.}
\newblock \bibinfo{title}{Host âˆ’guest superstructures on graphene-based
  kagome lattice}.
\newblock \emph{\bibinfo{journal}{Journal of Physical Chemistry C}}
  \textbf{\bibinfo{volume}{116}}, \bibinfo{pages}{11091âˆ’11095}
  (\bibinfo{year}{2012}).

\bibitem{zhou2013template}
\bibinfo{author}{Zhou, H.} \emph{et~al.}
\newblock \bibinfo{title}{Template-directed assembly of pentacene molecules on
  epitaxial graphene on {R}u(0001)}.
\newblock \emph{\bibinfo{journal}{Nano Research}} \textbf{\bibinfo{volume}{6}},
  \bibinfo{pages}{131--137} (\bibinfo{year}{2013}).

\bibitem{roos2011intermolecular}
\bibinfo{author}{Roos, M.} \emph{et~al.}
\newblock \bibinfo{title}{Intermolecular vs molecule--substrate interactions: a
  combined {STM} and theoretical study of supramolecular phases on
  graphene/{R}u(0001)}.
\newblock \emph{\bibinfo{journal}{Beilstein Journal of Nanotechnology}}
  \textbf{\bibinfo{volume}{2}}, \bibinfo{pages}{365--373}
  (\bibinfo{year}{2011}).

\bibitem{gao_apl_2011}
\bibinfo{author}{Zhou, H.~T.} \emph{et~al.}
\newblock \bibinfo{title}{Direct imaging of intrinsic molecular orbitals using
  two-dimensional, epitaxially-grown, nanostructured graphene for study of
  single molecule and interactions}.
\newblock \emph{\bibinfo{journal}{Applied Physics Letters}}
  \textbf{\bibinfo{volume}{99}}, \bibinfo{pages}{153101}
  (\bibinfo{year}{2011}).

\bibitem{Garnica2013}
\bibinfo{author}{Garnica, M.} \emph{et~al.}
\newblock \bibinfo{title}{Long-range magnetic order in a purely organic 2{D}
  layer adsorbed on epitaxial graphene}.
\newblock \emph{\bibinfo{journal}{Nature Physics}}
  \textbf{\bibinfo{volume}{9}}, \bibinfo{pages}{368} (\bibinfo{year}{2013}).

\bibitem{miranda_cm_2014}
\bibinfo{author}{Maccariello, D.} \emph{et~al.}
\newblock \bibinfo{title}{Spatially resolved, site-dependent charge transfer
  and induced magnetic moment in {TCNQ} adsorbed on graphene}.
\newblock \emph{\bibinfo{journal}{Chemistry of Materials}}
  \textbf{\bibinfo{volume}{26}}, \bibinfo{pages}{2883âˆ’2890}
  (\bibinfo{year}{2014}).

\bibitem{loh_acsn_2012}
\bibinfo{author}{Lu, J.} \emph{et~al.}
\newblock \bibinfo{title}{Using the graphene moir\'e pattern for the trapping
  of {C}$_{60}$ and homoepitaxy of graphene}.
\newblock \emph{\bibinfo{journal}{ACS Nano}} \textbf{\bibinfo{volume}{6}},
  \bibinfo{pages}{944} (\bibinfo{year}{2012}).

\bibitem{li2012self}
\bibinfo{author}{Li, G.} \emph{et~al.}
\newblock \bibinfo{title}{Self-assembly of {C}$_{60}$ monolayer on epitaxially
  grown, nanostructured graphene on {R}u(0001) surface}.
\newblock \emph{\bibinfo{journal}{Applied Physics Letters}}
  \textbf{\bibinfo{volume}{100}}, \bibinfo{pages}{013304}
  (\bibinfo{year}{2012}).

\bibitem{liljeroth_jpcc_2014}
\bibinfo{author}{J\"arvinen, P.} \emph{et~al.}
\newblock \bibinfo{title}{Self-assembly and orbital imaging of metal
  phthalocyanines on a graphene model surface}.
\newblock \emph{\bibinfo{journal}{Journal of Physical Chemistry C}}
  \textbf{\bibinfo{volume}{118}}, \bibinfo{pages}{13320--13325}
  (\bibinfo{year}{2014}).

\bibitem{Tsai2015}
\bibinfo{author}{Tsai, H.-Z.} \emph{et~al.}
\newblock \bibinfo{title}{Molecular self-assembly in a poorly screened
  environment: F$_4${TCNQ} on graphene/{BN}}.
\newblock \emph{\bibinfo{journal}{ACS Nano}} \textbf{\bibinfo{volume}{9}},
  \bibinfo{pages}{12168--12173} (\bibinfo{year}{2015}).

\bibitem{nilius_jpcc_2008}
\bibinfo{author}{Lin, X.} \emph{et~al.}
\newblock \bibinfo{title}{Self-assembly of {MgPc} molecules on polar {FeO} thin
  films}.
\newblock \emph{\bibinfo{journal}{Journal of Physical Chemistry C}}
  \textbf{\bibinfo{volume}{112}}, \bibinfo{pages}{15325--15328}
  (\bibinfo{year}{2008}).

\bibitem{Milian2004}
\bibinfo{author}{Milian, B.} \emph{et~al.}
\newblock \bibinfo{title}{On the electron affinity of {TCNQ}}.
\newblock \emph{\bibinfo{journal}{Chemical Physics Letters}}
  \textbf{\bibinfo{volume}{391}}, \bibinfo{pages}{148--151}
  (\bibinfo{year}{2004}).

\bibitem{groning_pccp_2014}
\bibinfo{author}{Iannuzzi, M.} \emph{et~al.}
\newblock \bibinfo{title}{Site-selective adsorption of phthalocyanine on
  {h-BN}/{R}h(11) nanomesh}.
\newblock \emph{\bibinfo{journal}{Physical Chemistry Chemical Physics}}
  \textbf{\bibinfo{volume}{16}}, \bibinfo{pages}{12374} (\bibinfo{year}{2014}).

\bibitem{liljeroth_acsn_2013}
\bibinfo{author}{Schulz, F.} \emph{et~al.}
\newblock \bibinfo{title}{Templated self-assembly and local doping of molecules
  on epitaxial hexagonal boron nitride}.
\newblock \emph{\bibinfo{journal}{ACS Nano}} \textbf{\bibinfo{volume}{7}},
  \bibinfo{pages}{11121--11128} (\bibinfo{year}{2013}).

\bibitem{auwarter_acsn_2014}
\bibinfo{author}{Joshi, S.} \emph{et~al.}
\newblock \bibinfo{title}{Control of molecular organization and energy level
  alignment by an electronically nanopatterned boron nitride template}.
\newblock \emph{\bibinfo{journal}{ACS Nano}} \textbf{\bibinfo{volume}{8}},
  \bibinfo{pages}{430} (\bibinfo{year}{2014}).

\bibitem{Ternes2013Mn12}
\bibinfo{author}{Kahle, S.} \emph{et~al.}
\newblock \bibinfo{title}{The quantum magnetism of individual
  manganese-12-acetate molecular magnets anchored at surfaces}.
\newblock \emph{\bibinfo{journal}{Nano Letters}} \textbf{\bibinfo{volume}{12}},
  \bibinfo{pages}{518--521} (\bibinfo{year}{2012}).

\bibitem{liljeroth_natphy_2015}
\bibinfo{author}{Schulz, F.} \emph{et~al.}
\newblock \bibinfo{title}{Many-body transitions in a single molecule visualized
  by scanning tunnelling microscopy}.
\newblock \emph{\bibinfo{journal}{Nature Physics}}
  \textbf{\bibinfo{volume}{11}}, \bibinfo{pages}{229--234}
  (\bibinfo{year}{2015}).

\bibitem{osterwalder_angc_2007}
\bibinfo{author}{Berner, S.} \emph{et~al.}
\newblock \bibinfo{title}{Boron nitride nanomesh: functionality from a
  corrugated monolayer}.
\newblock \emph{\bibinfo{journal}{Angewandte Chemie International Edition}}
  \textbf{\bibinfo{volume}{46}}, \bibinfo{pages}{5115--5119}
  (\bibinfo{year}{2007}).

\bibitem{Groning2014radical}
\bibinfo{author}{Dienel, T.} \emph{et~al.}
\newblock \bibinfo{title}{Dehalogenation and coupling of a polycyclic
  hydrocarbon on an atomically thin insulator}.
\newblock \emph{\bibinfo{journal}{ACS Nano}} \textbf{\bibinfo{volume}{8}},
  \bibinfo{pages}{6571--6579} (\bibinfo{year}{2014}).

\bibitem{groning_nanosc_2010}
\bibinfo{author}{Widmer, R.} \emph{et~al.}
\newblock \bibinfo{title}{Probing the selectivity of a nanostructured surface
  by xenon adsorption}.
\newblock \emph{\bibinfo{journal}{Nanoscale}} \textbf{\bibinfo{volume}{2}},
  \bibinfo{pages}{502--508} (\bibinfo{year}{2010}).

\bibitem{Urgel2015coord}
\bibinfo{author}{Urgel, J.~I.} \emph{et~al.}
\newblock \bibinfo{title}{Controlling coordination reactions and assembly on a
  {C}u(111) supported boron nitride monolayer}.
\newblock \emph{\bibinfo{journal}{Journal of the American Chemical Society}}
  \textbf{\bibinfo{volume}{137}}, \bibinfo{pages}{2420--2423}
  (\bibinfo{year}{2015}).

\bibitem{Dedkov2014}
\bibinfo{author}{Dedkov, Y.} \emph{et~al.}
\newblock \bibinfo{title}{Multichannel scanning probe microscopy and
  spectroscopy of graphene moir\'e structures}.
\newblock \emph{\bibinfo{journal}{Physical Chemistry Chemical Physics}}
  \textbf{\bibinfo{volume}{16}}, \bibinfo{pages}{3894--3908}
  (\bibinfo{year}{2014}).

\bibitem{Altenburg2014}
\bibinfo{author}{Altenburg, S.~J.} \emph{et~al.}
\newblock \bibinfo{title}{Local work function and {STM} tip-induced distortion
  of graphene on {I}r(111)}.
\newblock \emph{\bibinfo{journal}{New Journal of Physics}}
  \textbf{\bibinfo{volume}{16}}, \bibinfo{pages}{053036}
  (\bibinfo{year}{2014}).

\bibitem{liljeroth_prb_2011}
\bibinfo{author}{Sun, Z.} \emph{et~al.}
\newblock \bibinfo{title}{Topographic and electronic contrast of the graphene
  moir\'e on {I}r(111) probed by scanning tunneling microscopy and noncontact
  atomic force microscopy}.
\newblock \emph{\bibinfo{journal}{Physical Review B}}
  \textbf{\bibinfo{volume}{83}}, \bibinfo{pages}{081415}
  (\bibinfo{year}{2011}).

\bibitem{miranda_cc_2010}
\bibinfo{author}{Barja, S.} \emph{et~al.}
\newblock \bibinfo{title}{Self-organization of electron acceptor molecules on
  graphene}.
\newblock \emph{\bibinfo{journal}{Chemical Communications}}
  \textbf{\bibinfo{volume}{46}}, \bibinfo{pages}{8198} (\bibinfo{year}{2010}).

\bibitem{sainio_jpcc_2012}
\bibinfo{author}{H\"am\"al\"ainen, S.~K.} \emph{et~al.}
\newblock \bibinfo{title}{Self-assembly of cobalt-phthalocyanine molecules on
  epitaxial graphene on {I}r(111)}.
\newblock \emph{\bibinfo{journal}{Journal of Physical Chemistry C}}
  \textbf{\bibinfo{volume}{116}}, \bibinfo{pages}{20433}
  (\bibinfo{year}{2012}).

\bibitem{wang2010selective}
\bibinfo{author}{Wang, Y.-L.} \emph{et~al.}
\newblock \bibinfo{title}{Selective adsorption and electronic interaction of
  {F}$_{16}${CuPc} on epitaxial graphene}.
\newblock \emph{\bibinfo{journal}{Physical Review B}}
  \textbf{\bibinfo{volume}{82}}, \bibinfo{pages}{245420}
  (\bibinfo{year}{2010}).

\bibitem{liljeroth_nl_2013}
\bibinfo{author}{J\"arvinen, P.} \emph{et~al.}
\newblock \bibinfo{title}{Molecular self-assembly on graphene on {SiO}$_2$ and
  {h-BN} substrates}.
\newblock \emph{\bibinfo{journal}{Nano Letters}} \textbf{\bibinfo{volume}{13}},
  \bibinfo{pages}{3199} (\bibinfo{year}{2013}).

\bibitem{jung2014atomically}
\bibinfo{author}{Jung, M.} \emph{et~al.}
\newblock \bibinfo{title}{Atomically resolved orientational ordering of
  {C}$_{60}$ molecules on epitaxial graphene on {C}u(111)}.
\newblock \emph{\bibinfo{journal}{Nanoscale}} \textbf{\bibinfo{volume}{6}},
  \bibinfo{pages}{11835--11840} (\bibinfo{year}{2014}).

\bibitem{guisinger_nl_2012}
\bibinfo{author}{Cho, J.} \emph{et~al.}
\newblock \bibinfo{title}{Structural and electronic decoupling of {C}$_{60}$
  from epitaxial graphene on {SiC}}.
\newblock \emph{\bibinfo{journal}{Nano Letters}} \textbf{\bibinfo{volume}{12}},
  \bibinfo{pages}{3018--3024} (\bibinfo{year}{2012}).

\bibitem{vsvec2012van}
\bibinfo{author}{{\v{S}}vec, M.} \emph{et~al.}
\newblock \bibinfo{title}{van der {W}aals interactions mediating the cohesion
  of fullerenes on graphene}.
\newblock \emph{\bibinfo{journal}{Physical Review B}}
  \textbf{\bibinfo{volume}{86}}, \bibinfo{pages}{121407}
  (\bibinfo{year}{2012}).

\bibitem{rodriguez_jpcc_2014}
\bibinfo{author}{Mart{\'\i}nez-Galera, A.~J.} \emph{et~al.}
\newblock \bibinfo{title}{Imaging molecular orbitals of {PTCDA} on graphene on
  {P}t(111): electronic structure by {STM} and first-principles calculations}.
\newblock \emph{\bibinfo{journal}{Journal of Physical Chemistry C}}
  \textbf{\bibinfo{volume}{118}}, \bibinfo{pages}{12782}
  (\bibinfo{year}{2014}).

\bibitem{huang2009structural}
\bibinfo{author}{Huang, H.} \emph{et~al.}
\newblock \bibinfo{title}{Structural and electronic properties of {PTCDA} thin
  films on epitaxial graphene}.
\newblock \emph{\bibinfo{journal}{ACS Nano}} \textbf{\bibinfo{volume}{3}},
  \bibinfo{pages}{3431--3436} (\bibinfo{year}{2009}).

\bibitem{wang2009room}
\bibinfo{author}{Wang, Q.~H.} \emph{et~al.}
\newblock \bibinfo{title}{Room-temperature molecular-resolution
  characterization of self-assembled organic monolayers on epitaxial graphene}.
\newblock \emph{\bibinfo{journal}{Nature Chemistry}}
  \textbf{\bibinfo{volume}{1}}, \bibinfo{pages}{206--211}
  (\bibinfo{year}{2009}).

\bibitem{emery2011structural}
\bibinfo{author}{Emery, J.~D.} \emph{et~al.}
\newblock \bibinfo{title}{Structural analysis of {PTCDA} monolayers on
  epitaxial graphene with ultra-high vacuum scanning tunneling microscopy and
  high-resolution {X}-ray reflectivity}.
\newblock \emph{\bibinfo{journal}{Surface Science}}
  \textbf{\bibinfo{volume}{605}}, \bibinfo{pages}{1685--1693}
  (\bibinfo{year}{2011}).

\bibitem{alaboson2011seeding}
\bibinfo{author}{Alaboson, J.~M.} \emph{et~al.}
\newblock \bibinfo{title}{Seeding atomic layer deposition of high-k dielectrics
  on epitaxial graphene with organic self-assembled monolayers}.
\newblock \emph{\bibinfo{journal}{ACS Nano}} \textbf{\bibinfo{volume}{5}},
  \bibinfo{pages}{5223--5232} (\bibinfo{year}{2011}).

\bibitem{karmel2014self}
\bibinfo{author}{Karmel, H.~J.} \emph{et~al.}
\newblock \bibinfo{title}{Self-assembled organic monolayers on epitaxial
  graphene with enhanced structural and thermal stability}.
\newblock \emph{\bibinfo{journal}{Chemical Communications}}
  \textbf{\bibinfo{volume}{50}}, \bibinfo{pages}{8852--8855}
  (\bibinfo{year}{2014}).

\bibitem{martin_ns_2014}
\bibinfo{author}{Stradi, D.} \emph{et~al.}
\newblock \bibinfo{title}{Controlling the spatial arrangement of organic
  magnetic anions adsorbed on epitaxial graphene on {R}u(0001)}.
\newblock \emph{\bibinfo{journal}{Nanoscale}} \textbf{\bibinfo{volume}{6}},
  \bibinfo{pages}{15271} (\bibinfo{year}{2014}).

\bibitem{Decker2013}
\bibinfo{author}{Decker, R.} \emph{et~al.}
\newblock \bibinfo{title}{Atomic-scale magnetism of cobalt-intercalated
  graphene}.
\newblock \emph{\bibinfo{journal}{Physical Review B}}
  \textbf{\bibinfo{volume}{87}}, \bibinfo{pages}{041403}
  (\bibinfo{year}{2013}).

\bibitem{Decker2014}
\bibinfo{author}{Decker, R.} \emph{et~al.}
\newblock \bibinfo{title}{Local tunnel magnetoresistance of an iron
  intercalated graphene-based heterostructure}.
\newblock \emph{\bibinfo{journal}{Journal of Physics: Condensed Matter}}
  \textbf{\bibinfo{volume}{26}}, \bibinfo{pages}{394004}
  (\bibinfo{year}{2014}).

\bibitem{wiesendanger_acsn_2013}
\bibinfo{author}{Bazarnik, M.} \emph{et~al.}
\newblock \bibinfo{title}{Tailoring molecular self-assembly of magnetic
  phthalocyanine molecules on {F}e- and {C}o-intercalated graphene}.
\newblock \emph{\bibinfo{journal}{ACS Nano}} \textbf{\bibinfo{volume}{12}},
  \bibinfo{pages}{11341} (\bibinfo{year}{2013}).

\bibitem{roos2011hierarchical}
\bibinfo{author}{Roos, M.} \emph{et~al.}
\newblock \bibinfo{title}{Hierarchical interactions and their influence upon
  the adsorption of organic molecules on a graphene film}.
\newblock \emph{\bibinfo{journal}{Journal of the American Chemical Society}}
  \textbf{\bibinfo{volume}{133}}, \bibinfo{pages}{9208--9211}
  (\bibinfo{year}{2011}).

\bibitem{ahn_apl_2014}
\bibinfo{author}{Jung, W.} \emph{et~al.}
\newblock \bibinfo{title}{Influence of graphene-substrate interactions on
  configurations of organic molecules on graphene: pentacene/epitaxial
  graphene/{SiC}}.
\newblock \emph{\bibinfo{journal}{Applied Physics Letters}}
  \textbf{\bibinfo{volume}{105}}, \bibinfo{pages}{071606}
  (\bibinfo{year}{2014}).

\bibitem{Chen2008}
\bibinfo{author}{Chen, W.} \emph{et~al.}
\newblock \bibinfo{title}{Molecular orientation transition of organic thin
  films on graphite: the effect of intermolecular electrostatic and interfacial
  dispersion forces}.
\newblock \emph{\bibinfo{journal}{Chemical Communications}}
  \bibinfo{pages}{4276--4278} (\bibinfo{year}{2008}).

\bibitem{karmel2013self}
\bibinfo{author}{Karmel, H.~J.} \emph{et~al.}
\newblock \bibinfo{title}{Self-assembled two-dimensional heteromolecular
  nanoporous molecular arrays on epitaxial graphene}.
\newblock \emph{\bibinfo{journal}{The Journal of Physical Chemistry Letters}}
  \textbf{\bibinfo{volume}{5}}, \bibinfo{pages}{270--274}
  (\bibinfo{year}{2013}).

\bibitem{lackinger2009carboxylic}
\bibinfo{author}{Lackinger, M.} \emph{et~al.}
\newblock \bibinfo{title}{Carboxylic acids: versatile building blocks and
  mediators for two-dimensional supramolecular self-assembly}.
\newblock \emph{\bibinfo{journal}{Langmuir}} \textbf{\bibinfo{volume}{25}},
  \bibinfo{pages}{11307--11321} (\bibinfo{year}{2009}).

\bibitem{macleod2015substrate}
\bibinfo{author}{MacLeod, J.} \emph{et~al.}
\newblock \bibinfo{title}{Substrate effects in the supramolecular assembly of
  1, 3, 5-benzene tricarboxylic acid on graphite and graphene}.
\newblock \emph{\bibinfo{journal}{Langmuir}} \textbf{\bibinfo{volume}{31}},
  \bibinfo{pages}{7016--7024} (\bibinfo{year}{2015}).

\bibitem{zhou2014switchable}
\bibinfo{author}{Zhou, Q.} \emph{et~al.}
\newblock \bibinfo{title}{Switchable supramolecular assemblies on graphene}.
\newblock \emph{\bibinfo{journal}{Nanoscale}} \textbf{\bibinfo{volume}{6}},
  \bibinfo{pages}{8387--8391} (\bibinfo{year}{2014}).

\bibitem{banerjee2016flexible}
\bibinfo{author}{Banerjee, K.} \emph{et~al.}
\newblock \bibinfo{title}{Flexible self-assembled molecular templates on
  graphene}.
\newblock \emph{\bibinfo{journal}{The Journal of Physical Chemistry C}}
  \textbf{\bibinfo{volume}{120}}, \bibinfo{pages}{8772--8780}
  (\bibinfo{year}{2016}).

\bibitem{riss2014imaging}
\bibinfo{author}{Riss, A.} \emph{et~al.}
\newblock \bibinfo{title}{Imaging and tuning molecular levels at the surface of
  a gated graphene device}.
\newblock \emph{\bibinfo{journal}{ACS Nano}} \textbf{\bibinfo{volume}{8}},
  \bibinfo{pages}{5395--5401} (\bibinfo{year}{2014}).

\bibitem{Xi1992Ullmann}
\bibinfo{author}{Xi, M.} \emph{et~al.}
\newblock \bibinfo{title}{Iodobenzene on {C}u(111): formation and coupling of
  adsorbed phenyl groups}.
\newblock \emph{\bibinfo{journal}{Surface Science}}
  \textbf{\bibinfo{volume}{278}}, \bibinfo{pages}{19 -- 32}
  (\bibinfo{year}{1992}).

\bibitem{Hla2000Ullmann}
\bibinfo{author}{Hla, S.~W.} \emph{et~al.}
\newblock \bibinfo{title}{Inducing all steps of a chemical reaction with the
  scanning tunneling microscope tip: towards single molecule engineering}.
\newblock \emph{\bibinfo{journal}{Physical Review Letters}}
  \textbf{\bibinfo{volume}{85}}, \bibinfo{pages}{2777--2780}
  (\bibinfo{year}{2000}).

\bibitem{Grill2007covalent}
\bibinfo{author}{Grill, L.} \emph{et~al.}
\newblock \bibinfo{title}{Nano-architectures by covalent assembly of molecular
  building blocks}.
\newblock \emph{\bibinfo{journal}{Nature Nanotechnology}}
  \textbf{\bibinfo{volume}{2}}, \bibinfo{pages}{687--691}
  (\bibinfo{year}{2007}).

\bibitem{Cai2010GNR}
\bibinfo{author}{Cai, J.} \emph{et~al.}
\newblock \bibinfo{title}{Atomically precise bottom-up fabrication of graphene
  nanoribbons}.
\newblock \emph{\bibinfo{journal}{Nature}} \textbf{\bibinfo{volume}{466}},
  \bibinfo{pages}{470--473} (\bibinfo{year}{2010}).

\bibitem{Laffarenz2009polymer}
\bibinfo{author}{Lafferentz, L.} \emph{et~al.}
\newblock \bibinfo{title}{Conductance of a single conjugated polymer as a
  continuous function of its length}.
\newblock \emph{\bibinfo{journal}{Science}} \textbf{\bibinfo{volume}{323}},
  \bibinfo{pages}{1193--1197} (\bibinfo{year}{2009}).

\bibitem{Chen2013GNR}
\bibinfo{author}{Chen, Y.-C.} \emph{et~al.}
\newblock \bibinfo{title}{Tuning the band gap of graphene nanoribbons
  synthesized from molecular precursors}.
\newblock \emph{\bibinfo{journal}{ACS Nano}} \textbf{\bibinfo{volume}{7}},
  \bibinfo{pages}{6123--6128} (\bibinfo{year}{2013}).

\bibitem{Cai2014hetero}
\bibinfo{author}{Cai, J.} \emph{et~al.}
\newblock \bibinfo{title}{Graphene nanoribbon heterojunctions}.
\newblock \emph{\bibinfo{journal}{Nature Nanotechnology}}
  \textbf{\bibinfo{volume}{9}}, \bibinfo{pages}{896--900}
  (\bibinfo{year}{2014}).

\bibitem{Kimouche2015GNR}
\bibinfo{author}{Kimouche, A.} \emph{et~al.}
\newblock \bibinfo{title}{Ultra-narrow metallic armchair graphene nanoribbons}.
\newblock \emph{\bibinfo{journal}{Nature Communications}}
  \textbf{\bibinfo{volume}{6}}, \bibinfo{pages}{10177} (\bibinfo{year}{2015}).

\bibitem{Chen2015mixed}
\bibinfo{author}{Chen, Y.-C.} \emph{et~al.}
\newblock \bibinfo{title}{Molecular bandgap engineering of bottom-up
  synthesized graphene nanoribbon heterojunctions}.
\newblock \emph{\bibinfo{journal}{Nature Nanotechnology}}
  \textbf{\bibinfo{volume}{10}}, \bibinfo{pages}{156--160}
  (\bibinfo{year}{2015}).

\bibitem{Ruffiex2016zigzag}
\bibinfo{author}{Ruffieux, P.} \emph{et~al.}
\newblock \bibinfo{title}{On-surface synthesis of graphene nanoribbons with
  zigzag edge topology}.
\newblock \emph{\bibinfo{journal}{Nature}} \textbf{\bibinfo{volume}{531}},
  \bibinfo{pages}{489--492} (\bibinfo{year}{2016}).

\bibitem{Talirz2016review}
\bibinfo{author}{Talirz, L.} \emph{et~al.}
\newblock \bibinfo{title}{On-surface synthesis of atomically precise graphene
  nanoribbons}.
\newblock \emph{\bibinfo{journal}{Advanced Materials}}
  \textbf{\bibinfo{volume}{28}}, \bibinfo{pages}{6222--6231}
  (\bibinfo{year}{2016}).

\bibitem{Bieri2010CHP}
\bibinfo{author}{Bieri, M.} \emph{et~al.}
\newblock \bibinfo{title}{Two-dimensional polymer formation on surfaces:
  insight into the roles of precursor mobility and reactivity}.
\newblock \emph{\bibinfo{journal}{Journal of the American Chemical Society}}
  \textbf{\bibinfo{volume}{132}}, \bibinfo{pages}{16669--16676}
  (\bibinfo{year}{2010}).

\bibitem{Bacle2014I6CHP}
\bibinfo{author}{Bacle, P.} \emph{et~al.}
\newblock \bibinfo{title}{Chemical reactions on metal-supported hexagonal boron
  nitride investigated with density functional theory}.
\newblock \emph{\bibinfo{journal}{CHIMIA International Journal for Chemistry}}
  \textbf{\bibinfo{volume}{68}}, \bibinfo{pages}{596--601}
  (\bibinfo{year}{2014}).

\bibitem{Kern2015hBN}
\bibinfo{author}{Morchutt, C.} \emph{et~al.}
\newblock \bibinfo{title}{Covalent coupling via dehalogenation on {N}i(111)
  supported boron nitride and graphene}.
\newblock \emph{\bibinfo{journal}{Chemical Communications}}
  \textbf{\bibinfo{volume}{51}}, \bibinfo{pages}{2440--2443}
  (\bibinfo{year}{2015}).

\bibitem{Neel2016H2Pc}
\bibinfo{author}{N\'eel, N.} \emph{et~al.}
\newblock \bibinfo{title}{Depopulation of single-phthalocyanine molecular
  orbitals upon pyrrolic-hydrogen abstraction on graphene}.
\newblock \emph{\bibinfo{journal}{ACS Nano}} \textbf{\bibinfo{volume}{10}},
  \bibinfo{pages}{2010--2016} (\bibinfo{year}{2016}).

\bibitem{Auwarter20124level}
\bibinfo{author}{Auw\"arter, W.} \emph{et~al.}
\newblock \bibinfo{title}{A surface-anchored molecular four-level conductance
  switch based on single proton transfer}.
\newblock \emph{\bibinfo{journal}{Nature Nanotechnology}}
  \textbf{\bibinfo{volume}{7}}, \bibinfo{pages}{41--46} (\bibinfo{year}{2012}).

\bibitem{Berndt2015FePc}
\bibinfo{author}{Altenburg, S.~J.} \emph{et~al.}
\newblock \bibinfo{title}{Reaction of phthalocyanines with graphene on
  {I}r(111)}.
\newblock \emph{\bibinfo{journal}{Journal of the American Chemical Society}}
  \textbf{\bibinfo{volume}{137}}, \bibinfo{pages}{9452--9458}
  (\bibinfo{year}{2015}).

\bibitem{bligaard_pnas_2011}
\bibinfo{author}{Norskov, J.~K.} \emph{et~al.}
\newblock \bibinfo{title}{Density functional theory in surface chemistry and
  catalysis}.
\newblock \emph{\bibinfo{journal}{Proceedings of the National Academy of
  Sciences}} \textbf{\bibinfo{volume}{108}}, \bibinfo{pages}{937--943}
  (\bibinfo{year}{2011}).

\bibitem{moody_ssr_2004}
\bibinfo{author}{Nilsson, A.} \emph{et~al.}
\newblock \bibinfo{title}{Chemical bonding on surfaces probed by {X}-ray
  emission spectroscopy and density functional theory}.
\newblock \emph{\bibinfo{journal}{Surface Science Reports}}
  \textbf{\bibinfo{volume}{55}}, \bibinfo{pages}{49--167}
  (\bibinfo{year}{2004}).

\bibitem{altibelli_cpl_1993}
\bibinfo{author}{Chavy, C.} \emph{et~al.}
\newblock \bibinfo{title}{Interpretation of {STM} images: {C}$_{60}$ on the
  gold (110) surface}.
\newblock \emph{\bibinfo{journal}{Chemical Physics Letters}}
  \textbf{\bibinfo{volume}{214}}, \bibinfo{pages}{569--575}
  (\bibinfo{year}{1993}).

\bibitem{joachim_ss_2009}
\bibinfo{author}{Villagomez, C.~J.} \emph{et~al.}
\newblock \bibinfo{title}{Stm images of a large organic molecule adsorbed on a
  bare metal substrate or on a thin insulating layer: Visualization of homo and
  lumo}.
\newblock \emph{\bibinfo{journal}{Surface Science}}
  \textbf{\bibinfo{volume}{603}}, \bibinfo{pages}{1526--1532}
  (\bibinfo{year}{2009}).

\bibitem{Repp2005}
\bibinfo{author}{Repp, J.} \emph{et~al.}
\newblock \bibinfo{title}{Molecules on insulating films: Scanning-tunneling
  microscopy imaging of individual molecular orbitals}.
\newblock \emph{\bibinfo{journal}{Physical Review Letters}}
  \textbf{\bibinfo{volume}{94}}, \bibinfo{pages}{026803}
  (\bibinfo{year}{2005}).

\bibitem{persson_sci_2006}
\bibinfo{author}{Repp, J.} \emph{et~al.}
\newblock \bibinfo{title}{Imaging bond formation between a gold atom and
  pentacene on an insulating surface}.
\newblock \emph{\bibinfo{journal}{Science}} \textbf{\bibinfo{volume}{312}},
  \bibinfo{pages}{1196--1199} (\bibinfo{year}{2006}).

\bibitem{meyer_sci_2007}
\bibinfo{author}{Liljeroth, P.} \emph{et~al.}
\newblock \bibinfo{title}{Current-induced hydrogen tautomerization and
  conductance switching of naphthalocyanine molecules}.
\newblock \emph{\bibinfo{journal}{Science}} \textbf{\bibinfo{volume}{317}},
  \bibinfo{pages}{1203} (\bibinfo{year}{2007}).

\bibitem{ho_jcp_2004}
\bibinfo{author}{Liu, N.} \emph{et~al.}
\newblock \bibinfo{title}{Vibronic states in single molecules: {C}$_{60}$ and
  {C}$_{70}$ on ultrathin {Al}$_2${O}$_3$ films}.
\newblock \emph{\bibinfo{journal}{Journal of Chemical Physics}}
  \textbf{\bibinfo{volume}{120}}, \bibinfo{pages}{11371}
  (\bibinfo{year}{2004}).

\bibitem{kawai_prl_2009}
\bibinfo{author}{Tsukahara, N.} \emph{et~al.}
\newblock \bibinfo{title}{Adsorption-induced switching of magnetic anisotropy
  in a single iron(ii) phthalocyanine molecule on an oxidized {C}u(110)
  surface}.
\newblock \emph{\bibinfo{journal}{Physical Review Letters}}
  \textbf{\bibinfo{volume}{102}}, \bibinfo{pages}{167203}
  (\bibinfo{year}{2009}).

\bibitem{heinrich_sci_2007}
\bibinfo{author}{Hirjibehedin, C.~F.} \emph{et~al.}
\newblock \bibinfo{title}{Large magnetic anisotropy of a single atomic spin
  embedded in a surface molecular network}.
\newblock \emph{\bibinfo{journal}{Science}} \textbf{\bibinfo{volume}{317}},
  \bibinfo{pages}{1199--1203} (\bibinfo{year}{2007}).

\bibitem{pasa_jpcc_2013}
\bibinfo{author}{Zoldan, V.~C.} \emph{et~al.}
\newblock \bibinfo{title}{Coupling of cobalt-tetraphenylporphyrin molecules to
  a copper nitride layer}.
\newblock \emph{\bibinfo{journal}{Journal of Physical Chemistry C}}
  \textbf{\bibinfo{volume}{117}}, \bibinfo{pages}{15984--15990}
  (\bibinfo{year}{2013}).

\bibitem{joachim_nl_2009}
\bibinfo{author}{Bellec, A.} \emph{et~al.}
\newblock \bibinfo{title}{Imaging molecular orbitals by scanning tunneling
  microscopy on a passivated semiconductor}.
\newblock \emph{\bibinfo{journal}{Nano Letters}} \textbf{\bibinfo{volume}{9}},
  \bibinfo{pages}{144--147} (\bibinfo{year}{2009}).

\bibitem{szymonski_acsn_2013}
\bibinfo{author}{Godlewski, S.} \emph{et~al.}
\newblock \bibinfo{title}{Contacting a conjugated molecule with a surface
  dangling bond dimer on a hydrogenated {G}e(001) surface allows imaging of the
  hidden ground electronic state}.
\newblock \emph{\bibinfo{journal}{ACS Nano}} \textbf{\bibinfo{volume}{7}},
  \bibinfo{pages}{10105--10111} (\bibinfo{year}{2013}).

\bibitem{repp_prl_2013}
\bibinfo{author}{Pavlicek, N.} \emph{et~al.}
\newblock \bibinfo{title}{Symmetry dependence of vibration-assisted tunneling}.
\newblock \emph{\bibinfo{journal}{Physical Review Letters}}
  \textbf{\bibinfo{volume}{110}}, \bibinfo{pages}{136101}
  (\bibinfo{year}{2013}).

\bibitem{repp2010coherent}
\bibinfo{author}{Repp, J.} \emph{et~al.}
\newblock \bibinfo{title}{Coherent electron-nuclear coupling in oligothiophene
  molecular wires}.
\newblock \emph{\bibinfo{journal}{Nature Physics}}
  \textbf{\bibinfo{volume}{6}}, \bibinfo{pages}{975--979}
  (\bibinfo{year}{2010}).

\bibitem{millo_nat_1999}
\bibinfo{author}{Banin, U.} \emph{et~al.}
\newblock \bibinfo{title}{Identification of atomic-like electronic states in
  indium arsenide nanocrystal quantum dots}.
\newblock \emph{\bibinfo{journal}{Nature}} \textbf{\bibinfo{volume}{400}},
  \bibinfo{pages}{542--544} (\bibinfo{year}{1999}).

\bibitem{kahn_cpl_2002}
\bibinfo{author}{Tsiper, E.} \emph{et~al.}
\newblock \bibinfo{title}{Electronic polarization at surfaces and thin films of
  organic molecular crystals: {PTCDA}}.
\newblock \emph{\bibinfo{journal}{Chemical Physics Letters}}
  \textbf{\bibinfo{volume}{360}}, \bibinfo{pages}{47--52}
  (\bibinfo{year}{2002}).

\bibitem{zhu_prl_2005}
\bibinfo{author}{Zhao, J.} \emph{et~al.}
\newblock \bibinfo{title}{Single ${\mathrm{c}}_{59}\mathrm{N}$ molecule as a
  molecular rectifier}.
\newblock \emph{\bibinfo{journal}{Physical Review Letters}}
  \textbf{\bibinfo{volume}{95}}, \bibinfo{pages}{045502}
  (\bibinfo{year}{2005}).

\bibitem{zhu_jcp_2006}
\bibinfo{author}{Li, B.} \emph{et~al.}
\newblock \bibinfo{title}{Single-electron tunneling spectroscopy of single
  {C}$_{60}$ in double-barrier tunnel junction}.
\newblock \emph{\bibinfo{journal}{The Journal of Chemical Physics}}
  \textbf{\bibinfo{volume}{124}}, \bibinfo{pages}{064709}
  (\bibinfo{year}{2006}).

\bibitem{berndt_jpcc_2011}
\bibinfo{author}{Gopakumar, T.~G.} \emph{et~al.}
\newblock \bibinfo{title}{Coverage-driven electronic decoupling of
  fe-phthalocyanine from a ag(111) substrate}.
\newblock \emph{\bibinfo{journal}{Journal of Physical Chemistry C}}
  \textbf{\bibinfo{volume}{115}}, \bibinfo{pages}{12173--12179}
  (\bibinfo{year}{2011}).

\bibitem{ho_pnas_2005}
\bibinfo{author}{Nazin, G.} \emph{et~al.}
\newblock \bibinfo{title}{Tunneling rates in electron transport through
  double-barrier molecular junctions in a scanning tunneling microscope}.
\newblock \emph{\bibinfo{journal}{Proceedings of the National Academy of
  Sciences of the United States of America}} \textbf{\bibinfo{volume}{102}},
  \bibinfo{pages}{8832--8837} (\bibinfo{year}{2005}).

\bibitem{persson_prl_2005}
\bibinfo{author}{Repp, J.} \emph{et~al.}
\newblock \bibinfo{title}{Scanning tunneling spectroscopy of cl vacancies in
  nacl films: Strong electron-phonon coupling in double-barrier tunneling
  junctions}.
\newblock \emph{\bibinfo{journal}{Physical Review Letters}}
  \textbf{\bibinfo{volume}{95}}, \bibinfo{pages}{225503}
  (\bibinfo{year}{2005}).

\bibitem{pascal_cpl_2000}
\bibinfo{author}{Hill, I.} \emph{et~al.}
\newblock \bibinfo{title}{Charge-separation energy in films of $\pi$-conjugated
  organic molecules}.
\newblock \emph{\bibinfo{journal}{Chemical Physics Letters}}
  \textbf{\bibinfo{volume}{327}}, \bibinfo{pages}{181--188}
  (\bibinfo{year}{2000}).

\bibitem{van2010charge}
\bibinfo{author}{van~der Molen, S.~J.} \emph{et~al.}
\newblock \bibinfo{title}{Charge transport through molecular switches}.
\newblock \emph{\bibinfo{journal}{Journal of Physics: Condensed Matter}}
  \textbf{\bibinfo{volume}{22}}, \bibinfo{pages}{133001}
  (\bibinfo{year}{2010}).

\bibitem{seki_advmat_1999}
\bibinfo{author}{Ishii, H.} \emph{et~al.}
\newblock \bibinfo{title}{Energy level alignment and interfacial electronic
  structures at organic/metal and organic/organic interfaces}.
\newblock \emph{\bibinfo{journal}{Advanced Materials}}
  \textbf{\bibinfo{volume}{11}}, \bibinfo{pages}{605--625}
  (\bibinfo{year}{1999}).

\bibitem{louie_prl_2006}
\bibinfo{author}{Neaton, J.~B.} \emph{et~al.}
\newblock \bibinfo{title}{Renormalization of molecular electronic levels at
  metal-molecule interfaces}.
\newblock \emph{\bibinfo{journal}{Physical Review Letters}}
  \textbf{\bibinfo{volume}{97}}, \bibinfo{pages}{216405--}
  (\bibinfo{year}{2006}).

\bibitem{fahlman_advmat_2009}
\bibinfo{author}{Braun, S.} \emph{et~al.}
\newblock \bibinfo{title}{Energy-level alignment at organic/metal and
  organic/organic interfaces}.
\newblock \emph{\bibinfo{journal}{Advanced Materials}}
  \textbf{\bibinfo{volume}{21}}, \bibinfo{pages}{1450--1472}
  (\bibinfo{year}{2009}).

\bibitem{persson_prl_2011}
\bibinfo{author}{Gross, L.} \emph{et~al.}
\newblock \bibinfo{title}{High-resolution molecular orbital imaging using a
  $p$-wave {STM} tip}.
\newblock \emph{\bibinfo{journal}{Physical Review Letters}}
  \textbf{\bibinfo{volume}{107}}, \bibinfo{pages}{086101}
  (\bibinfo{year}{2011}).

\bibitem{kroger_jcp_2014}
\bibinfo{author}{Endlich, M.} \emph{et~al.}
\newblock \bibinfo{title}{Phthalocyanine adsorption to graphene on {I}r(111):
  evidence for decoupling from vibrational spectroscopy}.
\newblock \emph{\bibinfo{journal}{The Journal of Chemical Physics}}
  \textbf{\bibinfo{volume}{141}}, \bibinfo{pages}{184308}
  (\bibinfo{year}{2014}).

\bibitem{peter_jpcm_2009}
\bibinfo{author}{Xiao, J.} \emph{et~al.}
\newblock \bibinfo{title}{Changes in the adsorbate dipole layer with changing
  d-filling of the metal (ii) (co, ni, cu) phthalocyanines on au(111)}.
\newblock \emph{\bibinfo{journal}{Journal of Physics: Condensed Matter}}
  \textbf{\bibinfo{volume}{21}}, \bibinfo{pages}{052001--}
  (\bibinfo{year}{2009}).

\bibitem{miranda_nl_2014}
\bibinfo{author}{Garnica, M.} \emph{et~al.}
\newblock \bibinfo{title}{Probing the site-dependent kondo response of
  nanostructured graphene with organic molecules}.
\newblock \emph{\bibinfo{journal}{Nano Letters}} \textbf{\bibinfo{volume}{14}},
  \bibinfo{pages}{4560} (\bibinfo{year}{2014}).

\bibitem{zheng2016heterointerface}
\bibinfo{author}{Zheng, Y.~J.} \emph{et~al.}
\newblock \bibinfo{title}{Heterointerface screening effects between organic
  monolayers and monolayer transition metal dichalcogenides}.
\newblock \emph{\bibinfo{journal}{ACS Nano}} \textbf{\bibinfo{volume}{10}},
  \bibinfo{pages}{2476--2484} (\bibinfo{year}{2016}).

\bibitem{crommie_prb_2004}
\bibinfo{author}{Lu, X.} \emph{et~al.}
\newblock \bibinfo{title}{Charge transfer and screening in individual
  {C}$_{60}$ molecules on metal substrates: a scanning tunneling spectroscopy
  and theoretical study}.
\newblock \emph{\bibinfo{journal}{Physical Review B}}
  \textbf{\bibinfo{volume}{70}}, \bibinfo{pages}{115418--}
  (\bibinfo{year}{2004}).

\bibitem{scardamaglia2011metal}
\bibinfo{author}{Scardamaglia, M.} \emph{et~al.}
\newblock \bibinfo{title}{Metal-phthalocyanine array on the moir{\'e} pattern
  of a graphene sheet}.
\newblock \emph{\bibinfo{journal}{Journal of Nanoparticle Research}}
  \textbf{\bibinfo{volume}{13}}, \bibinfo{pages}{6013--6020}
  (\bibinfo{year}{2011}).

\bibitem{scardamaglia2013graphene}
\bibinfo{author}{Scardamaglia, M.} \emph{et~al.}
\newblock \bibinfo{title}{Graphene-induced substrate decoupling and ideal
  doping of a self-assembled iron-phthalocyanine single layer}.
\newblock \emph{\bibinfo{journal}{The Journal of Physical Chemistry C}}
  \textbf{\bibinfo{volume}{117}}, \bibinfo{pages}{3019--3027}
  (\bibinfo{year}{2013}).

\bibitem{bogani2008molecular}
\bibinfo{author}{Bogani, L.} \emph{et~al.}
\newblock \bibinfo{title}{Molecular spintronics using single-molecule magnets}.
\newblock \emph{\bibinfo{journal}{Nature materials}}
  \textbf{\bibinfo{volume}{7}}, \bibinfo{pages}{179--186}
  (\bibinfo{year}{2008}).

\bibitem{graham_sci_2003}
\bibinfo{author}{Ouyang, Z.} \emph{et~al.}
\newblock \bibinfo{title}{Preparing protein microarrays by soft-landing of
  mass-selected ions}.
\newblock \emph{\bibinfo{journal}{Science}} \textbf{\bibinfo{volume}{301}},
  \bibinfo{pages}{1351--1354} (\bibinfo{year}{2003}).

\bibitem{oshea_nanotech_2007}
\bibinfo{author}{Satterley, C.~J.} \emph{et~al.}
\newblock \bibinfo{title}{Electrospray deposition of fullerenes in ultra-high
  vacuum: inÃ‚Â situ scanning tunneling microscopy and photoemission
  spectroscopy}.
\newblock \emph{\bibinfo{journal}{Nanotechnology}}
  \textbf{\bibinfo{volume}{18}}, \bibinfo{pages}{455304--}
  (\bibinfo{year}{2007}).

\bibitem{erler2015highly}
\bibinfo{author}{Erler, P.} \emph{et~al.}
\newblock \bibinfo{title}{Highly ordered surface self-assembly of fe4 single
  molecule magnets}.
\newblock \emph{\bibinfo{journal}{Nano Letters}} \textbf{\bibinfo{volume}{15}},
  \bibinfo{pages}{4546--4552} (\bibinfo{year}{2015}).

\bibitem{groning_acsn_2015}
\bibinfo{author}{Liu, L.} \emph{et~al.}
\newblock \bibinfo{title}{Interplay between energy-level position and charging
  effect of manganese phthalocyanines on an atomically thin insulator}.
\newblock \emph{\bibinfo{journal}{ACS Nano}} \textbf{\bibinfo{volume}{9}},
  \bibinfo{pages}{10125--10132} (\bibinfo{year}{2015}).

\bibitem{wingreen1989inelastic}
\bibinfo{author}{Wingreen, N.~S.} \emph{et~al.}
\newblock \bibinfo{title}{Inelastic scattering in resonant tunneling}.
\newblock \emph{\bibinfo{journal}{Physical Review B}}
  \textbf{\bibinfo{volume}{40}}, \bibinfo{pages}{11834} (\bibinfo{year}{1989}).

\bibitem{gadzuk1991inelastic}
\bibinfo{author}{Gadzuk, J.}
\newblock \bibinfo{title}{Inelastic resonance scattering, tunneling, and
  desorption}.
\newblock \emph{\bibinfo{journal}{Physical Review B}}
  \textbf{\bibinfo{volume}{44}}, \bibinfo{pages}{13466} (\bibinfo{year}{1991}).

\bibitem{galperin2007molecular}
\bibinfo{author}{Galperin, M.} \emph{et~al.}
\newblock \bibinfo{title}{Molecular transport junctions: vibrational effects}.
\newblock \emph{\bibinfo{journal}{Journal of Physics: Condensed Matter}}
  \textbf{\bibinfo{volume}{19}}, \bibinfo{pages}{103201}
  (\bibinfo{year}{2007}).

\bibitem{galperin2008nuclear}
\bibinfo{author}{Galperin, M.} \emph{et~al.}
\newblock \bibinfo{title}{Nuclear coupling and polarization in molecular
  transport junctions: beyond tunneling to function}.
\newblock \emph{\bibinfo{journal}{Science}} \textbf{\bibinfo{volume}{319}},
  \bibinfo{pages}{1056--1060} (\bibinfo{year}{2008}).

\bibitem{ho_prl_2005}
\bibinfo{author}{Pradhan, N.~A.} \emph{et~al.}
\newblock \bibinfo{title}{Atomic scale conductance induced by single impurity
  charging}.
\newblock \emph{\bibinfo{journal}{Physical Review Letters}}
  \textbf{\bibinfo{volume}{94}}, \bibinfo{pages}{076801--}
  (\bibinfo{year}{2005}).

\bibitem{wiesendanger_prb_2008}
\bibinfo{author}{Marczinowski, F.} \emph{et~al.}
\newblock \bibinfo{title}{Effect of charge manipulation on scanning tunneling
  spectra of single {Mn} acceptors in {InAs}}.
\newblock \emph{\bibinfo{journal}{Physical Review B}}
  \textbf{\bibinfo{volume}{77}}, \bibinfo{pages}{115318--}
  (\bibinfo{year}{2008}).

\bibitem{wu2004control}
\bibinfo{author}{Wu, S.~W.} \emph{et~al.}
\newblock \bibinfo{title}{Control of relative tunneling rates in single
  molecule bipolar electron transport}.
\newblock \emph{\bibinfo{journal}{Physical Review Letters}}
  \textbf{\bibinfo{volume}{93}}, \bibinfo{pages}{236802}
  (\bibinfo{year}{2004}).

\bibitem{kondo_physrev_1968}
\bibinfo{author}{Kondo, J.}
\newblock \bibinfo{title}{Effect of ordinary scattering on exchange scattering
  from magnetic impurity in metals}.
\newblock \emph{\bibinfo{journal}{Physical Review}}
  \textbf{\bibinfo{volume}{169}}, \bibinfo{pages}{437-- 440}
  (\bibinfo{year}{1968}).

\bibitem{Jamneala2000}
\bibinfo{author}{Jamneala, T.} \emph{et~al.}
\newblock \bibinfo{title}{Scanning tunneling spectroscopy of transition-metal
  impurities at the surface of gold}.
\newblock \emph{\bibinfo{journal}{Physical Review B}}
  \textbf{\bibinfo{volume}{61}}, \bibinfo{pages}{9990--9993}
  (\bibinfo{year}{2000}).

\bibitem{Madhavan2001}
\bibinfo{author}{Madhavan, V.} \emph{et~al.}
\newblock \bibinfo{title}{Local spectroscopy of a kondo impurity: Co on
  {A}u(111)}.
\newblock \emph{\bibinfo{journal}{Physical Review B}}
  \textbf{\bibinfo{volume}{64}}, \bibinfo{pages}{165412}
  (\bibinfo{year}{2001}).

\bibitem{Wahl2004}
\bibinfo{author}{Wahl, P.} \emph{et~al.}
\newblock \bibinfo{title}{Kondo temperature of magnetic impurities at
  surfaces}.
\newblock \emph{\bibinfo{journal}{Physical Review Letters}}
  \textbf{\bibinfo{volume}{93}}, \bibinfo{pages}{176603}
  (\bibinfo{year}{2004}).

\bibitem{Zhao2005}
\bibinfo{author}{Zhao, A.} \emph{et~al.}
\newblock \bibinfo{title}{Controlling the kondo effect of an adsorbed magnetic
  ion through its chemical bonding}.
\newblock \emph{\bibinfo{journal}{Science}} \textbf{\bibinfo{volume}{309}},
  \bibinfo{pages}{1542--1544} (\bibinfo{year}{2005}).

\bibitem{Iancu2006}
\bibinfo{author}{Iancu, V.} \emph{et~al.}
\newblock \bibinfo{title}{Manipulating kondo temperature via single molecule
  switching}.
\newblock \emph{\bibinfo{journal}{Nano Letters}} \textbf{\bibinfo{volume}{6}},
  \bibinfo{pages}{820--823} (\bibinfo{year}{2006}).

\bibitem{Iancu2006a}
\bibinfo{author}{Iancu, V.} \emph{et~al.}
\newblock \bibinfo{title}{Manipulation of the kondo effect via two-dimensional
  molecular assembly}.
\newblock \emph{\bibinfo{journal}{Physical Review Letters}}
  \textbf{\bibinfo{volume}{97}}, \bibinfo{pages}{266603}
  (\bibinfo{year}{2006}).

\bibitem{Gao2007}
\bibinfo{author}{Gao, L.} \emph{et~al.}
\newblock \bibinfo{title}{Site-specific kondo effect at ambient temperatures in
  iron-based molecules}.
\newblock \emph{\bibinfo{journal}{Physical Review Letters}}
  \textbf{\bibinfo{volume}{99}}, \bibinfo{pages}{106402}
  (\bibinfo{year}{2007}).

\bibitem{Mugarza2011}
\bibinfo{author}{Mugarza, A.} \emph{et~al.}
\newblock \bibinfo{title}{Spin coupling and relaxation inside molecule-metal
  contacts}.
\newblock \emph{\bibinfo{journal}{Nature Communication}}
  \textbf{\bibinfo{volume}{2}}, \bibinfo{pages}{490} (\bibinfo{year}{2011}).

\bibitem{DiLullo2012}
\bibinfo{author}{DiLullo, A.} \emph{et~al.}
\newblock \bibinfo{title}{Molecular kondo chain}.
\newblock \emph{\bibinfo{journal}{Nano Letters}} \textbf{\bibinfo{volume}{12}},
  \bibinfo{pages}{3174--3179} (\bibinfo{year}{2012}).

\bibitem{Wu2015}
\bibinfo{author}{Wu, F.} \emph{et~al.}
\newblock \bibinfo{title}{Modulation of the molecular spintronic properties of
  adsorbed copper corroles}.
\newblock \emph{\bibinfo{journal}{Nature Communication}}
  \textbf{\bibinfo{volume}{6}}, \bibinfo{pages}{--} (\bibinfo{year}{2015}).

\bibitem{Fernandez-Torrente2008}
\bibinfo{author}{Fernadez-Torrente, I.} \emph{et~al.}
\newblock \bibinfo{title}{Vibrational kondo effect in pure organic
  charge-transfer assemblies}.
\newblock \emph{\bibinfo{journal}{Physical Review Letters}}
  \textbf{\bibinfo{volume}{101}}, \bibinfo{pages}{217203}
  (\bibinfo{year}{2008}).

\bibitem{vojta_rpp_2013}
\bibinfo{author}{Fritz, L.} \emph{et~al.}
\newblock \bibinfo{title}{The physics of kondo impurities in graphene}.
\newblock \emph{\bibinfo{journal}{Rep. Prog. Phys.}}
  \textbf{\bibinfo{volume}{76}}, \bibinfo{pages}{032501}
  (\bibinfo{year}{2013}).

\bibitem{Wehling2010}
\bibinfo{author}{Wehling, T.~O.} \emph{et~al.}
\newblock \bibinfo{title}{Orbitally controlled kondo effect of co adatoms on
  graphene}.
\newblock \emph{\bibinfo{journal}{Physical Review B}}
  \textbf{\bibinfo{volume}{81}}, \bibinfo{pages}{115427}
  (\bibinfo{year}{2010}).

\bibitem{Vojta2010}
\bibinfo{author}{Vojta, M.} \emph{et~al.}
\newblock \bibinfo{title}{Gate-controlled kondo screening in graphene: Quantum
  criticality and electron-hole asymmetry}.
\newblock \emph{\bibinfo{journal}{EPL (Europhysics Letters)}}
  \textbf{\bibinfo{volume}{90}}, \bibinfo{pages}{27006} (\bibinfo{year}{2010}).

\bibitem{baskaran_prb_2008}
\bibinfo{author}{Sengupta, K.} \emph{et~al.}
\newblock \bibinfo{title}{Tuning kondo physics in graphene with gate voltage}.
\newblock \emph{\bibinfo{journal}{Physical Review B}}
  \textbf{\bibinfo{volume}{77}}, \bibinfo{pages}{045417--}
  (\bibinfo{year}{2008}).

\bibitem{Zhu2010}
\bibinfo{author}{Zhu, Z.-G.} \emph{et~al.}
\newblock \bibinfo{title}{Single- or multi-flavor kondo effect in graphene}.
\newblock \emph{\bibinfo{journal}{EPL (Europhysics Letters)}}
  \textbf{\bibinfo{volume}{90}}, \bibinfo{pages}{67001} (\bibinfo{year}{2010}).

\bibitem{Chen2011}
\bibinfo{author}{Chen, J.-H.} \emph{et~al.}
\newblock \bibinfo{title}{Tunable kondo effect in graphene with defects}.
\newblock \emph{\bibinfo{journal}{Nature Physics}}
  \textbf{\bibinfo{volume}{7}}, \bibinfo{pages}{535--538}
  (\bibinfo{year}{2011}).

\bibitem{heiko_nphy_2012}
\bibinfo{author}{Jobst, J.} \emph{et~al.}
\newblock \bibinfo{title}{Origin of logarithmic resistance correction in
  graphene}.
\newblock \emph{\bibinfo{journal}{Nat Phys}} \textbf{\bibinfo{volume}{8}},
  \bibinfo{pages}{352--352} (\bibinfo{year}{2012}).

\bibitem{fuhrer_natphys_2012}
\bibinfo{author}{Chen, J.-H.} \emph{et~al.}
\newblock \bibinfo{title}{Reply to "origin of logarithmic resistance correction
  in graphene"}.
\newblock \emph{\bibinfo{journal}{Nat Phys}} \textbf{\bibinfo{volume}{8}},
  \bibinfo{pages}{353--353} (\bibinfo{year}{2012}).

\bibitem{miranda_ss_2014}
\bibinfo{author}{Garnica, M.} \emph{et~al.}
\newblock \bibinfo{title}{Mapping spin distributions in electron acceptor
  molecules adsorbed on nanostructured graphene by the kondo effect}.
\newblock \emph{\bibinfo{journal}{Surface Science}}
  \textbf{\bibinfo{volume}{630}}, \bibinfo{pages}{356â€“360}
  (\bibinfo{year}{2014}).

\bibitem{gao_nl_2014}
\bibinfo{author}{Ren, J.} \emph{et~al.}
\newblock \bibinfo{title}{Kondo effect of cobalt adatoms on a graphene
  monolayer controlled by substrate-induced ripples}.
\newblock \emph{\bibinfo{journal}{Nano Letters}} \textbf{\bibinfo{volume}{14}},
  \bibinfo{pages}{4011--4015} (\bibinfo{year}{2014}).

\bibitem{sarma2011electronic}
\bibinfo{author}{Sarma, S.~D.} \emph{et~al.}
\newblock \bibinfo{title}{Electronic transport in two-dimensional graphene}.
\newblock \emph{\bibinfo{journal}{Reviews of Modern Physics}}
  \textbf{\bibinfo{volume}{83}}, \bibinfo{pages}{407} (\bibinfo{year}{2011}).

\bibitem{schwierz2010graphene}
\bibinfo{author}{Schwierz, F.}
\newblock \bibinfo{title}{Graphene transistors}.
\newblock \emph{\bibinfo{journal}{Nature Nanotechnology}}
  \textbf{\bibinfo{volume}{5}}, \bibinfo{pages}{487--496}
  (\bibinfo{year}{2010}).

\bibitem{li2008chemically}
\bibinfo{author}{Li, X.} \emph{et~al.}
\newblock \bibinfo{title}{Chemically derived, ultrasmooth graphene nanoribbon
  semiconductors}.
\newblock \emph{\bibinfo{journal}{Science}} \textbf{\bibinfo{volume}{319}},
  \bibinfo{pages}{1229--1232} (\bibinfo{year}{2008}).

\bibitem{han2007energy}
\bibinfo{author}{Han, M.~Y.} \emph{et~al.}
\newblock \bibinfo{title}{Energy band-gap engineering of graphene nanoribbons}.
\newblock \emph{\bibinfo{journal}{Physical Review Letters}}
  \textbf{\bibinfo{volume}{98}}, \bibinfo{pages}{206805}
  (\bibinfo{year}{2007}).

\bibitem{eroms2009weak}
\bibinfo{author}{Eroms, J.} \emph{et~al.}
\newblock \bibinfo{title}{Weak localization and transport gap in graphene
  antidot lattices}.
\newblock \emph{\bibinfo{journal}{New Journal of Physics}}
  \textbf{\bibinfo{volume}{11}}, \bibinfo{pages}{095021}
  (\bibinfo{year}{2009}).

\bibitem{giesbers2012charge}
\bibinfo{author}{Giesbers, A.} \emph{et~al.}
\newblock \bibinfo{title}{Charge transport gap in graphene antidot lattices}.
\newblock \emph{\bibinfo{journal}{Physical Review B}}
  \textbf{\bibinfo{volume}{86}}, \bibinfo{pages}{045445}
  (\bibinfo{year}{2012}).

\bibitem{kosynkin2009longitudinal}
\bibinfo{author}{Kosynkin, D.~V.} \emph{et~al.}
\newblock \bibinfo{title}{Longitudinal unzipping of carbon nanotubes to form
  graphene nanoribbons}.
\newblock \emph{\bibinfo{journal}{Nature}} \textbf{\bibinfo{volume}{458}},
  \bibinfo{pages}{872--876} (\bibinfo{year}{2009}).

\bibitem{tao2011spatially}
\bibinfo{author}{Tao, C.} \emph{et~al.}
\newblock \bibinfo{title}{Spatially resolving edge states of chiral graphene
  nanoribbons}.
\newblock \emph{\bibinfo{journal}{Nature Physics}}
  \textbf{\bibinfo{volume}{7}}, \bibinfo{pages}{616--620}
  (\bibinfo{year}{2011}).

\bibitem{llinas2016short}
\bibinfo{author}{Llinas, J.~P.} \emph{et~al.}
\newblock \bibinfo{title}{Short-channel field effect transistors with 9-atom
  and 13-atom wide graphene nanoribbons}.
\newblock \emph{\bibinfo{journal}{arXiv preprint arXiv:1605.06730}}
  (\bibinfo{year}{2016}).

\bibitem{hunt2013massive}
\bibinfo{author}{Hunt, B.} \emph{et~al.}
\newblock \bibinfo{title}{Massive {D}irac fermions and {H}ofstadter butterfly
  in a van der {W}aals heterostructure}.
\newblock \emph{\bibinfo{journal}{Science}} \textbf{\bibinfo{volume}{340}},
  \bibinfo{pages}{1427--1430} (\bibinfo{year}{2013}).

\bibitem{woods2014commensurate}
\bibinfo{author}{Woods, C.} \emph{et~al.}
\newblock \bibinfo{title}{Commensurate-incommensurate transition in graphene on
  hexagonal boron nitride}.
\newblock \emph{\bibinfo{journal}{Nature Physics}}
  \textbf{\bibinfo{volume}{10}}, \bibinfo{pages}{451--456}
  (\bibinfo{year}{2014}).

\bibitem{nair2010fluorographene}
\bibinfo{author}{Nair, R.~R.} \emph{et~al.}
\newblock \bibinfo{title}{Fluorographene: a two-dimensional counterpart of
  teflon}.
\newblock \emph{\bibinfo{journal}{Small}} \textbf{\bibinfo{volume}{6}},
  \bibinfo{pages}{2877--2884} (\bibinfo{year}{2010}).

\bibitem{panchakarla2009synthesis}
\bibinfo{author}{Panchakarla, L.} \emph{et~al.}
\newblock \bibinfo{title}{Synthesis, structure, and properties of boron-and
  nitrogen-doped graphene}.
\newblock \emph{\bibinfo{journal}{Advanced Materials}}
  \textbf{\bibinfo{volume}{21}}, \bibinfo{pages}{4726} (\bibinfo{year}{2009}).

\bibitem{wei2009synthesis}
\bibinfo{author}{Wei, D.} \emph{et~al.}
\newblock \bibinfo{title}{Synthesis of n-doped graphene by chemical vapor
  deposition and its electrical properties}.
\newblock \emph{\bibinfo{journal}{Nano Letters}} \textbf{\bibinfo{volume}{9}},
  \bibinfo{pages}{1752--1758} (\bibinfo{year}{2009}).

\bibitem{telychko2014achieving}
\bibinfo{author}{Telychko, M.} \emph{et~al.}
\newblock \bibinfo{title}{Achieving high-quality single-atom nitrogen doping of
  graphene/{SiC} (0001) by ion implantation and subsequent thermal
  stabilization}.
\newblock \emph{\bibinfo{journal}{ACS Nano}} \textbf{\bibinfo{volume}{8}},
  \bibinfo{pages}{7318--7324} (\bibinfo{year}{2014}).

\bibitem{sforzini2016structural}
\bibinfo{author}{Sforzini, J.} \emph{et~al.}
\newblock \bibinfo{title}{Structural and electronic properties of
  nitrogen-doped graphene}.
\newblock \emph{\bibinfo{journal}{Physical Review Letters}}
  \textbf{\bibinfo{volume}{116}}, \bibinfo{pages}{126805}
  (\bibinfo{year}{2016}).

\bibitem{sque2007transfer}
\bibinfo{author}{Sque, S.~J.} \emph{et~al.}
\newblock \bibinfo{title}{The transfer doping of graphite and graphene}.
\newblock \emph{\bibinfo{journal}{Physica Status Solidi A}}
  \textbf{\bibinfo{volume}{204}}, \bibinfo{pages}{3078--3084}
  (\bibinfo{year}{2007}).

\bibitem{wehling2008molecular}
\bibinfo{author}{Wehling, T.} \emph{et~al.}
\newblock \bibinfo{title}{Molecular doping of graphene}.
\newblock \emph{\bibinfo{journal}{Nano Letters}} \textbf{\bibinfo{volume}{8}},
  \bibinfo{pages}{173--177} (\bibinfo{year}{2008}).

\bibitem{rochefort2008interaction}
\bibinfo{author}{Rochefort, A.} \emph{et~al.}
\newblock \bibinfo{title}{Interaction of substituted aromatic compounds with
  graphene}.
\newblock \emph{\bibinfo{journal}{Langmuir}} \textbf{\bibinfo{volume}{25}},
  \bibinfo{pages}{210--215} (\bibinfo{year}{2008}).

\bibitem{hu2013theoretical}
\bibinfo{author}{Hu, T.} \emph{et~al.}
\newblock \bibinfo{title}{Theoretical study of the interaction of electron
  donor and acceptor molecules with graphene}.
\newblock \emph{\bibinfo{journal}{The Journal of Physical Chemistry C}}
  \textbf{\bibinfo{volume}{117}}, \bibinfo{pages}{2411--2420}
  (\bibinfo{year}{2013}).

\bibitem{park2008new}
\bibinfo{author}{Park, C.-H.} \emph{et~al.}
\newblock \bibinfo{title}{New generation of massless {D}irac fermions in
  graphene under external periodic potentials}.
\newblock \emph{\bibinfo{journal}{Physical Review Letters}}
  \textbf{\bibinfo{volume}{101}}, \bibinfo{pages}{126804}
  (\bibinfo{year}{2008}).

\bibitem{romero2009adsorption}
\bibinfo{author}{Romero, H.~E.} \emph{et~al.}
\newblock \bibinfo{title}{Adsorption of ammonia on graphene}.
\newblock \emph{\bibinfo{journal}{Nanotechnology}}
  \textbf{\bibinfo{volume}{20}}, \bibinfo{pages}{245501}
  (\bibinfo{year}{2009}).

\bibitem{farmer2008chemical}
\bibinfo{author}{Farmer, D.~B.} \emph{et~al.}
\newblock \bibinfo{title}{Chemical doping and electron- hole conduction
  asymmetry in graphene devices}.
\newblock \emph{\bibinfo{journal}{Nano Letters}} \textbf{\bibinfo{volume}{9}},
  \bibinfo{pages}{388--392} (\bibinfo{year}{2008}).

\bibitem{dong2009doping}
\bibinfo{author}{Dong, X.} \emph{et~al.}
\newblock \bibinfo{title}{Doping single-layer graphene with aromatic
  molecules}.
\newblock \emph{\bibinfo{journal}{Small}} \textbf{\bibinfo{volume}{5}},
  \bibinfo{pages}{1422--1426} (\bibinfo{year}{2009}).

\bibitem{moser2008environment}
\bibinfo{author}{Moser, J.} \emph{et~al.}
\newblock \bibinfo{title}{The environment of graphene probed by electrostatic
  force microscopy}.
\newblock \emph{\bibinfo{journal}{Applied Physics Letters}}
  \textbf{\bibinfo{volume}{92}}, \bibinfo{pages}{123507}
  (\bibinfo{year}{2008}).

\bibitem{sato2011electrically}
\bibinfo{author}{Sato, Y.} \emph{et~al.}
\newblock \bibinfo{title}{Electrically controlled adsorption of oxygen in
  bilayer graphene devices}.
\newblock \emph{\bibinfo{journal}{Nano Letters}} \textbf{\bibinfo{volume}{11}},
  \bibinfo{pages}{3468--3475} (\bibinfo{year}{2011}).

\bibitem{sun2010linear}
\bibinfo{author}{Sun, J.} \emph{et~al.}
\newblock \bibinfo{title}{Linear tuning of charge carriers in graphene by
  organic molecules and charge-transfer complexes}.
\newblock \emph{\bibinfo{journal}{Physical Review B}}
  \textbf{\bibinfo{volume}{81}}, \bibinfo{pages}{155403}
  (\bibinfo{year}{2010}).

\bibitem{prado2010two}
\bibinfo{author}{Prado, M.~C.} \emph{et~al.}
\newblock \bibinfo{title}{Two-dimensional molecular crystals of phosphonic
  acids on graphene}.
\newblock \emph{\bibinfo{journal}{ACS Nano}} \textbf{\bibinfo{volume}{5}},
  \bibinfo{pages}{394--398} (\bibinfo{year}{2010}).

\bibitem{Coletti2010}
\bibinfo{author}{Coletti, C.} \emph{et~al.}
\newblock \bibinfo{title}{Charge neutrality and band-gap tuning of epitaxial
  graphene on {SiC} by molecular doping}.
\newblock \emph{\bibinfo{journal}{Physical Review B}}
  \textbf{\bibinfo{volume}{81}}, \bibinfo{pages}{235401}
  (\bibinfo{year}{2010}).

\bibitem{li2013toward}
\bibinfo{author}{Li, B.} \emph{et~al.}
\newblock \bibinfo{title}{Toward tunable doping in graphene {FET}s by molecular
  self-assembled monolayers}.
\newblock \emph{\bibinfo{journal}{Nanoscale}} \textbf{\bibinfo{volume}{5}},
  \bibinfo{pages}{9640--9644} (\bibinfo{year}{2013}).

\bibitem{lauffer2008molecular}
\bibinfo{author}{Lauffer, P.} \emph{et~al.}
\newblock \bibinfo{title}{Molecular and electronic structure of {PTCDA} on
  bilayer graphene on {SiC}(0001) studied with scanning tunneling microscopy}.
\newblock \emph{\bibinfo{journal}{Physica Status Solidi B}}
  \textbf{\bibinfo{volume}{245}}, \bibinfo{pages}{2064--2067}
  (\bibinfo{year}{2008}).

\bibitem{das2008monitoring}
\bibinfo{author}{Das, A.} \emph{et~al.}
\newblock \bibinfo{title}{Monitoring dopants by raman scattering in an
  electrochemically top-gated graphene transistor}.
\newblock \emph{\bibinfo{journal}{Nature Nanotechnology}}
  \textbf{\bibinfo{volume}{3}}, \bibinfo{pages}{210--215}
  (\bibinfo{year}{2008}).

\bibitem{meissner2012highly}
\bibinfo{author}{Meissner, M.} \emph{et~al.}
\newblock \bibinfo{title}{Highly ordered growth of {PTCDA} on epitaxial bilayer
  graphene}.
\newblock \emph{\bibinfo{journal}{Surface Science}}
  \textbf{\bibinfo{volume}{606}}, \bibinfo{pages}{1709--1715}
  (\bibinfo{year}{2012}).

\bibitem{jee2009pentacene}
\bibinfo{author}{Jee, H.-g.} \emph{et~al.}
\newblock \bibinfo{title}{Pentacene as protection layers of graphene on {SiC}
  surfaces}.
\newblock \emph{\bibinfo{journal}{Applied Physics Letters}}
  \textbf{\bibinfo{volume}{95}}, \bibinfo{pages}{093107}
  (\bibinfo{year}{2009}).

\bibitem{Chen2007}
\bibinfo{author}{Chen, W.} \emph{et~al.}
\newblock \bibinfo{title}{Surface transfer p-type doping of epitaxial
  graphene}.
\newblock \emph{\bibinfo{journal}{Journal of the American Chemical Society}}
  \textbf{\bibinfo{volume}{129}}, \bibinfo{pages}{10418--10422}
  (\bibinfo{year}{2007}).

\bibitem{jnawali2015observation}
\bibinfo{author}{Jnawali, G.} \emph{et~al.}
\newblock \bibinfo{title}{Observation of ground-and excited-state charge
  transfer at the {C}$_{60}$/graphene interface}.
\newblock \emph{\bibinfo{journal}{ACS Nano}} \textbf{\bibinfo{volume}{9}},
  \bibinfo{pages}{7175--7185} (\bibinfo{year}{2015}).

\bibitem{pinto2009p}
\bibinfo{author}{Pinto, H.} \emph{et~al.}
\newblock \bibinfo{title}{p-type doping of graphene with {F}$_4${TCNQ}}.
\newblock \emph{\bibinfo{journal}{Journal of Physics: Condensed Matter}}
  \textbf{\bibinfo{volume}{21}}, \bibinfo{pages}{402001}
  (\bibinfo{year}{2009}).

\bibitem{manna2009tuning}
\bibinfo{author}{Manna, A.~K.} \emph{et~al.}
\newblock \bibinfo{title}{Tuning the electronic structure of graphene by
  molecular charge transfer: a computational study}.
\newblock \emph{\bibinfo{journal}{Chemistry--An Asian Journal}}
  \textbf{\bibinfo{volume}{4}}, \bibinfo{pages}{855--860}
  (\bibinfo{year}{2009}).

\bibitem{kanai2009determination}
\bibinfo{author}{Kanai, K.} \emph{et~al.}
\newblock \bibinfo{title}{Determination of electron affinity of electron
  accepting molecules}.
\newblock \emph{\bibinfo{journal}{Applied Physics A}}
  \textbf{\bibinfo{volume}{95}}, \bibinfo{pages}{309--313}
  (\bibinfo{year}{2009}).

\bibitem{choudhury2010xps}
\bibinfo{author}{Choudhury, D.} \emph{et~al.}
\newblock \bibinfo{title}{{XPS} evidence for molecular charge-transfer doping
  of graphene}.
\newblock \emph{\bibinfo{journal}{Chemical Physics Letters}}
  \textbf{\bibinfo{volume}{497}}, \bibinfo{pages}{66--69}
  (\bibinfo{year}{2010}).

\bibitem{voggu2008effects}
\bibinfo{author}{Voggu, R.} \emph{et~al.}
\newblock \bibinfo{title}{Effects of charge transfer interaction of graphene
  with electron donor and acceptor molecules examined using raman spectroscopy
  and cognate techniques}.
\newblock \emph{\bibinfo{journal}{Journal of Physics: Condensed Matter}}
  \textbf{\bibinfo{volume}{20}}, \bibinfo{pages}{472204}
  (\bibinfo{year}{2008}).

\bibitem{xu2012investigating}
\bibinfo{author}{Xu, H.} \emph{et~al.}
\newblock \bibinfo{title}{Investigating the mechanism of hysteresis effect in
  graphene electrical field device fabricated on sio2 substrates using raman
  spectroscopy}.
\newblock \emph{\bibinfo{journal}{Small}} \textbf{\bibinfo{volume}{8}},
  \bibinfo{pages}{2833--2840} (\bibinfo{year}{2012}).

\bibitem{wang2011quantitative}
\bibinfo{author}{Wang, X.} \emph{et~al.}
\newblock \bibinfo{title}{Quantitative analysis of graphene doping by organic
  molecular charge transfer}.
\newblock \emph{\bibinfo{journal}{The Journal of Physical Chemistry C}}
  \textbf{\bibinfo{volume}{115}}, \bibinfo{pages}{7596--7602}
  (\bibinfo{year}{2011}).

\bibitem{kim2014infrared}
\bibinfo{author}{Kim, N.~W.} \emph{et~al.}
\newblock \bibinfo{title}{Infrared spectroscopy of large scale single layer
  graphene on self assembled organic monolayer}.
\newblock \emph{\bibinfo{journal}{Applied Physics Letters}}
  \textbf{\bibinfo{volume}{104}}, \bibinfo{pages}{041904}
  (\bibinfo{year}{2014}).

\bibitem{lafkioti2010graphene}
\bibinfo{author}{Lafkioti, M.} \emph{et~al.}
\newblock \bibinfo{title}{Graphene on a hydrophobic substrate: doping reduction
  and hysteresis suppression under ambient conditions}.
\newblock \emph{\bibinfo{journal}{Nano Letters}} \textbf{\bibinfo{volume}{10}},
  \bibinfo{pages}{1149--1153} (\bibinfo{year}{2010}).

\bibitem{lee2011control}
\bibinfo{author}{Lee, W.~H.} \emph{et~al.}
\newblock \bibinfo{title}{Control of graphene field-effect transistors by
  interfacial hydrophobic self-assembled monolayers}.
\newblock \emph{\bibinfo{journal}{Advanced Materials}}
  \textbf{\bibinfo{volume}{23}}, \bibinfo{pages}{3460--3464}
  (\bibinfo{year}{2011}).

\bibitem{park2011work}
\bibinfo{author}{Park, J.} \emph{et~al.}
\newblock \bibinfo{title}{Work-function engineering of graphene electrodes by
  self-assembled monolayers for high-performance organic field-effect
  transistors}.
\newblock \emph{\bibinfo{journal}{The Journal of Physical Chemistry Letters}}
  \textbf{\bibinfo{volume}{2}}, \bibinfo{pages}{841--845}
  (\bibinfo{year}{2011}).

\bibitem{yokota2011carrier}
\bibinfo{author}{Yokota, K.} \emph{et~al.}
\newblock \bibinfo{title}{Carrier control of graphene driven by the proximity
  effect of functionalized self-assembled monolayers}.
\newblock \emph{\bibinfo{journal}{Nano Letters}} \textbf{\bibinfo{volume}{11}},
  \bibinfo{pages}{3669--3675} (\bibinfo{year}{2011}).

\bibitem{yan2011controlled}
\bibinfo{author}{Yan, Z.} \emph{et~al.}
\newblock \bibinfo{title}{Controlled modulation of electronic properties of
  graphene by self-assembled monolayers on {SiO}$_2$ substrates}.
\newblock \emph{\bibinfo{journal}{ACS Nano}} \textbf{\bibinfo{volume}{5}},
  \bibinfo{pages}{1535--1540} (\bibinfo{year}{2011}).

\bibitem{park2012single}
\bibinfo{author}{Park, J.} \emph{et~al.}
\newblock \bibinfo{title}{Single-gate bandgap opening of bilayer graphene by
  dual molecular doping}.
\newblock \emph{\bibinfo{journal}{Advanced Materials}}
  \textbf{\bibinfo{volume}{24}}, \bibinfo{pages}{407--411}
  (\bibinfo{year}{2012}).

\bibitem{zhang2009direct}
\bibinfo{author}{Zhang, Y.} \emph{et~al.}
\newblock \bibinfo{title}{Direct observation of a widely tunable bandgap in
  bilayer graphene}.
\newblock \emph{\bibinfo{journal}{Nature}} \textbf{\bibinfo{volume}{459}},
  \bibinfo{pages}{820--823} (\bibinfo{year}{2009}).

\bibitem{joshi2010intrinsic}
\bibinfo{author}{Joshi, P.} \emph{et~al.}
\newblock \bibinfo{title}{Intrinsic doping and gate hysteresis in graphene
  field effect devices fabricated on sio2 substrates}.
\newblock \emph{\bibinfo{journal}{Journal of Physics: Condensed Matter}}
  \textbf{\bibinfo{volume}{22}}, \bibinfo{pages}{334214}
  (\bibinfo{year}{2010}).

\bibitem{zhang2011opening}
\bibinfo{author}{Zhang, W.} \emph{et~al.}
\newblock \bibinfo{title}{Opening an electrical band gap of bilayer graphene
  with molecular doping}.
\newblock \emph{\bibinfo{journal}{ACS Nano}} \textbf{\bibinfo{volume}{5}},
  \bibinfo{pages}{7517--7524} (\bibinfo{year}{2011}).

\bibitem{lee2015chemically}
\bibinfo{author}{Lee, S.~Y.} \emph{et~al.}
\newblock \bibinfo{title}{Chemically modulated band gap in bilayer graphene
  memory transistors with high on/off ratio}.
\newblock \emph{\bibinfo{journal}{ACS Nano}} \textbf{\bibinfo{volume}{9}},
  \bibinfo{pages}{9034--9042} (\bibinfo{year}{2015}).

\end{thebibliography}

\clearpage

\end{document}